\documentclass[fleqn,usenatbib,useAMS]{mnras}

\usepackage{graphicx}	
\usepackage{amsmath}	
\usepackage{multicol}        
\usepackage{bm}		
\usepackage{pdflscape}	
\usepackage{orcidlink}

\usepackage{gensymb}
\usepackage{subfig}
\usepackage{multirow}

\usepackage[T1]{fontenc}
\usepackage{ae,aecompl}

\usepackage{newtxtext,newtxmath}

\title[Steeper radio spectral slopes in dusty QSOs]{Connection between steep radio spectral slopes and dust extinction in QSOs: evidence for outflow-driven shocks in dusty QSOs}

\author[V. A. Fawcett et al.]
{V. A. Fawcett$^{\orcidlink{0000-0003-1251-532X}}$,$^{1}$\thanks{E-mail: vicky.fawcett@newcastle.ac.uk}
{C. M. Harrison$^{\orcidlink{0000-0001-8618-4223}}$,$^{1}$}
{D. M. Alexander$^{\orcidlink{0000-0002-5896-6313}}$,$^{2}$}
{L. K. Morabito$^{\orcidlink{0000-0003-0487-6651}}$,$^{2}$}
{P. Kharb$^{\orcidlink{0000-0003-3203-1613}}$,$^{3}$}
\newauthor
{D. J. Rosario$^{\orcidlink{0000-0002-0001-3587}}$,$^{1}$}
{Janhavi Baghel$^{\orcidlink{0000-0002-0367-812X}}$,$^{3}$}
{Salmoli Ghosh$^{\orcidlink{0009-0000-1447-5419}}$,$^{3}$}
{Silpa S.$^{\orcidlink{0000-0003-0667-7074}}$,$^{4}$}
{J. Petley$^{\orcidlink{0000-0002-4496-0754}}$,$^{2,5}$}
{C. Sargent$^{2}$}
\newauthor
{and G. Calistro Rivera$^{\orcidlink{0000-0003-0085-6346}}$$^{6}$}
\\
$^{1}$School of Mathematics, Statistics and Physics, Newcastle University, NE1 7RU, UK\\
$^{2}$Centre for Extragalactic Astronomy, Department of Physics, Durham University, South Road, Durham, DH1 3LE, UK\\
$^{3}$National Centre for Radio Astrophysics (NCRA) - Tata Institute of Fundamental Research (TIFR), S. P. Pune University Campus, Post Bag 3,\\ Ganeshkhind, Pune 411007, India\\
$^{4}$Departamento de Astronomía, Avenida Esteban Iturra s/n, Casilla 160-C, Universidad de Concepción, Concepción, Chile\\
$^{5}$Leiden Observatory, Leiden University, PO Box 9513, 2300 RA Leiden, The Netherlands\\
$^{6}$German Aerospace Center (DLR), Institute of Communications and Navigation, Wessling, Germany\\
}
\date{}

\pubyear{2024}

\defcitealias{rosario_21}{R21}
\defcitealias{klindt}{K19}

\begin{document}
\label{firstpage}
\pagerange{\pageref{firstpage}--\pageref{lastpage}}
\maketitle

\begin{abstract}
Recent studies have found a striking positive correlation between the amount of dust obscuration and enhanced radio emission in quasi-stellar objects (QSOs). However, what causes this connection remains unclear. In this paper we analyse uGMRT Band-3 (400\,MHz) and Band-4 (650\,MHz) data of a sample of 38 $1.0$\,$<$\,$z$\,$<$\,$1.5$ QSOs with existing high-resolution $0\farcs2$ e-MERLIN 1.4\,GHz imaging. In combination with archival radio data, we have constructed sensitive 4--5 band radio SEDs across $0.144$--$3$\,GHz to further characterize the radio emission in dusty QSOs. We find that the dusty QSOs (those with $E(B-V)$\,$>$\,0.1\,mag) are more likely to exhibit steep spectral slopes ($\alpha$\,$<$\,$-0.5$; $S$\,$\propto$\,$\nu^{\alpha}$) than the non-dusty QSOs ($E(B-V)$\,$<$\,0.1\,mag), with fractions of 46$\pm$12  and 12$\pm$4 per~cent, respectively. A higher fraction of the non-dusty QSOs have peaked radio SEDs (48$\pm$9 per~cent) compared to the dusty QSOs (23$\pm$8 per~cent). We discuss the origin of the radio emission, finding that the majority of the peaked, predominantly non-dusty, QSOs have consistent sizes and luminosities with compact jetted radio galaxies. However, the connection between steepness and dust obscuration implies an outflow-driven shock origin for the enhanced radio more commonly found in dusty QSOs. These results add to the emerging picture whereby dusty QSOs are in an earlier blow-out phase, with shocks that heat and destroy the surrounding dust, eventually revealing a typical non-dusty QSO.
\end{abstract}


\section{Introduction}

Quasi-stellar objects (QSOs) are the most powerful class of active galactic nuclei (AGN), and are often classified by their extremely high bolometric luminosities ($10^{45-48}$\,erg\,s$^{-1}$). The majority of QSOs are optically very blue due to an unobscured view of the accretion disc which has emission that peaks in the optical/ultra-violet (UV). However, a subset have been found to display much redder colours (``red QSOs''; \citealt{Webster1995}). There are several different explanations for the red colours in these QSOs, such as a moderate viewing angle through the dusty torus \citep{wilkes,rose}, as suggested by the unification model of AGN \citep{urry}, an intrinsically red continuum \citep{whiting,young}, an unusual covering factor of hot dust \citep{rose_2014}, a strong synchrotron component \citep{whiting}, or dust within the host-galaxy extinguishing the emission at shorter wavelengths \citep{glik12,kim18,fawcett22,kim_2024_dust}. Interestingly, simulations of galaxy formation predict that these red QSOs could be in a short-lived ``blow-out'' phase, whereby powerful outflows interact and shock the surrounding dust and gas, eventually clearing it and revealing a typical unobscured (blue) QSO \citep{hop6,hop}. If red QSOs do indeed represent a transitional phase in galaxy evolution then they would provide excellent laboratories to study AGN ``feedback'' in action (see review by \citealt{harrison_2024}). 

Radio observations provide a powerful tool for distinguishing between different red QSO scenarios, since radio emission is not affected by dust. Therefore, if red and blue QSOs are intrinsically the same objects (as suggested by the orientation scenario) then we would not expect to observe any differences in the radio emission (apart from potentially enhanced radio emission in blue QSOs due to Doppler boosting of a more face-on radio jet; e.g., \citealt{lahteenm}). Indeed, evidence supporting the red QSO blow-out phase scenario over the orientation scenario has largely come from differences found in the radio properties compared to blue QSOs. For example, studies have found that red QSOs are more likely to be detected in the radio compared to redshift and luminosity-matched blue QSOs, which has been observed with radio data across a large range of frequencies, spatial resolutions, and sensitivities \citep{richards,georg,ban12,klindt,glik12,glikman22,rosario,fawcett20,fawcett21}. This enhanced radio emission in red QSOs has been found to be predominantly compact, on host-galaxy scales (i.e., $\lesssim$\,10\,kpc) and tends to be radio-quiet/intermediate (i.e., they don't host large-scale powerful radio jets; \citealt{klindt,fawcett20,rosario_21}). Previous work has also found that star formation is unlikely the dominant mechanism, especially for more luminous QSOs \citep{fawcett20,rosario,calistro,andonie,bohan}. These results point towards either low-powered radio jets (e.g., \citealt{girdhar}) or outflow-driven shocks (e.g., \citealt{stepney,haidar}) as the origin of the radio emission in red QSOs (see \cite{pan} for a review on different radio emission mechanisms).

Building on this previous work, \citet{fawcett23} explored the radio detection rate of QSOs from the Dark Energy Spectroscopic Instrument (DESI; \citealt{desi,desiII}) as a function of dust extinction and found a striking positive correlation (confirmed by \citealt{calistro_2024,petley_2024}). This suggests a causal link between opacity and the production of radio emission in QSOs, which is likely due to shocks from outflows, which may be driven by radiation pressure, accretion disc winds, and/or jets on the surrounding interstellar medium (ISM); see review by \cite{harrison_2024}. These shocks will heat and eventually destroy the dust, reducing the obscuration and, therefore, make the QSO appear bluer. This scenario is further supported by \cite{calistro}, who found a hot dust excess and stronger [\ion{O}{iii}]$\lambda$5007 outflows in the red QSOs, suggesting dust is getting heated by outflow-driven shocks. Indeed, this may be a more extreme example of the spatially-resolved connection observed between outflows, dust, and shocks in nearby AGN (e.g., \citealt{haidar}). If synchrotron radiation from a shocked dusty environment was found to be the driving mechanism behind the radio emission in red QSOs, then this would present strong evidence for the dusty red QSO blow-out phase.

One of the most effective ways to determine the radio emission mechanism is high spatial resolution radio imaging (e.g., Very Long Baseline Interferometry; VLBI). For example, a highly collimated structure and/or hot spots would likely be associated with a radio jet, whereas a less collimated, more diffuse structure is likely to be an outflow or star formation \citep{pan,Kharb2021,njeri,chen_24}. However, this technique has its limitations. For example, if the jet is inclined towards the accretion disc then this can drastically alter the morphology of the resulting radio emission, making the origin more unclear \citep{mukherjee,meenakshi}. To compare the $\sim$\,kpc scale radio morphology of red and blue QSOs, \cite{rosario_21} obtained $0\farcs2$ e-MERLIN 1.4\,GHz images of 20 red and 20 blue (selected based on $g-i$ colours) luminous ($L_{\rm bol}\approx10^{46-47}$\,erg\,s$^{-1}$) QSOs at intermediate redshifts ($1.0$\,$<$\,$z$\,$<$\,$1.55$) with unresolved, detected radio emission in the Faint Images of the Radio Sky at Twenty-centimeters (FIRST; \citealt{becker}; $5''$ resolution; $L_{\rm 1.4\,GHz}\approx10^{25-26}$\,W\,Hz$^{-1}$). At these redshifts the host-galaxy scale radio emission could be probed; they found a statistically significant difference in the incidence of $\sim$\,2--10\,kpc-scale extended radio emission in the red QSOs compared to the blue QSOs which indicated that the enhanced radio emission in red QSOs was likely due to outflows or low-powered jets that are confined within the host galaxy \citep{rosario_21}. However, despite the impressive angular resolution of e-MERLIN (25$\times$ better than FIRST) the majority of the sample ($\sim$\,$85$ per~cent) remained unresolved. 

To understand what mechanism drives these $\sim$kpc-scale unresolved radio structures (and to gain additional insight into the nature of the resolved sources), analyzing radio spectral energy distributions (SEDs) can be an effective method. For example, older radio emission produced by shocks from jets or winds (optically thin synchrotron emission) would result in a steep radio spectral slope \citep{faucher,nims,laor} and a young, self-absorbed radio nucleus (optically thick synchrotron emission) would result in an inverted or flat spectrum \citep{blandford_79}. 
In particular, a well-studied class of objects known as compact steep spectrum (CSS) and gigahertz-peaked spectrum (GPS) sources can be identified by a spectral turnover in their SED around $\sim$\,100\,MHz and $\sim$\,1\,GHz, respectively (see reviews by \citealt{odea_1997,o_dea_21}). These systems likely host compact jets that are thought to be either ``frustrated'' (i.e., confined by the galaxy ISM) on scales of $\sim$0.1--2\,kpc \citep{van_bregel,odea_1991,orienti_16}, in a young evolutionary phase \citep{phillips,carvalho,bicknell_18}, or both \citep{Patil_2020}. The frequency of the spectral turnover has been found to strongly anti-correlate with the projected linear size of the radio emission ($\nu$\,$\propto$\,$l^{-0.65}$; e.g., \citealt{fanti,odea_1997,gps}) which is expected for synchrotron self-absorption \citep{snellen}.\footnote{Note: free-free absorption by ionised gas surrounding the radio emission has also been suggested to explain this correlation (e.g., \citealt{Bicknell_1997,stawarz}). Additionally, the location of this peak can also be affected by the magnetic field strength \citep{duffy}.} Therefore, a peak in the radio SED of a QSO would likely suggest the presence of a compact radio jet.

Many studies have explored the radio spectral slopes of different classes of QSOs (e.g., \citealt{callingham,laor,radio_slope,shao,kukreti,hayashi}), including between red and blue QSOs (\citealt{georg12,rosario,glikman22}; Sargent et~al. \textit{in prep}). For example, utilizing data from FIRST and the VLA Sky Survey (VLASS; \citealt{vlass}), \cite{glikman22} explored the 1.4--3\,GHz spectral slopes and found that infrared (IR)-selected red QSOs were on average steeper compared to blue QSOs. On the other hand, utilizing data from the LOw-Frequency ARray (LOFAR; \citealt{lofar}) Two-metre Sky Survey (LoTSS; \citealt{lotss}) and FIRST, \cite{rosario} found no differences between the lower frequency 0.144--1.4\,GHz radio spectral slopes of optically-selected red and blue QSOs. This could suggest that the radio spectral slope of red QSOs only deviates at higher frequencies or that differences only arise in the more heavily reddened QSOs studied in \cite{glikman22}. However, both of these studies are limited to two data points in frequency. Without additional data points information can be lost; e.g., whether or not there is a spectral turnover and, if so, the exact location of this turnover \citep{Patil_2022}. Radio observations at additional frequencies can be acquired, but usually only for small samples. Therefore, a combination of both approaches is needed to fully understand the nature of the radio emission in different QSO populations.

In this paper we construct sensitive radio SEDs from 144\,MHz--3\,GHz of the QSOs that have been observed at $0\farcs2$ with e-MERLIN \citep{rosario_21}, by combining dedicated observations at 400\,MHz and 650\,MHz from the upgraded Giant Metrewave Radio Telescope (uGMRT; \citealt{gmrt,ugmrt}) and archival radio data. Using these data we can carefully explore the differences, if any, between the radio SEDs of dusty red and typical blue QSOs. If found, this would indicate differences in either the driving mechanism (i.e., outflow-driven shocks, frustrated jets, etc.) or evolution (i.e., young/old) of the radio emission. The presence of a spectral turnover can also determine whether the QSOs are similar to GPS/CSS-like sources, by comparing their peak frequency and measured e-MERLIN sizes to the known anti-correlation for GPS/CSS sources. Finally, finding steep radio spectral slopes in the dusty QSOs would be consistent with the shocked dust scenario as the origin of the enhanced radio emission found in red QSOs. In Section~\ref{sec:method} we describe the sample selection, data utilized in this paper, and SED fitting procedure. In Section~\ref{sec:results} we present our results and in Section~\ref{sec:discussion} we discuss the origin of the radio emission. In this paper we define spectral index $\alpha$ as $S_{\nu}$\,$\propto$\,$\nu^{\alpha}$, where $S_{\nu}$ is the flux density at frequency $\nu$. Throughout our work we adopt a standard flat $\uplambda$-cosmology with $H_0$\,$=$\,70~km\,s$^{-1}$Mpc$^{-1}$, $\Omega$\textsubscript{M}\,$=$\,0.3~and~$\Omega_{\uplambda}$\,$=$\,0.7 (e.g., \citealt{planck_2020}).

\section{Methods}\label{sec:method}
In this paper we explore the radio SEDs of SDSS-selected QSOs with complementary e-MERLIN 1.4\,GHz imaging (\citealt{rosario_21}; hereafter, \citetalias{rosario_21}). To construct the radio SEDs, we combined observations from the uGMRT in addition to archival radio datasets. In the following sections we describe the sample selection (Section~\ref{sec:sample}), the uGMRT observations and data reduction process (Section~\ref{sec:reduction}), the various archival radio datasets utilized (Sections~\ref{sec:archival} and \ref{sec:archival2}), how radio-loudness and radio luminosity are calculated (Section~\ref{sec:rad_lum}), the radio SED fitting method (Section~\ref{sec:model_characterisation}), and the dust extinction fitting method (Section~\ref{sec:dust_method}).

\subsection{Sample selection}\label{sec:sample}
\begin{table*}
\centering
\caption{Table displaying the basic properties of our rQSO and cQSO samples. The columns from left to right display the: (1) SDSS name, (2) whether the QSO is part of the red (rQSO) or blue (cQSO) sample, (3) RA and Dec from Gaia DR3 \citep{gaia_dr3}, (4) redshift from SDSS, (5) 6$\upmu$m luminosity ($L_{\rm 6\umu m}$), (6) 1.4\,GHz luminosity from FIRST ($L_{\rm 1.4\,GHz}$), (7) radio-loudness ($\mathcal{R}$), (8) measured dust extinction ($E(B-V)$), (9) $\Delta(g-i)$, and (10) whether the source shows extended radio emission in the $0\farcs2$ e-MERLIN imaging (e-MERLIN Ext.?). \newline \textsuperscript{\textdagger}Using a representative $\alpha$\,$=$\,$-0.5$ for the \textit{K}-correction.}
\begin{tabular}{|c|c|c|c|c|c|c|c|c|c|c|}
\hline
\hline
Name & Sample & RA & Dec & $z$ & $L_{\rm 6\upmu m}$ & $L_{\rm 1.4 GHz}$\textsuperscript{\textdagger} & $\mathcal{R}$ & $E(B-V)$ & $\Delta(g-i)$ & e-MERLIN Ext.?  \\
& & & & & log[erg\,s$^{-1}$] & log[W\,Hz$^{-1}$] & & [mag] & &  \\
\hline
0823+5609 & rQSO & 08 23 14.7 & +56 09 48.9 & 1.44 & 45.1 & 25.7 & -3.3 & 0.13$\pm$0.02 & 0.46 &  \\
0828+2731 & rQSO & 08 28 37.7 & +27 31 36.9 & 1.48 & 45.4 & 25.5 & -3.7 & 0.08$\pm$0.01 & 0.33 &  \\
0946+2548 & rQSO & 09 46 15.0 & +25 48 42.0 & 1.19 & 45.8 & 25.6 & -4.0 & 0.27$\pm$0.03 & 0.84 & Y \\
0951+5253 & rQSO & 09 51 14.3 & +52 53 16.7 & 1.43 & 45.9 & 26.0 & -3.7 & 0.27$\pm$0.03 & 0.95 &  \\
1007+2853 & rQSO & 10 07 13.6 & +28 53 48.4 & 1.05 & 46.1 & 25.6 & -4.4 & 0.59$\pm$0.06 & 1.58 & Y \\
1057+3119 & rQSO & 10 57 05.1 & +31 19 07.8 & 1.33 & 45.8 & 26.0 & -3.6 & 0.06$\pm$0.01 & 0.27 &  \\
1122+3124 & rQSO & 11 22 20.4 & +31 24 40.9 & 1.45 & 45.7 & 26.0 & -3.6 & 0.13$\pm$0.02 & 0.36 & Y \\
1140+4416 & rQSO & 11 40 46.8 & +44 16 09.8 & 1.41 & 45.4 & 25.8 & -3.4 & 0.06$\pm$0.01 & 0.36 &  \\
1153+5651 & rQSO & 11 53 13.0 & +56 51 26.3 & 1.20 & 45.9 & 26.2 & -3.5 & 0.21$\pm$0.03 & 0.83 & Y \\
1159+2151 & rQSO & 11 59 24.0 & +21 51 03.0 & 1.05 & 45.4 & 25.5 & -3.8 & 0.17$\pm$0.02 & 0.36 &  \\
1202+6317 & rQSO & 12 02 01.9 & +63 17 59.4 & 1.48 & 45.5 & 25.6 & -3.7 & 0.05$\pm$0.01 & 0.35 &  \\
1211+2221 & rQSO & 12 11 01.8 & +22 21 06.7 & 1.30 & 46.0 & 25.5 & -4.4 & 0.09$\pm$0.01 & 0.36 &  \\
1251+4317 & rQSO & 12 51 46.3 & +43 17 29.7 & 1.45 & 45.3 & 26.1 & -3.0 & 0.13$\pm$0.02 & 0.32 &  \\
1315+2017 & rQSO & 13 15 56.3 & +20 17 01.6 & 1.43 & 45.8 & 25.6 & -4.1 & 0.07$\pm$0.01 & 0.45 &  \\
1323+3948 & rQSO & 13 23 04.2 & +39 48 55.0 & 1.28 & 45.5 & 25.8 & -3.5 & 0.11$\pm$0.01 & 0.32 &  \\
1342+4326 & rQSO & 13 42 36.9 & +43 26 32.1 & 1.04 & 45.6 & 26.0 & -3.4 & 0.42$\pm$0.05 & 1.19 &  \\
1410+4016 & rQSO & 14 10 53.1 & +40 16 18.5 & 1.04 & 45.0 & 25.6 & -3.2 & 0.11$\pm$0.01 & 0.36 &  \\
1531+4528 & rQSO & 15 31 33.5 & +45 28 41.6 & 1.02 & 45.5 & 25.4 & -3.9 & 0.22$\pm$0.03 & 0.62 &  \\
1535+2434 & rQSO & 15 35 55.2 & +24 34 28.6 & 1.08 & 45.4 & 25.7 & -3.6 & 0.18$\pm$0.02 & 0.86 & Y \\
\hline
0748+2200 & cQSO & 07 48 15.4 & +22 00 59.4 & 1.06 & 46.0 & 25.6 & -4.3 & 0.01$\pm$0.003 & -0.01 &  \\
1003+2727 & cQSO & 10 03 18.9 & +27 27 34.3 & 1.29 & 45.9 & 25.6 & -4.1 & 0.03$\pm$0.01 & 0.05  &   \\
1019+2817 & cQSO & 10 19 35.2 & +28 17 38.9 & 1.01 & 45.5 & 25.9 & -3.5 & 0.08$\pm$0.01 & 0.07  &   \\
1038+4155 & cQSO & 10 38 50.8 & +41 55 12.7 & 1.47 & 46.0 & 25.6 & -4.3 & 0.01$\pm$0.01 & 0.01  &   \\
1042+4834 & cQSO & 10 42 40.1 & +48 34 03.4 & 1.04 & 45.6 & 25.9 & -3.5 & 0.05$\pm$0.01 & 0.00   &  \\
1046+3427 & cQSO & 10 46 20.1 & +34 27 08.4 & 1.20 & 45.8 & 26.3 & -3.3 & 0.02$\pm$0.004 & -0.01 &  \\
1057+3315 & cQSO & 10 57 36.1 & +33 15 45.9 & 1.47 & 45.3 & 25.4 & -3.8 & 0.02$\pm$0.003 & 0.06  &  \\
1103+5849 & cQSO & 11 03 52.4 & +58 49 23.5 & 1.33 & 45.7 & 25.6 & -3.9 & 0.01$\pm$0.01 & -0.04 &   \\
1203+4510 & cQSO & 12 03 35.3 & +45 10 49.5 & 1.08 & 45.3 & 26.2 & -3.0 & 0.06$\pm$0.01 & 0.09  &   \\
1222+3723 & cQSO & 12 22 21.3 & +37 23 35.8 & 1.26 & 45.3 & 25.8 & -3.4 & 0.02$\pm$0.004 & -0.03 &  \\
1304+3206 & cQSO & 13 04 33.4 & +32 06 35.5 & 1.34 & 45.7 & 25.9 & -3.7 & 0.06$\pm$0.01 & -0.01 &   \\
1410+2217 & cQSO & 14 10 27.5 & +22 17 02.6 & 1.42 & 45.5 & 26.3 & -3.1 & 0.04$\pm$0.01 & 0.01  &  Y \\
1428+2916 & cQSO & 14 28 24.7 & +29 16 06.7 & 1.04 & 45.2 & 26.2 & -2.9 & 0.04$\pm$0.01 & 0.00   &  \\
1432+2925 & cQSO & 14 32 49.5 & +29 25 05.7 & 1.04 & 45.4 & 25.8 & -3.4 & 0.05$\pm$0.01 & -0.01 &   \\
1530+2310 & cQSO & 15 30 44.0 & +23 10 13.4 & 1.41 & 46.1 & 25.6 & -4.4 & 0.02$\pm$0.004 & -0.07 &  \\
1554+2859 & cQSO & 15 54 36.6 & +28 59 42.5 & 1.19 & 46.0 & 25.5 & -4.3 & 0.02$\pm$0.005 & 0.01  &  \\
1602+4530 & cQSO & 16 02 45.9 & +45 30 50.3 & 1.41 & 45.6 & 25.6 & -3.9 & 0.02$\pm$0.005 & 0.12  &  \\
1630+3847 & cQSO & 16 30 23.1 & +38 47 00.7 & 1.53 & 45.1 & 26.0 & -3.0 & 0.05$\pm$0.01 & 0.09  &   \\
1657+2045 & cQSO & 16 57 24.8 & +20 45 59.5 & 1.47 & 45.5 & 25.9 & -3.5 & 0.01$\pm$0.003 & 0.00   &  \\
\hline
\hline
\end{tabular}
\label{tab:obs}
\end{table*}

\begin{figure}
    \centering
    \includegraphics[width=0.43\textwidth]{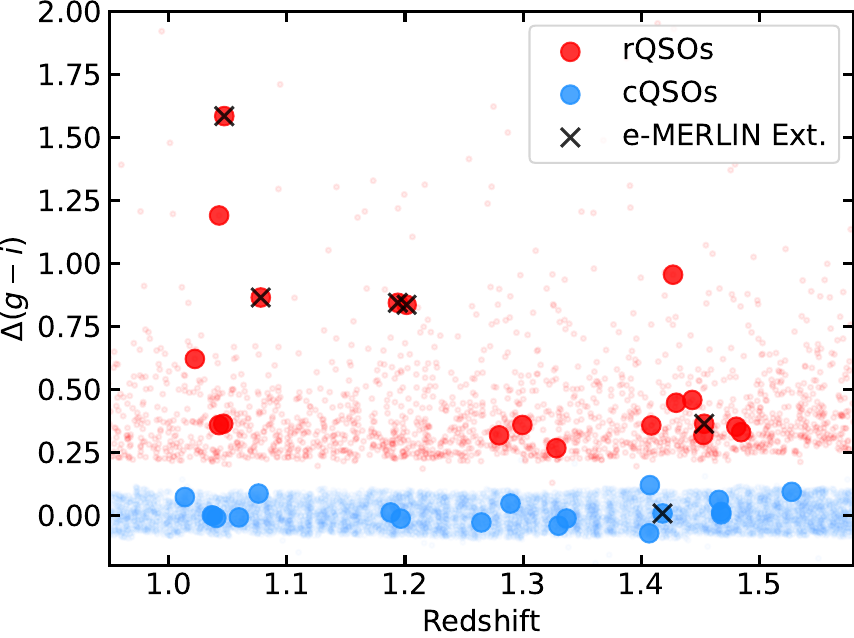}
    \includegraphics[width=0.43\textwidth]{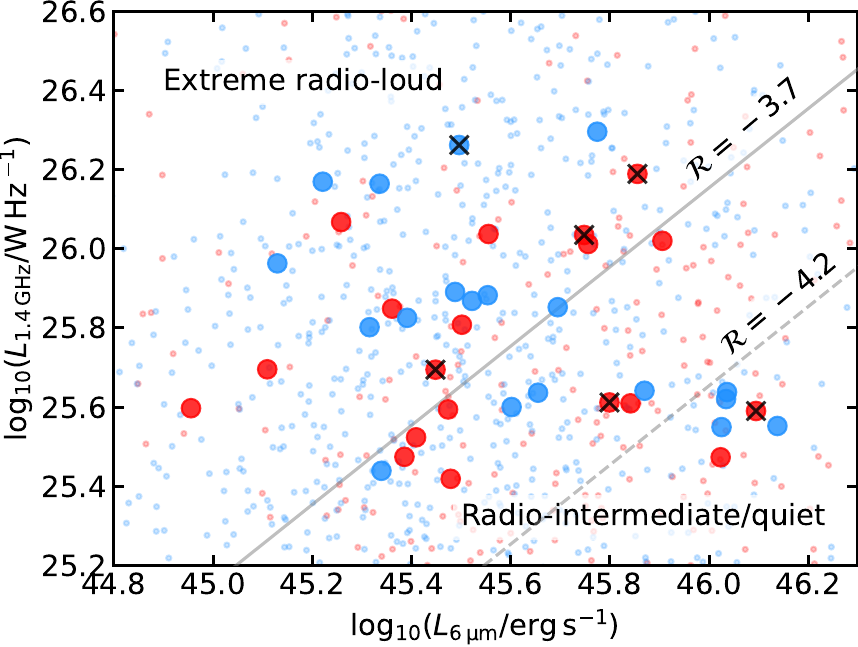}
    \caption{(Top) $\Delta(g-i)$ versus redshift and (bottom) $L_{\rm 1.4\,GHz}$ versus $L_{\rm 6\upmu m}$ for our rQSO and cQSO samples. The underlying parent rQSO and cQSO samples are displayed by faint dots (red and blue, respectively). In the bottom panel the solid grey line displays the divide between radio-quiet/intermediate sources ($\mathcal{R}$\,$<$\,$-3.7$) and extreme radio-loud sources ($\mathcal{R}$\,$>$\,$-3.7$), as defined in \protect\citetalias{klindt}; $\sim$\,47 per~cent (9/19) and $\sim$\,42 per~cent (8/19) of the rQSOs and cQSOs lie below $\mathcal{R}$\,$<$\,$-3.7$, respectively. The classical radio-loud/radio-quiet divide is $\mathcal{R}$\,$=$\,$-4.2$ which is indicated by the grey dashed line. The QSOs with extended radio emission in the $0\farcs2$ e-MERLIN imaging are indicated by black crosses.}
    \label{fig:context}
\end{figure}

The sample of red QSOs (hereafter, ``rQSOs'') and control blue QSOs (hereafter, ``cQSOs'') observed with uGMRT were chosen to have existing $0\farcs2$ e-MERLIN observations at 1.4\,GHz (\citetalias{rosario_21}). The sample selection is presented in detail in \citetalias{rosario_21}, which we describe briefly below.

Starting with the SDSS DR7 \citep{dr7} QSO catalogue, a sample of rQSOs and cQSOs were chosen as the top 10 percentile and middle 50 percentile of the $g-i$ distribution, respectively, in consecutive redshift bins of 1000 sources (see top panel of Fig.~\ref{fig:context}), following the same method as \cite{klindt}; hereafter, \citetalias{klindt}. The QSOs were also selected to be detected in FIRST, with no visibly resolved radio emission beyond the $5''$ resolution. In order to robustly determine the radio morphology via visual inspection, additional cuts were applied: a 1.4\,GHz flux density of $>$\,3\,mJy, a peak flux S/N\,$>$\,15, and a flux ratio between FIRST and the 20\,cm NRAO VLA Sky Survey (NVSS; 1.4\,GHz at $45''$ resolution; \citealt{NVSS}) of less than two. These criteria ensured a robust detection in FIRST and the removal of any sources with radio lobes detected in NVSS but resolved out by the higher resolution FIRST imaging \citetalias{klindt}.\footnote{Note: two QSOs (1046+3427 and 1057+3119) are found to have a single large scale extended radio lobe in FIRST that was incorrectly not associated with the core in the \citetalias{rosario_21} study. With the addition of lower frequency radio data, either a diffuse connection between the lobe and core or an additional second symmetric lobe is revealed. We now conclude that there is large scale extended emission associated with the unresolved cores in these two objects; we comment on this in Appendix~\ref{sec:large_scale}.} 

A narrow redshift range ($1.0$\,$<$\,$z$\,$<$\,$1.55$) was chosen to minimize Malmquist bias. A radio luminosity range of $25.5$\,$<$\,log$L_{\rm 1.4\,GHz}$\,$<$\,$26.5$\,W\,Hz$^{-1}$ was chosen to probe the fainter end of FIRST-detected QSOs at the chosen redshift. Finally, \citetalias{rosario_21} selected 20 rQSOs and 20 cQSOs from this final QSO sample, matched in redshift and $L_{\rm 6\upmu m}$, using a matching tolerance of 0.05 and 0.2\,dex, respectively (following the method from \citetalias{klindt}; see Section~2.3 therein). One cQSO (1511+3428) was later found to have FIRST radio emission slightly offset from the optical centre, which therefore should have been classified as ``extended''. Furthermore, a reanalysis of the Galactic extinction of one rQSO (0842+5804) placed its intrinsic colour outside of the red colour selections. In this paper we remove both of these QSOs from our analyses, resulting in a final sample of 19 rQSOs and 19 cQSOs. The redshift, $L_{\rm 1.4\,GHz}$, and $L_{\rm 6\upmu m}$ distributions of the final samples are displayed in Fig.~\ref{fig:context}.

\subsection{uGMRT observations and data reduction}\label{sec:reduction}

In this paper, we present the upgraded GMRT (uGMRT; \citealt{gmrt2,ugmrt}) Band-3 (250--500\,MHz; central frequency 400\,MHz) and Band-4 (550--850\,MHz; central frequency 650\,MHz) observations taken in February 2020 (Proposal ID: 37\_064, PI: V. Fawcett). 

Due to two and three antennas not working for the Band-4 and Band-3 observations, respectively, 28 and 27 antennas were used for the observations with the full available bandwidth (200\,MHz for Band-3 and 400\,MHz for Band-4) and 4096 channels. The flux density calibrators 3C147 and 3C286 were observed for 10\,mins at the beginning and end of each observation. The phase calibrators (listed in Table~\ref{tab:ugmrt_obs}) were observed for 5\,mins every 2--8 targets, depending on their on-sky position. The QSO targets were observed for 5\,mins in 6 observing blocks, using the same grouping as that used for the e-MERLIN observations (see Section~2.4 of \citetalias{rosario_21}). For one rQSO (1315+2017) there is no Band-3 observation. 

The data were reduced in \texttt{CASA}-6\footnote{Common Astronomy Software Applications; \url{https://casa.nrao.edu}.} using \texttt{CAPTURE}\footnote{CAsa Pipeline-cum-Toolkit for Upgraded Giant Metrewave Radio Telescope data REduction; \url{https://github.com/ruta-k/CAPTURE-CASA6}.} \citep{capture}, an automated pipeline to produce images from interferometric data obtained with the uGMRT. The LTA data files produced by the uGMRT were first converted into FITS format by the \texttt{LISTSCAN} and \texttt{GVFITS} functions. A measurement set was then produced using the \texttt{IMPORTUVFITS} task. The standard steps of flagging, calibration, and imaging were then followed in \texttt{CAPTURE}. Self-calibration with both phase-only and amplitude and phase solutions was carried out in eight iterations. The flux scale of \cite{Perley_2017} was used for the absolute flux calibration. For all sources, imaging was carried out with \texttt{robust\,=\,0}, a pixel scale of $1.5''$ and $1''$ for Band-3 and 4, respectively, and an image size of 8000 pixels (88.9 arcmin and 133.3 arcmin for Band-3 and 4, respectively). Due to the low flux density of some of the targets, the mean flux cut off for flagging bad antennas in Band-3 was reduced from the default value of 0.3\,Jy to 0.08\,Jy in the \texttt{CAPTURE} \texttt{ugfunctions} script. This was required since for some sources $>$\,70 per~cent of the antennas were getting flagged. We note that although including noisier data in the imaging might result in lower signal-to-noise (SNR) images, all our sources have a final SNR\,$>$\,6 in both Band-3 and Band-4. The median flagging fraction for all the QSOs in Band-3 and Band-4 is 24 per~cent and 9 per~cent, respectively. 

Finally, flux density measurements were obtained using the \texttt{IMFIT} task in \texttt{CASA}, which fits 2D elliptical Gaussians to the image. For sources with extended radio lobes only the flux in the radio core was extracted. A typical root mean square (RMS) of $\sim$\,0.17\,mJy and $\sim$\,0.06\,mJy was achieved for Band-3 and 4, respectively. The online Supplementary material and Table~\ref{tab:ugmrt_obs} contains the observation details and extracted flux values for the Band-3 and 4 data for our sample. 

\subsection{Archival radio data used in SED fitting}\label{sec:archival}

In order to robustly measure the radio SEDs of the QSOs, we combined our 2-band uGMRT data with various archival radio data. At the redshifts of our sample, all the radio data used in the SED fitting are for galaxy-wide emission ($\sim$\,20--50\,kpc) and sources with extended low-frequency emission are treated carefully (see Section~\ref{sec:sed}).\footnote{We do not match the angular resolution of the radio data used in the SED fitting in Section~\ref{sec:results} (i.e., LoTSS; $6''$, FIRST; $5''$, uGMRT Band-3 and 4; $\sim$\,$6$--$10''$, and VLASS; $2\farcs5$) since it has been shown that the flux biases introduced from these different resolutions only becomes an issue with extended sources \citep{kukreti}. Our sample were selected to be unresolved in FIRST and Fig.~\ref{fig:peak_int} displays very little variation between the FIRST integrated and peak fluxes, demonstrating there is very little extended emission at 1.4\,GHz. We do note that some of the sources have large-scale extended emission in LoTSS; we treat these separately, refitting the SED without the LoTSS data point (see Section~\ref{sec:sed}).} The archival radio data used in this paper is displayed in Table~\ref{tab:radio} and summarized below.

For the final radio SEDs, we utilized the integrated flux density, rather than the peak flux density, in order to better capture the flux density for sources that might be extended at lower radio frequencies (a comparison between the FIRST peak and integrated flux densities is shown in Fig.~\ref{fig:peak_int}). We carefully treat sources that are only extended in LoTSS separately (see Section~\ref{sec:sed}). 

\begin{table*}
    \centering
    \caption{Table displaying the archival radio data used in this study. The columns from left to right display the: (1) name of the radio survey, (2) survey frequency, (3) spatial resolution, (3) sensitivity, (4) survey area, (5) matching radius used, (6)-(7) number of rQSOs and cQSOs covered by the survey, and (8)-(9) number of rQSOs and cQSOs detected. All the sources are detected in FIRST and NVSS, by selection. The e-MERLIN details are from dedicated observations presented in \citetalias{rosario_21}. \newline\textdagger Included in SED fitting. \textdaggerdbl Not included in SED fitting. $^*$Matching to optical position.}
    \begin{tabular}{cccccccccc}
        \hline \hline
        Radio survey & Frequency  & Res.  & Sensitivity & Area & Match radius & \# rQSOs Cov. & \# cQSOs Cov. & \# rQSOs Det. & \# cQSOs Det.  \\
         &  (MHz) &  (arcsec) &  (mJy\,bm$^{-1}$) & (deg$^2$) & (arcsec) \\
        \hline 
        LoLSS DR1\textdagger & 41--66 & 15 & 1.55 & 650 & 10 & 1 & 0 & 1 & 0\\
        TGSS ADR1\textdagger & 140--156 & 25 & 5 & 36\,900 & 10 & 19 & 19 & 4 & 2\\
        LoTSS DR2\textdagger & 120--168 & 6 & 0.083 & 5740 & 0.2$^*$ & 14 & 13 & 12 & 13 \\
        FIRST\textdagger & 1400 & 5 & 0.65 & 10\,575 & 10 & 19 & 19 & 19 & 19\\
        VLASS\textdagger & 3000 & 2.5 & 1 & 33\,885 & 1 & 19 & 19 & 19 & 19\\
        \hline
        WENNS\textdaggerdbl & 323--328  & 54 & 18 & 27\,500 & 25 & 14 & 14 & 5 & 3\\
        RACS\textdaggerdbl & 600--1175 & 25 & 0.2--0.4 & 34\,240 & 15 & 7 & 7 & 7 & 7 \\
        NVSS\textdaggerdbl & 1400 & 45 & 2.5 & 32\,259 & 25 & 19 & 19 & 19 & 19\\
        \hline
        e-MERLIN\textdaggerdbl & 1230--1740 & 0.2 & 0.08 & - & - & 19 & 19 & 19 & 19\\
        \hline \hline
    \end{tabular}
    \label{tab:radio}
\end{table*}

\subsubsection{LoTSS DR2}
The LOFAR Two Metre Sky Survey DR2 (LoTSS; \citealt{lotssdr2}) is a 120--168\,MHz LOFAR radio survey that aims to observe the whole northern sky at a resolution of $6''$. The second data release covers 5740\,deg$^{2}$ down to a sensitivity limit of $\sim$\,83\,$\upmu$Jy\,beam$^{-1}$. In this paper we utilize the catalogue with associated optical and/or near-infrared counterparts from \cite{lotss_opt}, which contains 4,116,934 sources. 14 rQSOs and 13 cQSOs lie in the LoTSS DR2 sky coverage, and 12/14 rQSOs and 13/13 cQSOs are detected ($0\farcs2$ matching radius with optical positions, consistent with \citealt{rosario}).

\subsubsection{FIRST}
The Faint Images of the Radio Sky at Twenty-centimeters (FIRST; \citealt{becker}) is a 1.4\,GHz VLA radio survey, covering 10,575\,deg$^{2}$ of the sky in the SDSS region at a resolution of $5''$. The final catalogue \citep{helfand} contains 946,432 sources down to a sensitivity limit of $\sim$\,0.65\,mJy\,beam$^{-1}$. As our sample selection requires a robust FIRST detection, all 38 QSOs have FIRST flux densities ($10''$ matching radius, consistent with \citetalias{klindt}).

\subsubsection{VLASS}\label{sec:vlass}
The VLA Sky Survey (VLASS; \citealt{vlass}) is a 2--4\,GHz VLA radio survey, covering 33,885\,deg$^{2}$ of the sky at a resolution of $2\farcs5$. In this paper, we utilize the Quick Look Epoch 2 catalogue, which consists of 2,995,025 sources down to a sensitivity of 1\,mJy\,beam$^{-1}$.
We included an additional error of 10 per~cent of the integrated flux density to account for the known flux underestimation \citep{lacy_vlass,Gordon_2021}. All 38 QSOs are detected ($1''$ matching radius). We note that only five QSOs have a $F_{\rm 3\,GHz}$\,$<$\,3\,mJy, which are known to have more unreliable flux densities \citep{Gordon_2021}; for these sources, we add instead an additional systematic error of 20 per~cent of the integrated flux density and flag these sources in Table~\ref{tab:radio_tab}. We also note that in our SED fitting we utilize the integrated flux densities from the Epoch 2 catalogue, which are known to be more reliable than those from Epoch 1 and more reliable than the peak flux densities.\footnote{For more details, see \url{https://cirada.ca/vlasscatalogueql0}.}

In order to test the impact of radio variability on our study, we compared the integrated flux densities from both the Epoch 1 and 2 Quick Look tables; we found no significant variation in the VLASS integrated fluxes for any of our sample (Fig.~\ref{fig:vlass_variability}). For more discussion on radio variability, see Appendix~\ref{sec:variability}.

\subsubsection{TGSS ADR1}
The TIFR Giant Meterwave Radio Telescope Sky Survey Alternative Data Release (TGSS ADR1; \citealt{tgss}) is a 140--156\,MHz radio survey covering 36,900\,deg$^{2}$ of the sky at a resolution of $25''$. The $7\sigma$ catalogue contains 623,604 sources down to a sensitivity limit of $\sim$\,5\,mJy\,beam$^{-1}$. Using a $10''$ matching radius we found that four rQSOs and two cQSOs are detected. Despite the low angular resolution of this survey, there are two sources that lie outside of the LoTSS coverage but are detected by TGSS; including these additional data points greatly improves the fitting constraints (we indicate these sources in Table~\ref{tab:radio_tab}). There are a further nine sources that lie outside of the LoTSS coverage, but have upper limits in TGSS. The majority of these upper limits do not affect the fitting due to the low sensitivity limit of TGSS; however, there is one source for which the best fitting model \textit{without} the TGSS upper limit is inconsistent with the upper limit (1003+2727).

\subsubsection{LoLSS DR1}
The LOFAR\footnote{\url{https://lofar-surveys.org/surveys.html}} LBA Sky Survey DR1 (LoLSS; \citealt{lolss}) is a 41--66\,MHz LOFAR radio survey that aims to observe the whole northern sky above declination 24\degree~at a resolution of $15''$. The first data release covers 650\,deg$^{2}$ in the HETDEX spring field and contains 42,463 sources down to a sensitivity limit of $\sim$\,1.55\,mJy\,beam$^{-1}$. Only one rQSO (1153+5651) lies in the LoLSS DR1 sky coverage and is also detected ($10''$ matching radius); despite the low angular resolution, the LoLSS data point is in agreement with the higher frequency data for this source and so we used this data in the fitting for an additional constraint.

\subsection{Supplementary archival radio data}\label{sec:archival2}
For our final radio SEDs (see Section~\ref{sec:results}) we also plot additional archival radio data including the e-MERLIN $0\farcs2$ radio fluxes (Section~\ref{sec:e-merlin}) and radio data from surveys with a much lower angular resolution compared to those listed above (Section~\ref{sec:low_res}). Although part of the selection criteria ensured that the flux offset between FIRST (at $5''$) and NVSS (at $45''$) was less than two, biases may still be introduced in the SED fitting if lower resolution data is introduced due to potential large scale radio lobes missed in the higher resolution surveys or additional contributions from additional background sources within the beam. Furthermore, the e-MERLIN observations have a considerably higher angular resolution ($>$\,12$\times$) compared to the other radio data utilized in this paper, and so might miss diffuse extended structures. Therefore, we do not include the following radio data in the SED fitting. Instead, these data provide another constraint on radio variability, which we discuss in Appendix~\ref{sec:variability}, in addition to providing radio sizes (from e-MERLIN) and helping to visually asses the reliability of the SED fits.

\subsubsection{e-MERLIN L-band}\label{sec:e-merlin}
The sample used in this paper was previously observed using the e-MERLIN interferometer with the L-band receivers (1.23--1.74\,GHz) at a $0\farcs2$ resolution (PI: D. Rosario; Project ID: CY7220).
The e-MERLIN flux densities were obtained from \citetalias{rosario_21}, who fit either one or multiple Gaussian components to the central $2''$ of the e-MERLIN image, depending on a visual examination of the fitting residuals. For the first round of fitting, a single Gaussian model was used which was initialized to match the shape of the restoring beam. After inspecting the resulting residuals, seven sources (six used in this paper, see Section~\ref{sec:sample}) were found to have emission extended beyond the beam which were then subject to a second round of fitting with $>$\,2 Gaussian components (e-MERLIN extended sources; see Tables~\ref{tab:obs} and \ref{tab:radio_tab}). For the remaining 32, a further analysis was performed, comparing the semi-major axis of the single-component Gaussian fits with that of the restoring beam to search for any additional extended source structure at the resolution limit of the e-MERLIN images. \citetalias{rosario_21} found no statistical difference between the semi-major axis of the Gaussian component and the restoring beam, demonstrating that these sources are unresolved in the e-MERLIN images (e-MERLIN unresolved sources).

The flux ratio between FIRST and e-MERLIN can provide information on whether diffuse extended emission (at the same frequency) from radio lobes that has been resolved out; \citetalias{rosario_21} found no significant differences between the flux densities for either the cQSOs or rQSOs (see Section 3.2 in \citetalias{rosario_21}). They also note that for the flux estimates for diffuse e-MERLIN sources are likely to have larger errors and may be overestimated, which might explain the $\approx$\,2.5\,$\times$ larger e-MERLIN flux compared to the FIRST flux for QSO 0946+2548. We plot the e-MERLIN flux densities as open stars on the SEDs (see Figs.~\ref{fig:example_seds}, \ref{fig:SED_rQSO}, and \ref{fig:SED_cQSO}).

The e-MERLIN radio sizes were calculated by taking the maximum of either the largest separation between the two Gaussian sub-components or the major-axis width of a the largest single component. For sources that only have a single core component, its major-axis is used as a limit on the size (i.e., e-MERLIN unresolved sources). These were then converted to physical sizes using the angular diameter distances \citetalias{rosario_21}.

\subsubsection{Lower angular resolution radio surveys}\label{sec:low_res}

The Westerbork Northern Sky Survey (WENNS; \citealt{wenss}) is a 322.5--327.5\,MHz radio survey, covering the whole of the sky above a declination 30$\degree$ at a resolution of $54''$. The $5\sigma$ combined catalogue contains 229,420 sources from the mini and main surveys down to a sensitivity limit of 18\,mJy. Using a $25''$ matching radius we find that five rQSOs and three cQSOs are detected.

The Rapid ASKAP Continuum Survey (RACS; \citealt{racs}) is the first large-area survey with the Australian Square Kilometer Array Pathfinder (ASKAP; \citealt{askap1,askap2}) which aims to observe the entire southern sky at a frequency of 700--1800\,MHz with $\sim$\,$25''$ resolution. The first release $5\sigma$ catalogue \citep{Hale_2021} contains 2,123,638 sources at a central frequency of 887.5\,MHz (288\,MHz bandwidth) down to a sensitivity limit of 0.25--0.3\,mJy\,beam$^{-1}$. Using a $15''$ matching radius we find that seven rQSOs and seven cQSOs are detected. 

The 20\,cm NRAO VLA Sky Survey (NVSS; \citealt{NVSS}) is a 1.4\,GHz at radio survey covering the whole sky above a declination of $-40\degree$ at $45''$ resolution. The $5\sigma$ catalogue contains 1,773,484 sources down to a sensitivity limit of $\sim$\,2.5\,mJy\,beam$^{-1}$. By selection (see Section~\ref{sec:sample}), all the QSOs in our sample are detected in NVSS.

\subsection{Radio-loudness and luminosity}\label{sec:rad_lum}
In this paper we utilize $L_{\rm 1.4\,GHz}$, which is calculated from the FIRST integrated fluxes, using the methodology described in \cite{alex2003}, assuming a uniform radio spectral index of $\alpha$\,$=$\,$-0.5$ for the $K$-correction (Fig.~\ref{fig:context}). We also explored how $L_{\rm 1.4\,GHz}$ changes once adopting our more robust individual values of $\alpha$, obtained from our SED fitting (Section~\ref{sec:sed}; median $\alpha_{\rm PL}$\,$=$\,$-0.4$), and found a median absolute difference of 0.19\,dex for the power-law sources.\footnote{The maximum offset in radio luminosity was for QSO 0823+5609, which had a difference of 0.26\,dex due to a flat PL slope of $\alpha_{\rm PL}$\,$=$\,0.23.} To explore the ``radio-loudness'' of our samples, we adopted the same parameter as that first used in \citetalias{klindt}, defined as the dimensionless quantity:
\begin{equation}\label{eq:RL}
    \mathcal{R}=\rm log_{10} \frac{\textit{L}_{1.4\,GHz}}{\textit{L}_{\rm 6\upmu m}}
\end{equation}
By this definition, the radio-loud/radio-quiet threshold is $\mathcal{R}=-4.2$, which is broadly consistent with the canonical definition often defined as the 5\,GHz-to-2500\,\AA~ratio (e.g., \citealt{kellermann}). We utilize the 6\,$\upmu$m luminosity rather than the optical luminosity since this is less susceptible to obscuration from dust (see \citetalias{klindt} for full details).\footnote{It was demonstrated in \cite{fawcett23} that even for an extreme QSO with an $E(B-V)$\,$=$\,1\,mag at $z$\,$=$\,1.5, the differences in $L_{\rm 6\upmu m}$ due to the loss of flux by dust extinction at rest-frame $6$\,$\upmu$m is $\sim$\,0.15\,dex.} \citetalias{klindt} showed that the differences in the radio properties between rQSOs and cQSOs arose around radio-loudness values of $-5$\,$<$\,$\mathcal{R}$\,$<$\,$-3.7$ (i.e., radio-quiet/radio-intermediate values; also demonstrated in \citealt{rosario,fawcett20,fawcett21}). Therefore, in this study we explored our results when splitting at $\mathcal{R}$\,$=$\,$-3.7$, referring to QSOs with a $\mathcal{R}$\,$<$\,$-3.7$ as ``radio-quiet/radio-intermediate'' and $\mathcal{R}$\,$>$\,$-3.7$ as ``extreme radio-loud'' (see the lower panel of Fig.~\ref{fig:context}). 

\subsection{Radio spectral fitting and characterization}\label{sec:model_characterisation}
\begin{figure*}
    \centering
    \includegraphics[width=0.33\textwidth]{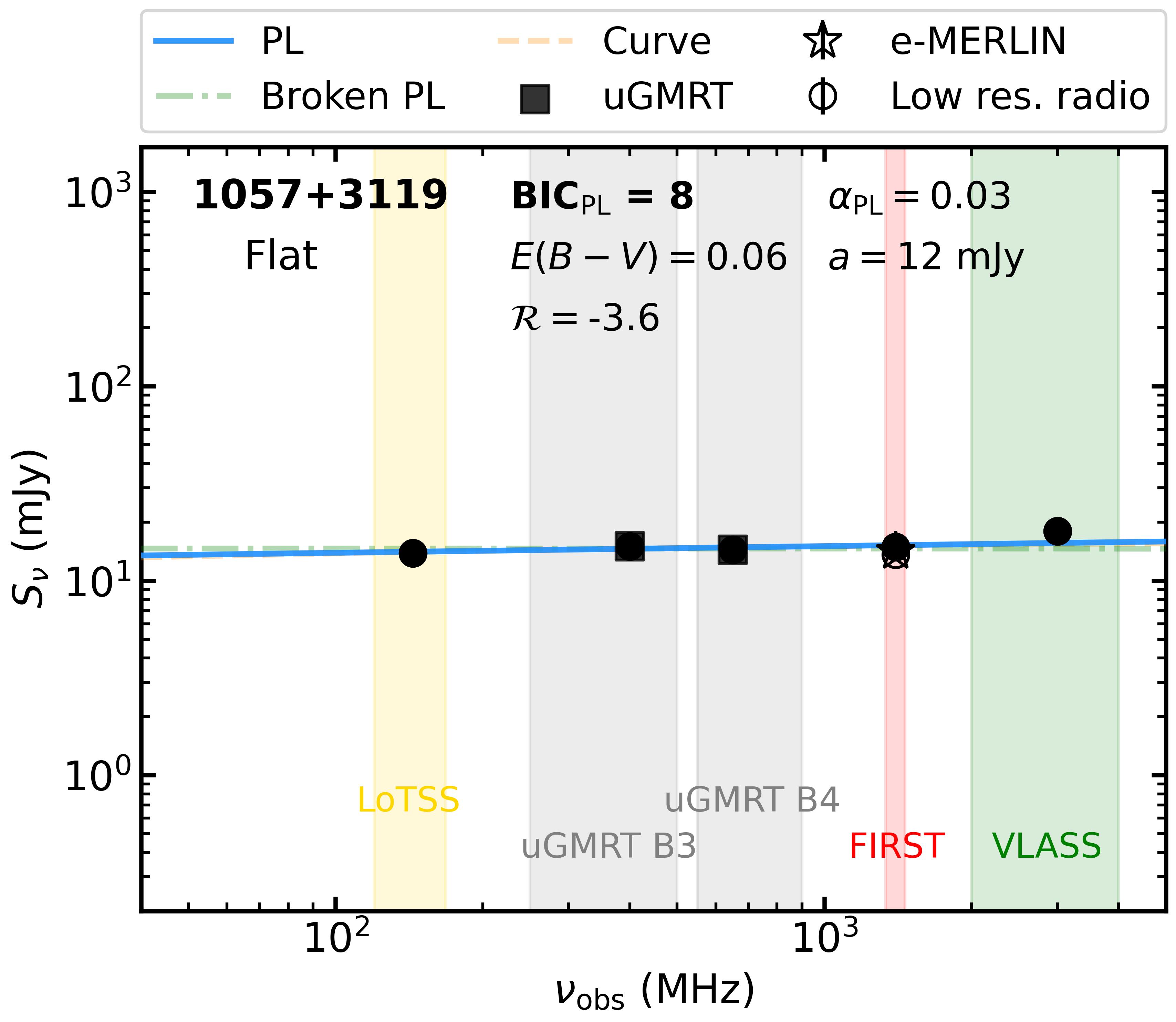}
    \includegraphics[width=0.33\textwidth]{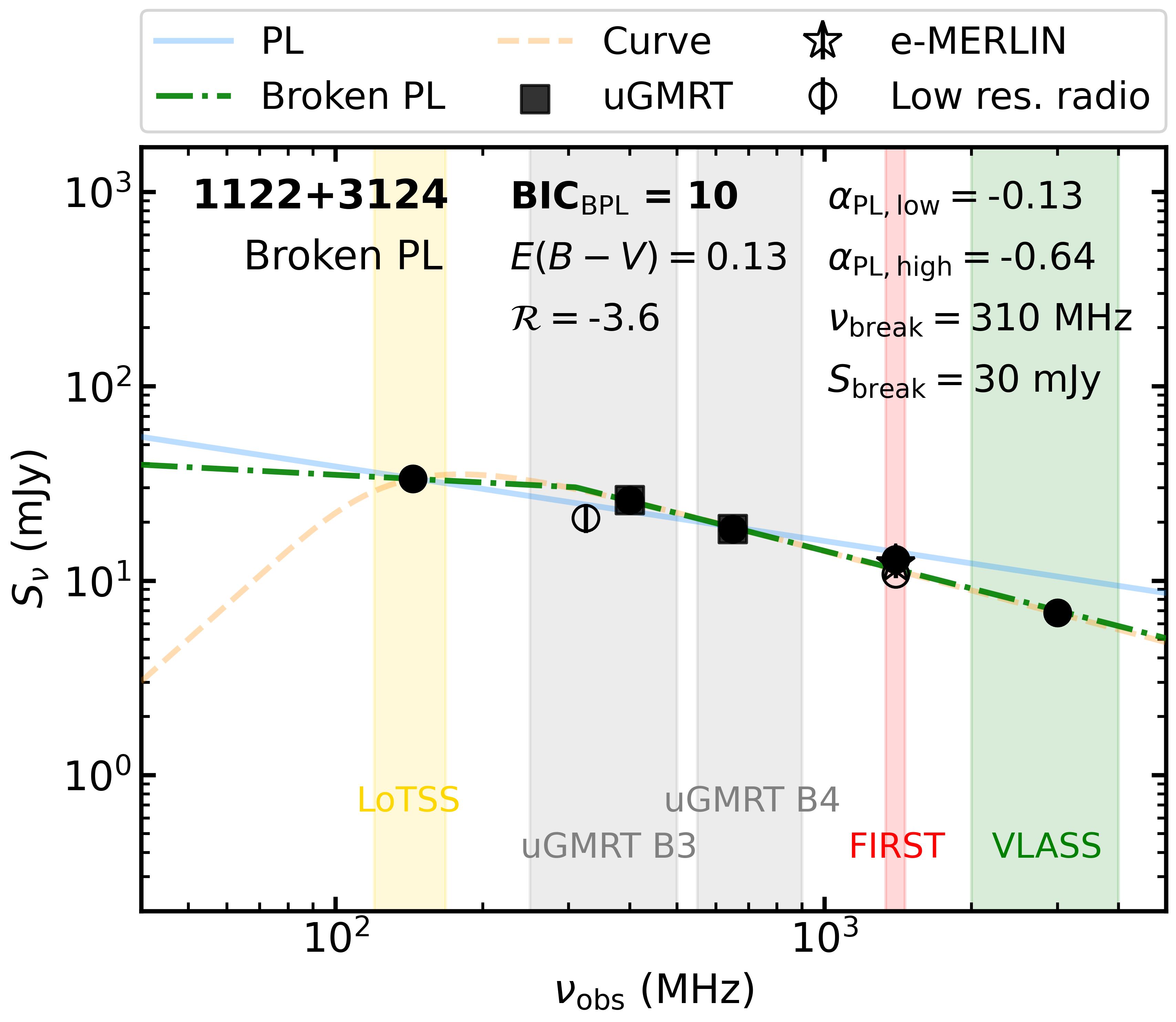}
    \includegraphics[width=0.33\textwidth]{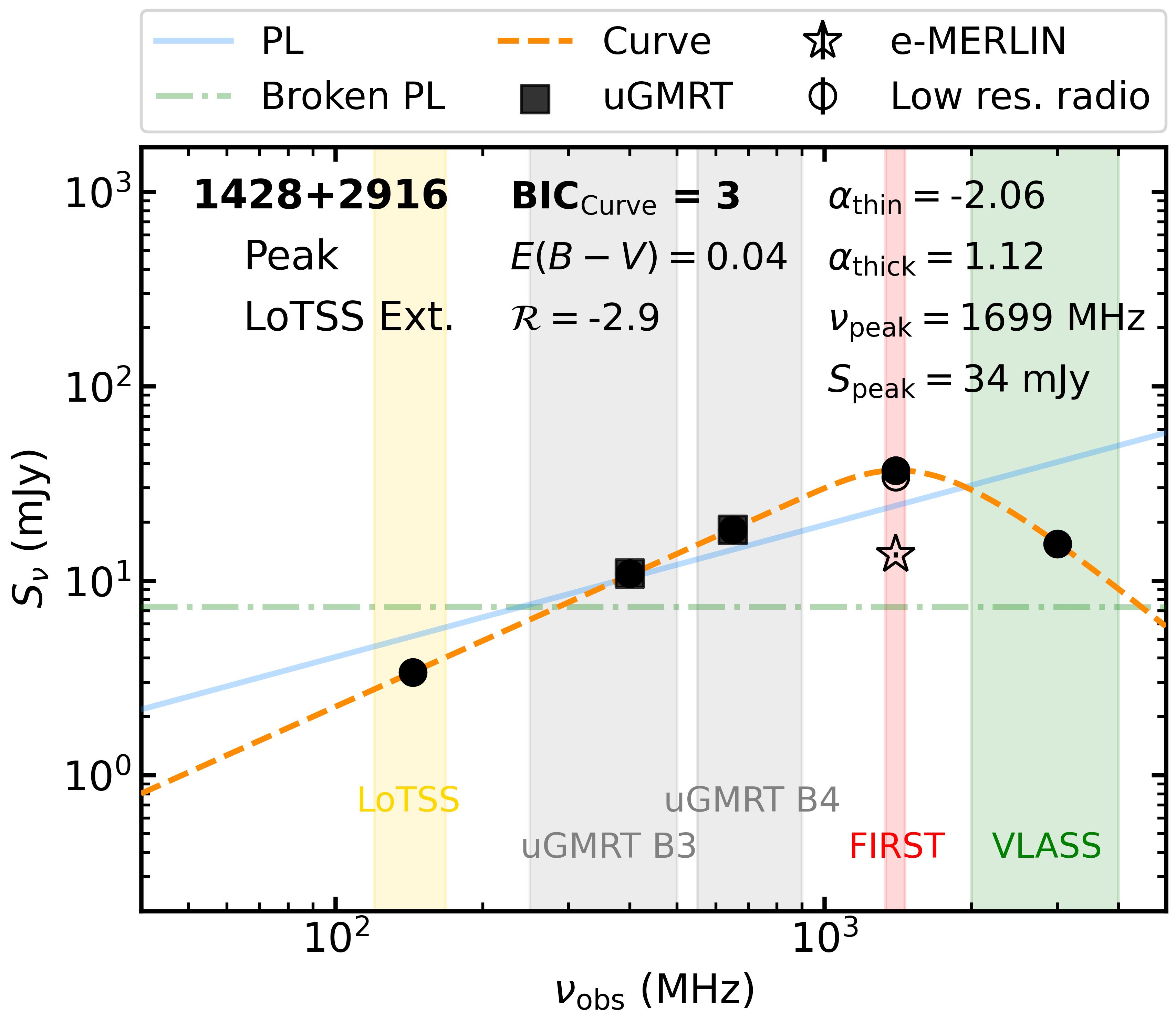}
    \includegraphics[width=0.67\linewidth]{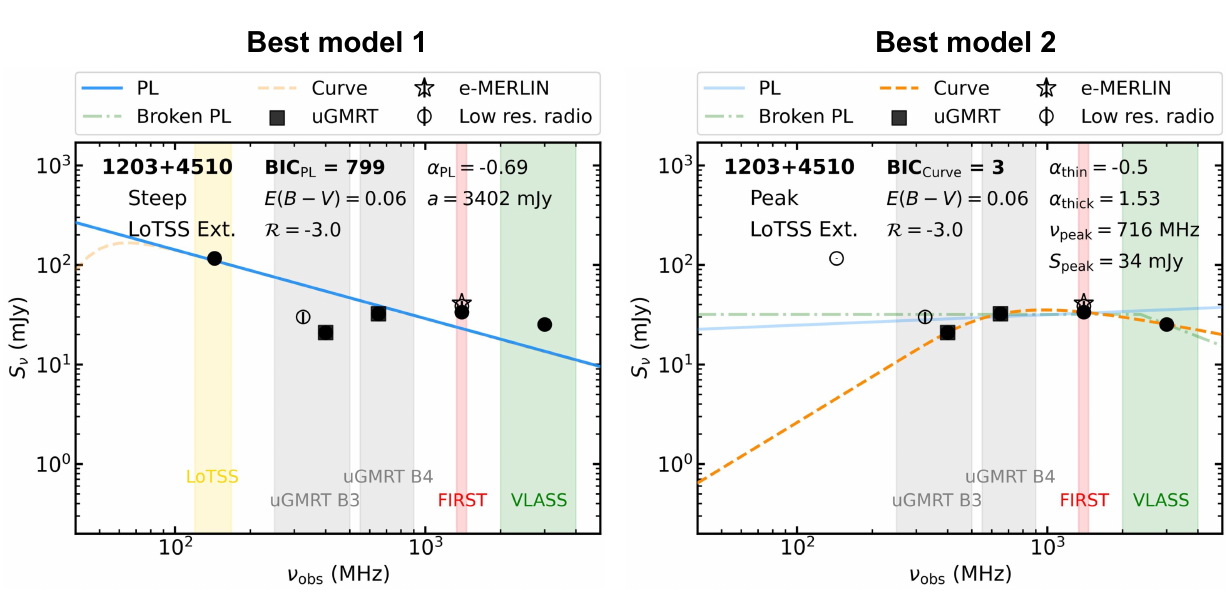}
    \caption{(Top) Three radio SED examples, displaying a flat PL (left), a steep BPL (middle), and a peaked (right) best-fitting model. (Bottom) Example SEDs of a QSO that was visually classified as upturned due to no model producing a good fit to the data and extended LoTSS emission identified in the image. After removing the LoTSS data point, the SED was refitted with the peak model providing a good fit to the data (right). For this source there may be two radio components; an older, steeper component and a younger peaked component, potentially the signature of a restarted radio source (see Section~\ref{sec:evo}). In both panels the radio data used in the fitting are shown by the solid black marker, with the radio survey wavebands indicated by the coloured regions. Our uGMRT data is indicated by the squares. The other empty markers indicate the additional archival data that was not included in the SED fitting (see Section~\ref{sec:archival2}). The model BIC, $E(B-V)$, radio-loudness values, whether the source is extended in LoTSS (LoTSS Ext.), and best fitting model parameters are indicated on each SED. All the SED fits are displayed in Figs.~\ref{fig:SED_rQSO} and \ref{fig:SED_cQSO}.}
    \label{fig:example_seds}
\end{figure*}

In this paper we aim to characterise the radio SEDs of our sample of QSOs, in order to determine how many are best fit by a typical synchrotron power-law (either continuous or broken) or display a curved, peaked SED. Furthermore, we can compare the dusty red QSOs and typical blue QSOs to understand whether they have different radio spectral properties. Three example SED fits are displayed in Fig.~\ref{fig:example_seds}, with one source favouring a flat power-law model, one source favouring a steep broken power-law model, and the other favouring a peaked model.

Therefore, in order to model the radio spectral properties of our sample, we adopted three different spectral models to fit the uGMRT plus archival radio data (e.g., \citealt{Patil_2022,kerrison}). To fit the three models to the data we used the \texttt{emcee}\footnote{\url{https://emcee.readthedocs.io/en/v2.2.1/}} package in Python \citep{emcee}.

The first model we used is a standard non-thermal power-law model:
\begin{equation}\label{eq:pl}
    S_{\nu} = a\nu^{\alpha_{\rm PL}} ,\
\end{equation}
where $a$ is the amplitude of the synchrotron spectrum, $\nu$ is the frequency, and $\alpha_{\rm PL}$ is the spectral index. The bounds for $\alpha_{\rm PL}$ were set at $\pm3$. 

The second model we used is a broken power-law (BPL) model to characterize a standard optically-thin synchrotron source with additional energy losses at higher frequencies (due to synchrotron and inverse-Compton losses; e.g., \citealt{panessa_22}):
\begin{equation}\label{eq:bpl}
    S_{\nu} = a
    \begin{cases} 
    \bigg(\frac{\nu}{\nu_{\rm break}}\bigg)^{\alpha_{\rm low}} & \text{for $\nu$\,$<$\,$\nu_{\rm break}$} \\
    \bigg(\frac{\nu}{\nu_{\rm break}}\bigg)^{\alpha_{\rm high}} & \text{for $\nu$\,$>$\,$\nu_{\rm break}$} \\
    \end{cases} ,\
\end{equation}
where $\alpha_{\rm low}$ and $\alpha_{\rm high}$ are the spectral indices of the power-law below and above the break frequency $\nu_{\rm break}$, respectively. Since energy losses steepen the spectrum, we required that $\alpha_{\rm high}$\,$<$\,$\alpha_{\rm low}$ and set the bounds for both power-low indices to be $-3$\,$<$\,$\alpha_{\rm PL}$\,$<$\,0. We required $\nu_{\rm break}$ to be within the frequency bounds of the radio data utilized in the SED fitting. There is some degeneracy between the two power-law indices and $\nu_{\rm break}$, which is reflected in the resulting Monte-Carlo errors.

The third model we used is the commonly adopted generic curved model (e.g., \citealt{snellen_98,callingham,shao_20,wolowska,kerrison}), which is able to characterize a peaked-spectrum source:
\begin{equation}\label{eq:curve}
    S_{\nu} = \frac{S_{\rm peak}}{(1-e^{-1})}\Big(1-e^{-(\nu/\nu_{\rm peak})^{{(\alpha_{\rm thin}}-\alpha_{\rm thick})}}\Big)\bigg(\frac{\nu}{\nu_{\rm peak}}\bigg)^{\alpha_{\rm thick}} ,\
\end{equation}
where $S_{\rm peak}$ is the flux density at the peak frequency, $\nu_{\rm peak}$ and $\alpha_{\rm thick}$ and $\alpha_{\rm thin}$ are spectral indices in the optically thick and thin regime, respectively (e.g., \citealt{snellen_98}). The bounds for $\alpha_{\rm thick}$ and $\alpha_{\rm thin}$ were set at $<$\,4 and $>$\,$-4$, respectively, following \cite{kerrison}. Therefore, $\alpha_{\rm thick}$ corresponds to the lower frequency ($\nu$\,$<$\,$\nu_{\rm peak}$), inverted part of the peaked spectrum and $\alpha_{\rm thin}$ corresponds to the higher frequency ($\nu$\,$>$\,$\nu_{\rm peak}$), steep part of the spectrum. When $\alpha_{\rm thick}$\,$=$\,$2.5$ this model reduces to the homogeneous, synchrotron self-absorbed case. It should be noted that this function sometimes fails to accurately fit the SED far away from the peak \citep{snellen_98}. However, due to our limited frequency range, this is unlikely to significantly affect the fits. 

The best fitting model was selected by utilizing the Bayesian information criterion (BIC; \citealt{bic}). Nominally, the fit with the minimum BIC value is taken to be the best fitting model. However, to favour the curve model over the other models, we required strong evidence that neither the PL nor BPL models were a sufficient fit to the data by imposing a $\Delta$(BIC)\,$>$\,$10$ between the curve model and other models. Furthermore, due to the degeneracy between the BPL model and both the curve and PL models, we required a $\Delta$(BIC)\,$>$\,$5$ between the BPL and other models to favour the BPL model. The final radio SEDs fits were further categorized as one of the following (for the BPL model, the $\alpha$ adopted is $\alpha_{\rm high}$ if the majority of data points have a $\nu$\,$>$\,$\nu_{\rm break}$ and $\alpha_{\rm low}$ vice versa):

\begin{enumerate}
    \item \textit{Steep}: Best fit with the PL or BPL models (Eqs.~\ref{eq:pl} and \ref{eq:bpl}) with $\alpha$\,$<$\,$-0.5$, 
    \item \textit{Flat}: Best fit with the PL or BPL models with $-0.5<$\,$\alpha$\,$<$\,$0.5$ (consistent with e.g., \citealt{urry,kerrison}),
    \item \textit{Inverted}: Best fit with the PL model with $\alpha_{\rm PL}$\,$>0.5$,
    \item \textit{Peaked}: Best fit with the curve model (Eq.~\ref{eq:curve}) and a spectral turnover from the fitted model within the frequency range of the data utilized (0.144--3\,GHz),
    \item \textit{Curved}: Best fit with the curve model and  a spectral turnover from the model that is predicted to be outside of frequency range of the data utilized.
\end{enumerate}

After visually inspecting the resulting SED fits and BIC values, we found that four QSOs (0951+5253, 1046+3427, 1203+4510, and 1630+3847) have an ``upturned'' radio SED, meaning that none of the models were a good fit to the data. For all four QSOs, these upturned SEDs were found to be driven by large-scale extended radio emission in LoTSS that is either not detected or not taken into account in the flux extraction for the higher frequency data. For example, QSO 1046+3427 displays large-scale radio emission in all frequency bands (see Fig.~\ref{fig:extended_radio}) that was not associated with the core emission in the FIRST catalogue but was associated as one source in LoTSS (see Appendix~\ref{sec:large_scale} for more discussion on missed large-scale radio emission in two QSOs). Some studies attempt to model these sources by using a linear combination of Eqs.~\ref{eq:pl} and \ref{eq:curve} (e.g., \citealt{kerrison}). However, in this study we are restricted to only five data points which would be less than the number of free parameters when combining these two equations. Therefore, for the four upturned QSOs we removed the LoTSS data point and refitted their radio SEDs in order to explore the compact emission ($\lesssim$ few kpc), consistent with the rest of the sources (an example is displayed in the bottom panel of Fig.~\ref{fig:example_seds}). We found that all upturned sources have a good fit without the LoTSS data point, demonstrating that their SEDs are comprised of two components; a steep low frequency component and a flat/peaked high frequency component. We comment on these sources further in Section~\ref{sec:evo}.

An additional QSO (1251+4317) has a good SED fit, apart from the VLASS data point which is over a magnitude brighter than that predicted by the best fitting model to the other data. Interestingly, this source also has the largest offset between the measured e-MERLIN flux ($0\farcs2$) and FIRST flux ($5''$), with the e-MERLIN flux 2.4$\times$ brighter than FIRST, despite the much higher angular resolution of e-MERLIN. This could suggest that this sources is undergoing rapid core variability. However, in Appendix~\ref{sec:variability} we explore the VLASS variability on a two-year timescale and do not find any significant differences for any of the sources. Other possibilities include the superposition of multiple radio components or a young restarted core (e.g., \citealt{Nyland2020}).

If the number of data points in the fitting is less than or equal to the number of parameters in the best fitting model then we indicate the parameters on these fits as ``unconstrained''. However, we are still able to distinguish which is the best-fitting model and, therefore, include these sources in the raw numbers for each best-fitting model (these sources are indicated in Table~\ref{tab:radio_tab}). We note that our fitting is restricted to the frequency range of the radio data used. Therefore, a radio SED best fit by a PL or BPL may be better fit by a curve model if lower/higher frequency radio data was utilized. We take this into account when exploring trends with $\nu_{\rm peak}$ in Section~\ref{sec:jets}. We present the results of this fitting approach in Section~\ref{sec:sed}.

\subsection{Dust extinction fitting}\label{sec:dust_method}

In order to explore the differences in the radio SEDs as a function of dust extinction, we quantified the amount of dust extinction in each of the QSOs. To do this we fit a blue QSO spectral template with varying amounts of dust extinction to the SDSS optical spectra, following the same method as \cite{fawcett22}. We briefly outline the approach here.

We used the blue QSO VLT/\textit{X-shooter} composite from \cite{fawcett22} as our unreddened template. We then masked the emission lines and smoothed the composite with a Gaussian filter. This blue template was then fitted to the SDSS spectra using a least-squares minimization code, which varied the dust extinction ($E(B-V)$), using a simple PL dust extinction curve ($A_V$\,$\propto$\,$\lambda^{-1}$; $R_V$\,$=$\,4) ranging from $-0.5$\,$<$\,$E(B-V)$\,$<$\,$2.5$\footnote{Note: since the blue template was constructed from real data, some QSOs in our sample may have a bluer spectrum than the template which would result in a negative value of dust extinction.}, and also the normalization, avoiding emission line regions. The resulting $E(B-V)$ values for each QSO is displayed in Tables~\ref{tab:obs} and \ref{tab:radio_tab}. We note that using the standard extinction law of the Small Magellanic Cloud (SMC; \citealt{SMC}) results in similar $E(B-V)$ values, with a median difference of 0.015\,mag and 0.008\,mag for the cQSOs and rQSOs, respectively. The maximum difference in fitted $E(B-V)$ between the two curves is 0.04\,mag. Due to the fitting often underestimating the error, an additional 10 per~cent error on the $E(B-V)$ has been included.

The top left panel of Fig.~\ref{fig:uGMRT_alpha} displays the overall distribution of $E(B-V)$ values, demonstrating a small overlap in the rQSOs and cQSOs at an $E(B-V)$\,$<$\,0.1\,mag (see \citealt{fawcett23} for a comparison between optical colour and dust extinction) and a tail towards dustier QSOs at an $E(B-V)$\,$>$\,0.1\,mag. Therefore, despite the initial sample selection based on $g-i$ colour, for the rest of the paper we instead focus on the more physical quantity of dust extinction, displaying both the $E(B-V)$ and $R_V$-dependent $A_V$ values on the relevant figures. We also compare the ``dusty'' QSOs to the ``non-dusty'' QSOs by applying a separation at an $E(B-V)$\,$=$\,0.1\,mag ($A_V$\,$=$\,0.4\,mag). Due to previous work that found no significant differences in the accretion properties between SDSS-selected red and blue QSOs, we do not expect varying accretion rates across the sample to significantly affect the results (\citealt{fawcett22}; see also Fig.~\ref{fig:Redd}).

\section{Results}\label{sec:results}
With the combination of dedicated uGMRT observations and archival radio data, we aim to explore the differences (if any) between the radio SEDs of SDSS QSOs as a function of obscuration. In Section~\ref{sec:gmrt_alpha} we first present the uGMRT Band-3--4 radio spectral slopes and in Section~\ref{sec:sed} we describe the basic results from the SED fitting, exploring trends with dust extinction and radio-loudness.

\subsection{uGMRT spectral slopes}\label{sec:gmrt_alpha}
\begin{figure*}
    \centering
    \includegraphics[width=0.76\textwidth]{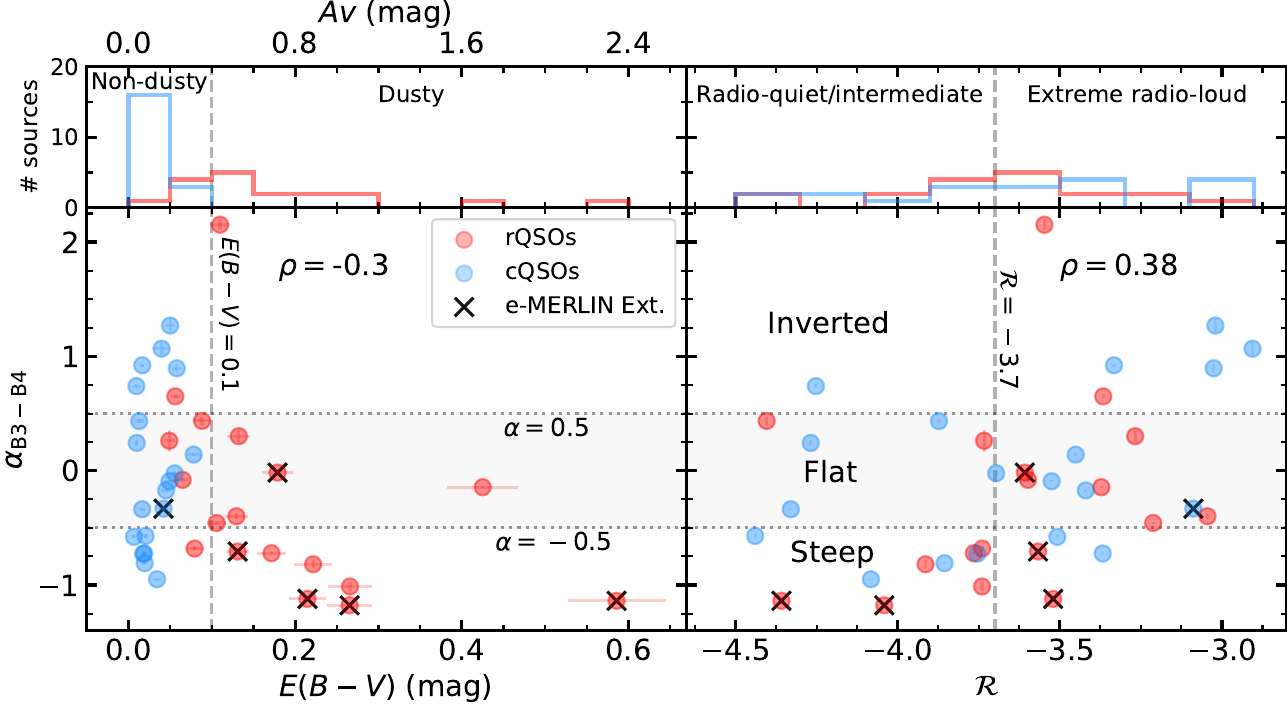}
    \caption{The uGMRT Band-3--4 (400--650\,MHz) radio spectral slope versus (left) $E(B-V)$ and (right) radio-loudness ($\mathcal{R}$) for the rQSOs (red) and cQSOs (blue). The QSOs with extended $0\farcs2$ e-MERLIN radio emission are indicated by the black crosses. The Spearman's rank correlation coefficients displayed on both panels reveal a weak negative and positive trend between the radio spectral slope and $E(B-V)$ and $\mathcal{R}$, respectively. The vertical dashed line in the left and right panels indicate the the boundary between non-dusty and dusty QSOs ($E(B-V)$\,$=$\,0.1\,mag) and radio-quiet/intermediate and extreme radio-loud QSOs ($\mathcal{R}$\,$=$\,$-3.7$), respectively. Due to the small overlap in the more physical $E(B-V)$ parameter for the rQSOs and cQSOs displayed in the top left panel, for the rest of the paper we explore trends as a function of dust extinction.}
    \label{fig:uGMRT_alpha}
\end{figure*}

To characterize the radio spectral slopes of our sample to first order, we can calculate the 2-point radio spectral slope utilizing only the uGMRT Band-3 (400\,MHz) and Band-4 (650\,MHz) data. This provides an empirical measurement without any model assumptions. Additionally, since these data were taken during the same week and from the same instrument, any variability issues will be minimized. Furthermore, comparing the radio spectral slopes obtained with only two data points to our fully characterized SEDs (Section~\ref{sec:sed}), we can determine what information is gained from the inclusion of additional radio data.

The left and right panels of Fig.~\ref{fig:uGMRT_alpha} display the uGMRT spectral slope versus $E(B-V)$ (see Section~\ref{sec:dust_method}) and radio-loudness ($\mathcal{R}$; see Section~\ref{sec:rad_lum}), respectively. There appears to be a tentative negative trend between the spectral slope and the amount of dust extinction (i.e., the redder the QSO, the steeper the radio spectral slope); the Spearman's rank correlation coefficient reveals a weak trend ($\rho$\,$=$\,$-0.3$, $p$\,$=$\,$0.07$). Splitting the sample by dust extinction, we find that the average spectral slope for the dusty QSOs ($E(B-V)$\,$>$\,0.1\,mag; $A_V$\,$>$\,$0.4$\,mag) is much steeper than that for the non-dusty QSOs ($E(B-V)$\,$<$\,0.1\,mag), with an $\alpha$\,$=$\,$-0.71$ and $\alpha$\,$=$\,$-0.05$, respectively.\footnote{This result also holds when comparing the rQSOs and cQSOs, with a median spectral slope of $\alpha$\,$=$\,$-0.43$ and $\alpha$\,$=$\,$-0.09$, respectively.} We also find that 6/25 (24$\pm$6 per~cent) of the non-dusty QSOs are inverted ($\alpha$\,$>$\,0.5) compared to zero dusty QSOs.\footnote{In Section~\ref{sec:sed} we further explore these sources with the additional archival radio data and find them to in fact have curved radio SEDs.} A positive trend ($\rho$\,$=$\,$0.38$, $p$\,$=$\,$0.02$) is found between the spectral slope and radio-loudness (i.e., more radio-quiet QSOs appear to be slightly steeper). Splitting the sample by radio-loudness, we find that the average radio spectral slope for the radio-quiet/intermediate QSOs ($\mathcal{R}$\,$<$\,$-3.7$) is much steeper than that for the extreme radio-loud QSOs ($\mathcal{R}$\,$>$\,$-3.7$; $\alpha$\,$=$\,$-0.70$ and $\alpha$\,$=$\,$-0.08$, respectively).

Utilizing only the uGMRT data we have found trends between the steepness of the radio spectral slope with dust extinction and radio-loudness. Therefore, this two-point fitting shows some initial tentative differences between the underlying shape of the radio SEDs in the QSOs as a function of dust extinction. In the following sections, we include the additional archival data (see Section~\ref{sec:archival}) in order to model and characterize the full radio SED from 0.144\,MHz--3\,GHz.

\subsection{Radio SEDs of red and blue QSOs}\label{sec:sed}

\begin{table*}
\centering
\caption{Table displaying the radio SED fitting results. The columns from left to right display the: (1) SDSS name, (2) measured dust extinction ($E(B-V)$), (3) uGMRT Band-3 to Band-4 spectral index ($\alpha_{\rm B3-B4}$), (4) best-fitting model (PL, BPL, or curve) for each QSO in our sample, including the sub-classification (steep, flat, inverted, peaked, or curved; see Section~\ref{sec:model_characterisation}), (5) BIC value for the best fitting model, (6)-(11) spectral index for the PL model ($\alpha_{\rm PL}$; see Eq.~\ref{eq:pl}), spectral index for the BPL model ($\alpha_{\rm BPL}$; see Eq.~\ref{eq:bpl}), peak frequency ($\nu_{\rm peak}$) and spectral indices in the optically thick and thin regime for the curve model ($\alpha_{\rm thick}$ and $\alpha_{\rm thin}$; see Eq.~\ref{eq:curve}), depending on the best-fitting model (for the BPL model, the $\alpha_{\rm BPL}$ displayed corresponds to $\alpha_{\rm high}$ in Eq.~\ref{eq:bpl} if the majority of data points have a $\nu$\,$>$\,$\nu_{\rm break}$ and $\alpha_{\rm low}$ vice versa), (12) whether the source shows extended radio emission in the $0\farcs2$ e-MERLIN imaging (E) or the $6''$ LoTSS imaging (L), and (13) which survey data is included in the SED fitting: FIRST (F), LoTSS (L), TGSS (T), VLASS (V), uGMRT Band-3 (B3), and uGMRT Band-4 (B4). An electronic table containing the SED fitting classifications and best fitting parameters can be found in the online Supplementary material. \newline \textsuperscript{\textdagger}VLASS integrated flux density $<$\,3\,mJy. \textsuperscript{\textdaggerdbl}Detected in TGSS and not covered by LoTSS. $^*$Originally classified as upturned (see Section~\ref{sec:model_characterisation}). $^\alpha$Unconstrained model fit parameters due to lack of degrees of freedom (see Section~\ref{sec:model_characterisation}). \textsuperscript{\S}No Band-3 data. }
\begin{tabular}{|c|c|c|c|c|c|c|c|c|c|c|c|c|c}
\hline
\hline
Name & $E(B-V)$ & $\alpha_{\rm B3-B4}$ & Best model & BIC$_{\rm best}$ & $\alpha_{\rm PL}$ & $\alpha_{\rm BPL}$ & $S_{\rm peak}$ & $\nu_{\rm peak}$ & $\alpha_{\rm thin}$ & $\alpha_{\rm thick}$ & Ext.? & Data \\
& [mag] & & & & & & [mJy] & [MHz] & & & & \\
\hline

1657+2045 & 0.01 & -0.58 & PL (Flat) & 16 & -0.41 & & & & & & & F,\,V,\,B3,\,B4 \\
1038+4155 & 0.01 & 0.74 & PL (Flat) & 92 & 0.18 & & & & & & & F,\,L,\,V,\,B3,\,B4 \\
0748+2200 & 0.01 & 0.24 & Curve (Peak) & 3 & & & 7 & 1925 & -3.21 & 0.24 & & F,\,V\textsuperscript{\textdagger},\,B3,\,B4 \\
1103+5849 & 0.01 & 0.43 & Curve (Curve) & 5 & & & 6 & 757 & -0.02 & 0.83 & & F,\,L,\,V,\,B3,\,B4 \\
1554+2859 & 0.02 & -0.34 & BPL (Flat) & 4 & & -0.33 & & & & & & F,\,L,\,V\textsuperscript{\textdagger},\,B3,\,B4 \\
1046+3427 & 0.02 & 0.92 & Curve (Peak*) & 3 & & & 24 & 504 & -0.01 & 2.19 & L & F,\,L,\,V,\,B3,\,B4 \\
1057+3315 & 0.02 & -0.73 & PL (Flat) & 16 & -0.42 & & & & & & & F,\,L,\,V\textsuperscript{\textdagger},\,B3,\,B4 \\
1222+3723 & 0.02 & -0.72 & Curve (Peak) & 15 & & & 22 & 167 & -0.61 & 3.15 & & F,\,L,\,V,\,B3,\,B4 \\
1602+4530 & 0.02 & -0.81 & BPL (Steep) & 22 & & -0.72 & & & & & & F,\,L,\,V,\,B3,\,B4 \\
1530+2310 & 0.02 & -0.57 & PL (Flat) & 30 & -0.23 & & & & & & & F,\,V,\,B3,\,B4 \\
1003+2727$^\alpha$ & 0.03 & -0.95 & BPL (Steep) & 5 & & -1.05 & & & & & & F,\,V,\,B3,\,B4 \\
1428+2916 & 0.04 & 1.07 & Curve (Peak) & 3 & & & 34 & 1713 & -2.09 & 1.12 & L & F,\,L,\,V,\,B3,\,B4 \\
1410+2217\textsuperscript{\textdaggerdbl} & 0.04 & -0.33 & PL (Flat) & 4 & -0.32 & & & & & & E & F,\,T,\,V,\,B3,\,B4 \\
1432+2925 & 0.05 & -0.17 & Curve (Peak) & 3 & & & 24 & 392 & -0.61 & 0.79 & & F,\,L,\,V,\,B3,\,B4 \\
1042+4834 & 0.05 & -0.09 & Curve (Peak) & 94 & & & 9 & 2918 & -3.80 & 0.11 & & F,\,L,\,V,\,B3,\,B4 \\
1630+3847 & 0.05 & 1.27 & Curve (Peak*) & 2 & & & 11 & 2018 & -0.23 & 1.28 & L & F,\,L,\,V,\,B3,\,B4 \\

1202+6317 & 0.05 & 0.26 & Curve (Peak) & 7 & & & 4 & 1323 & -0.82 & 1.32 & & F,\,L,\,V,\,B3,\,B4 \\

1304+3206 & 0.06 & -0.02 & PL (Flat) & 51 & -0.20 & & & & & & L & F,\,L,\,V,\,B3,\,B4 \\
1203+4510 & 0.06 & 0.90 & Curve (Peak*) & 2 & & & 33 & 713 & -0.50 & 1.54 & L & F,\,L,\,V,\,B3,\,B4 \\

1140+4416 & 0.06 & 0.65 & Curve (Peak) & 41 & & & 5 & 2890 & -3.89 & 0.33 & & F,\,L,\,V,\,B3,\,B4 \\
1057+3119 & 0.06 & -0.08 & PL (Flat) & 7 & 0.03 & & & & & & & F,\,L,\,V,\,B3,\,B4 \\
1315+2017\textsuperscript{\S}$^\alpha$ & 0.07 &  & PL (Flat) & 4 & 0.24 & & & & & & & F,\,V,\,B4 \\
0828+2731 & 0.08 & -0.68 & PL (Steep) & 8 & -0.71 & & & & & & & F,\,L,\,V\textsuperscript{\textdagger},\,B3,\,B4 \\

1019+2817$^\alpha$ & 0.08 & 0.14 & Curve (Peak) & 3 & & & 14 & 2610 & -3.83 & 0.16 & & F,\,V,\,B3,\,B4 \\

1211+2221 & 0.09 & 0.44 & PL (Flat) & 10 & 0.23 & & & & & & & F,\,V\textsuperscript{\textdagger},\,B3,\,B4 \\
\hline
1410+4016 & 0.11 & -0.46 & Curve (Peak) & 32 & & & 8 & 145 & -0.27 & 3.96 & & F,\,L,\,V,\,B3,\,B4 \\
1323+3948 & 0.11 & 2.15 & Curve (Peak) & 3 & & & 11 & 1664 & -0.63 & 2.15 & & F,\,L,\,V,\,B3,\,B4 \\
1251+4317 & 0.13 & -0.40 & PL (Flat) & 113 & -0.34 & & & & & & & F,\,L,\,V,\,B3,\,B4 \\
1122+3124 & 0.13 & -0.71 & BPL (Steep) & 9 & & -0.64 & & & & & E & F,\,L,\,V,\,B3,\,B4 \\
0823+5609 & 0.13 & 0.30 & PL (Flat) & 3 & 0.23 & & & & & & & F,\,L,\,V,\,B3,\,B4 \\
1159+2151 & 0.17 & -0.72 & PL (Steep) & 42 & -0.58 & & & & & & & F,\,V,\,B3,\,B4 \\
1535+2434$^\alpha$ & 0.18 & -0.02 & BPL (Flat) & 7 & & 0.0 & & & & & E & F,\,V,\,B3,\,B4 \\
1153+5651 & 0.21 & -1.12 & BPL (Steep) & 515 & & -0.82 & & & & & E & F,\,L,\,V,\,B3,\,B4 \\
1531+4528 & 0.22 & -0.82 & PL (Steep) & 62 & -0.80 & & & & & & L & F,\,L,\,V,\,B3,\,B4 \\
0946+2548\textsuperscript{\textdaggerdbl} & 0.27 & -1.18 & BPL (Steep) & 5 & & -1.02 & & & & & E & F,\,T,\,V,\,B3,\,B4 \\
0951+5253 & 0.27 & -1.01 & PL (Flat*) & 22 & -0.07 & & & & & & L & F,\,L,\,V,\,B3,\,B4 \\
1342+4326 & 0.42 & -0.14 & Curve (Peak) & 27 & & & 20 & 178 & -0.13 & 2.78 & & F,\,L,\,V,\,B3,\,B4 \\
1007+2853 & 0.59 & -1.14 & BPL (Steep) & 5 & & -1.00 & & & & & E & F,\,L,\,V,\,B3,\,B4 \\

\hline
\hline
\end{tabular}
\label{tab:radio_tab}
\end{table*}

We fit the radio SEDs of the QSOs with a PL model (Eq.~\ref{eq:pl}), a broken PL (BPL) model (Eq.~\ref{eq:bpl}), and a curve model (Eq.~\ref{eq:curve}) in order to characterize the radio spectral shape, following the methodology described in Section~\ref{sec:model_characterisation}. Additionally, each QSO was assigned a sub-classification depending on either the value of $\alpha$ in the case of a PL or BPL model best fit, or whether there was a spectral turnover in the model within the frequency range of the radio data utilized (see Section~\ref{sec:model_characterisation} for sub-classification definitions). Example radio SED fits are display in Fig.~\ref{fig:example_seds}, with the full sample displayed in Figs.~\ref{fig:SED_rQSO} and \ref{fig:SED_cQSO}. A breakdown of the fitting parameters for each QSO is displayed in Table~\ref{tab:radio_tab}. 

We found that the majority of QSOs are best fit by the either the PL or BPL model (23/38; 61$\pm$8 per~cent\footnote{All the percentage errors in this paper were calculated using the method described in \cite{cam} and correspond to $1\sigma$ binomial uncertainties.}; 15 PL and 8 BPL). Out of these PL/BPL sources, 9/23 (39$\pm$9 per~cent) have steep ($\alpha_{\rm PL}$\,$<$\,$-0.5$) radio spectral indices, indicating optically-thin and evolved synchrotron emission from either radio jets or shocks caused by outflows and/or jets interacting with the surrounding ISM \citep{faucher,nims}. The other PL/BPL sources 14/23 (61$\pm$11 per~cent) have a flat radio spectrum ($-0.5$\,$<$\,$\alpha_{\rm PL}$\,$<$\,$0.5$), indicating either a self-absorbed radio nucleus or an unresolved jet-base, both on very compact scales \citep{rybicki,eckart}. No QSOs displayed a rare inverted PL spectrum.\footnote{Note: two curved sources (i.e., those not well described by a PL or BPL) display extremely inverted slopes (1103+5839 and 1323+3948), indicating a very absorbed radio nucleus or a GPS-like source that peaks at higher frequencies than our data covers.} Of the sources not well-described by a PL or BPL model, 15/38 (39$\pm$7 per~cent), were best fit with the curve model, with 14 displaying a clear peak within the frequency of data utilized in the fitting. A peak in the radio SED is expected to be caused by either synchrotron self-absorption (e.g., \citealt{snellen}) or free-free absorption by ionized gas surrounding the radio emission (e.g., \citealt{stawarz}). 

We find that dusty QSOs (those with $E(B-V)$\,$>$\,$0.1$\,mag) are more likely to exhibit steep spectral slopes ($\alpha$\,$<$\,$-0.5$) than the non-dusty QSOs ($E(B-V)$\,$<$\,0.1\,mag), with 6/13 (46$\pm$12 per~cent) and 3/25 (12$\pm$4 per~cent), respectively. On the other hand, a higher fraction of the non-dusty sources are curved sources compared to the dusty QSOs; 12/25 (48$\pm$9 per~cent) and 3/13 (23$\pm$8 per~cent), respectively. There are no significant differences in the fraction of sources that are flat; 4/13 (31$\pm$10 per~cent) and 10/25 (40$\pm$9 per~cent) for the dusty and non-dusty QSOs, respectively. This could suggest that the radio emission in non-dusty QSOs is more likely to be core dominated, potentially due to compact jets, whereas the radio emission in dusty QSOs is more likely to be extended, either due to outflow-driven shocks and/or more extended jets. We explore this further in Section~\ref{sec:discussion}. Isolating the eight BPL sources, which indicate high frequency synchrotron losses likely due to evolved radio emission, we find a higher fraction of the dusty QSOs are best fit by a BPL compared to the non-dusty QSOs; 5/13 (38$\pm$11 per~cent) and 3/25 (12$\pm$4 per~cent), respectively. All five of the dusty QSOs best fit with a BPL model are also extended in the $0\farcs2$ e-MERLIN 1.4\,GHz images, which is expected for an evolved, optically-thin synchrotron component (e.g., \citealt{laor}). The other extended e-MERLIN source is non-dusty ($E(B-V)$\,$=$\,0.04\,mag) and is best fit by a flat PL.  

\begin{figure}
    \centering
    \includegraphics[width=0.45\textwidth]{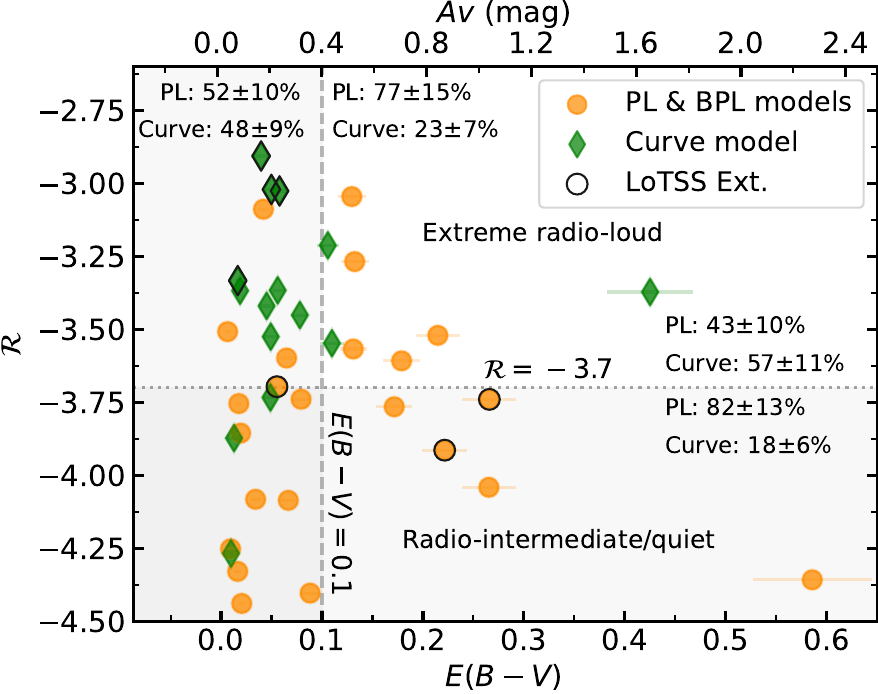}
    \caption{Radio-loudness versus dust extinction for the QSOs, split into sources best fit with the PL or BPL models (orange circles) or the curve model (green diamonds). The dashed vertical line indicates the split between radio-quiet/radio-intermediate ($\mathcal{R}$\,$<$\,$-3.7$) and extreme radio-loud ($\mathcal{R}$\,$>$\,$-3.7$) sources. The dotted horizontal line indicates $E(B-V)$\,$=$\,0.1\,mag. The majority of the QSOs at $\mathcal{R}$\,$<$\,$-3.7$ are best fit by the PL or BPL models (82$\pm$13 per~cent). A higher percentage of dusty QSOs at $E(B-V)$\,$>$\,0.1\,mag are best fit by the PL or BPL models (77$\pm$15 per~cent) compared to the non-dusty QSOs (52$\pm$10). This indicates that both radio-loudness and dust extinction may play a role in determining the radio spectral shape of QSOs.}
    \label{fig:ebv_R}
\end{figure}

Exploring the fitting results further, Fig.~\ref{fig:ebv_R} displays radio-loudness versus $E(B-V)$ for the QSOs, highlighting the sources best fit by either the PL/BPL or curve model. Splitting the samples at a radio-loudness value of $\mathcal{R}$\,$=$\,$-3.7$ (motivated by \citetalias{klindt}, who found red QSOs showed differences in their radio properties at a $\mathcal{R}$\,$<$\,$-3.7$) we find that the majority of radio-quiet/radio-intermediate QSOs ($\mathcal{R}$\,$<$\,$-3.7$) are best fit by the PL/BPL models (14/17; 82$\pm$13 per~cent) compared to the curve model (3/17; 18$\pm$6 per~cent). For the extreme radio-loud QSOs ($\mathcal{R}$\,$>$\,$-3.7$), there is an even split between QSOs best fit by the PL/BPL models (9/21; 43$\pm$10 per~cent) or the curve model (12/21; 57$\pm$11 per~cent), although we find 80$\pm$14 per~cent of the curved sources lie in this region. Overall, this implies that the curved radio SEDs are more likely to be associated with radio-loud QSOs.

Splitting by dust extinction ($E(B-V)$\,$=$\,0.1\,mag), we find a higher fraction of dusty sources are best fit by the PL or BPL model (10/13; 77$\pm$15 per~cent) compared to the curve model (3/13; 23$\pm$7 per~cent). We find no significant differences between the fraction of the non-dusty sources best fit by either the PL/BPL models or the curve model; 13/25 (52$\pm$10 per~cent) and 12/25 (48$\pm$10 per~cent), respectively. Therefore, radio-loudness appears to be the most important parameter in determining whether the radio SED of a QSO is peaked or not, with more radio-quiet and dusty QSOs preferring a PL/BPL fit. 

\begin{figure}
    \centering
    \includegraphics[width=0.48\textwidth]{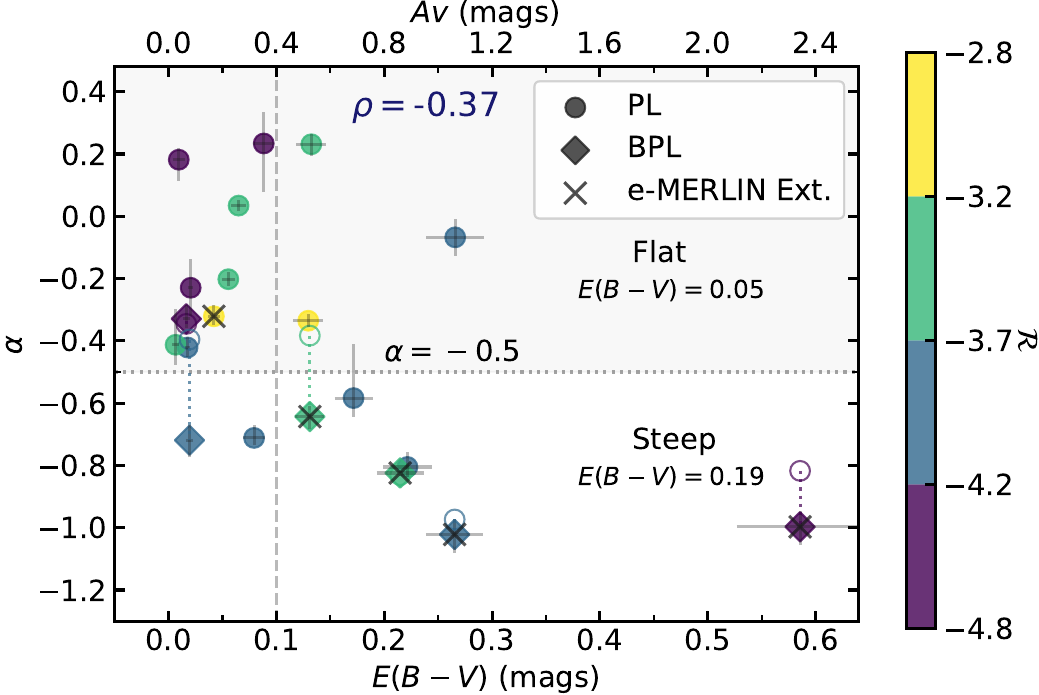}
    \caption{Radio spectral index obtained from the PL (circles) and BPL (diamonds) models versus $E(B-V)$, removing the unconstrained sources (see Table~\ref{tab:radio_tab}). The QSOs with extended $0\farcs2$ e-MERLIN radio emission are indicated by the black crosses. The best PL fit for the BPL sources are shown in the open circles, indicating that the radio spectral slope obtained with either model is fairly similar. The horizontal dotted line indicates the boundary between steep ($\alpha_{\rm PL}$\,$<$\,$-0.5$) and flat ($-0.5$\,$<$\,$\alpha_{\rm PL}$\,$<$\,$0.5$) radio spectral slopes. The colour bar indicates the radio loudness. The Spearman's rank correlation coefficient ($\rho$) is displayed, demonstrating a weak negative trend between $\alpha_{\rm PL}$ and $E(B-V)$.}
    \label{fig:ebv_alpha}
\end{figure}

Fig.~\ref{fig:ebv_alpha} displays the radio spectral index from the PL or BPL model versus the measured dust extinction (for the BPL model, the $\alpha$ adopted is $\alpha_{\rm high}$ if the majority of data points have a $\nu$\,$>$\,$\nu_{\rm break}$ and $\alpha_{\rm low}$ vice versa). We removed the three PL/BPL QSOs with unconstrained parameters due to lack of degrees of freedom (see Section~\ref{sec:model_characterisation} and Table~\ref{tab:radio_tab}). The horizontal dashed line indicates the boundary in $\alpha_{\rm PL}$ between the steep and flat sub-classifications. The steep spectrum PL/BPL sources are, on average, considerably redder compared to the flat sources (median $E(B-V)$\,$=$\,$0.19$ and $E(B-V)$\,$=$\,$0.05$\,mag for steep and flat, respectively). Furthermore, applying the Spearman's rank correlation coefficient reveals a tentative negative trend between $\alpha_{\rm PL}$ and $E(B-V)$ ($\rho$\,$=$\,$-0.37$, $p$\,$=$\,$0.11$; also found in Fig.~\ref{fig:uGMRT_alpha}). This suggests that there is a connection between the steepness of the radio spectral slope and the dust extinction in QSOs. Removing the QSO with the highest dust extinction ($E(B-V)$\,$\sim$\,0.6\,mag) reduces the significance of the trend ($\rho$\,$=$\,$-0.27$, $p$\,$=$\,$0.26$), demonstrating the need for a larger sample of QSOs with high values of dust extinction to robustly test this result. However, even after removing the extreme dusty QSO, the median dust extinction for the steep sources remains considerably higher than that of the flat sources ($E(B-V)$\,$=$\,$0.17$ and $E(B-V)$\,$=$\,$0.05$\,mag for steep and flat, respectively). We note that the non-dusty inverted sources found in Fig.~\ref{fig:uGMRT_alpha} are no longer present after the inclusion of the archival data due to these sources now best fit by the curve model. The majority of the e-MERLIN extended sources tend to be dusty and steep, suggesting that the tentative connection between $E(B-V)$ and $\alpha_{\rm PL}$ might be driven by small-scale extended radio emission (i.e., $\lesssim$2\,kpc). We discuss the potential mechanisms behind this connection in Sections~\ref{sec:jets} and \ref{sec:winds}.

\section{Discussion}\label{sec:discussion}

In this paper, we have used a combination of dedicated uGMRT observations (Band-3 and 4) and archival radio data to explore the radio SEDs of a sample of QSOs across a range of dust extinctions, with additional size and morphological information from previous $0\farcs2$ e-MERLIN 1.4\,GHz imaging (presented in \citetalias{rosario_21}). On the basis of our analyses, we have found evidence that dusty QSOs are more likely to exhibit a steep synchrotron radio spectrum compared to non-dusty QSOs, which are more likely to be peaked. In the following sections, we discuss this key result (Section~\ref{sec:steep_PL}), what is the most likely origin of the radio emission in dusty QSOs and whether this is different to typical blue QSOs (Sections~\ref{sec:jets} and \ref{sec:winds}), and explore any evidence that dusty QSOs are in an earlier evolutionary phase (Section~\ref{sec:evo}).

\subsection{Steep radio power-laws are more common in dusty QSOs}\label{sec:steep_PL}
Splitting the QSOs into two bins of reddening (boundary: $E(B-V)$\,$=$\,0.1\,mag), we found that a higher fraction of dusty QSOs are steep compared to the non-dusty QSOs; 6/13 (46$\pm$12 per~cent) and 3/25, 12$\pm$4 per~cent, respectively. Furthermore, comparing the steepness of the radio spectral slope to the dust extinction, we found a weak but significant correlation ($\rho$\,$=$\,$-0.37$, $p$\,$=$\,0.11), where the dusty QSOs are more likely to have steep spectral slopes (Fig.~\ref{fig:ebv_alpha}), also confirmed from only exploring the uGMRT Band-3--4 radio spectral slopes (Fig.~\ref{fig:uGMRT_alpha}). We also found a much higher fraction of radio-quiet/radio-intermediate sources ($\mathcal{R}$\,$<$\,$-3.7$) are fit by the PL/BPL models compared to the curve model (82$\pm$13 per~cent and 18$\pm$6 per~cent, respectively; Fig.~\ref{fig:ebv_R}). This suggests a causal connection between the radio spectral index and the dust extinction in QSOs, whereby dusty QSOs are more likely to be radio-quiet/intermediate and display a steep spectrum, which could arise from the interaction between winds and/or jets with the surrounding dusty ISM. We found a higher fraction of the non-dusty QSOs ($E(B-V)$\,$<$\,0.1\,mag) had peaked radio spectra compared to the dusty QSOs; 12/25 (48$\pm$9 per~cent) and 3/13 (23$\pm$8 per~cent), respectively, which indicates a young or frustrated radio jet. Additionally, the majority of the peaked sources are extremely radio-loud ($\mathcal{R}$\,$>$\,$-3.7$; Fig.~\ref{fig:ebv_R}; 80$\pm$14 per~cent). We found no difference in the fraction of flat spectrum sources for the dusty and non-dusty QSOs.

Overall, these results suggest that the radio emission in non-dusty blue QSOs is more likely due to radio-loud compact jets. On the other hand, dusty QSOs are more likely to host outflows which shock the surrounding dusty medium, resulting in more radio-quiet, small-scale extended radio emission. However, both radio jets and shocks due to winds and/or jets interacting with the ISM can cause extended radio emission with steep spectral slopes. Therefore, it is still not clear whether jets or winds (or both) are the origin of the radio emission in radio-quiet/intermediate dusty QSOs; we discuss this further in Sections~\ref{sec:jets} and \ref{sec:winds}. Although radio emission from star-formation can also produce a steep spectral slope ($\alpha$\,$<$\,$-0.5$; e.g., \citealt{Kimball_2011,condon,calistro_17}), we do not discuss this in the following sections due to the significant amount of evidence that suggests star-formation is not responsible for the enhanced radio emission observed in dusty QSOs in this luminosity regime, i.e., $L_{\rm 1.4\,GHz}\approx10^{25-26}$\,W\,Hz$^{-1}$ \citep{fawcett20,rosario,calistro,calistro_2024,bohan}.

\subsection{Jets as the origin of the radio emission}\label{sec:jets}

Traditionally, radio jets have been associated with the most radio-loud QSOs (e.g., \citealt{kellermann}). However, recent work has shown that radio jets may be more ubiquitous in QSOs, with observations of low-powered radio jets in traditionally radio-quiet sources, often associated with driving multi-phase outflows \citep{jarvis,venturi,macfarlane,girdhar,bohan,calistro_2024}. In our sample, the QSOs with a peaked radio SED tend to be radio-loud (Fig.~\ref{fig:ebv_R}) and don't follow a strong trend with dust extinction compared to the PL/BPL sources. This could be evidence that the origin of the radio emission in these QSOs is small-scale jets, since a peaked spectrum is often associated with either a jet ``frustrated'' by the ISM or a young jet \citep{odea_1991}. One way to test this theory is to compare the peaked QSOs to known jetted objects, such as GPS and CSS sources \citep{odea_1997}. GPS and CSS sources are also characterized by a peaked radio SED and have compact radio emission which is thought to be jet-dominated; whether the radio jets are young or frustrated is still debated \citep{van_bregel,odea_1991,bicknell_18}.

\begin{figure*}
    \centering
    \includegraphics[width=0.75\textwidth]{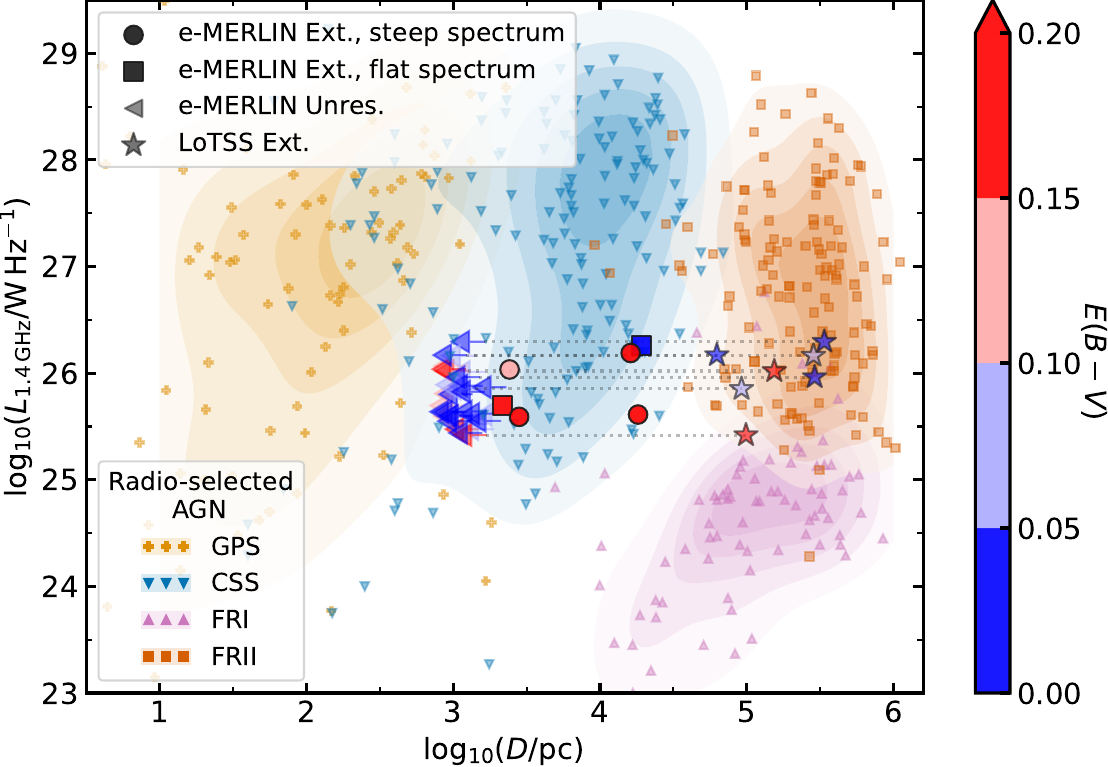}
    \caption{$L_{\rm 1.4\,GHz}$ versus projected linear size for the full sample, with the colour bar indicating dust extinction. The e-MERLIN extended sources with a radio SED best fit by either a PL/BPL with $\alpha$\,$>$\,0 or a curve model are indicated by the circles and squares, respectively. The unresolved sources are displayed as upper limits on the projected size. The coloured regions indicate where typical radio-selected AGN populations lie in this space, compiled by \protect\cite{An_2012}. We find that all of the QSOs have luminosities that are consistent with CSS-like sources, but require better constraints on the linear size for the e-MERLIN unresolved sources to determine if their sizes are more consistent with the GPS region. The LoTSS linear projected size is shown for the seven QSOs with extended LoTSS emission (stars); they are all in agreement with the FRII region and are more commonly blue. Figure adapted from \protect\cite{jarvis}.}
    \label{fig:radio_size}
\end{figure*}

To test whether our QSOs are consistent with GPS/CSS-like objects, we plotted the full QSO sample on the radio luminosity versus linear size plot (Fig.~\ref{fig:radio_size}), utilizing the size constraints from e-MERLIN. We can then compare how the radio sizes compare to common radio populations: GPS, CSS, and Fanaroff-Riley Class~I (FRI; brighter radio core compared to the lobes) or Class~II (FRII; brighter radio lobes compared to the core; \citealt{fr}). All of the e-MERLIN extended QSOs have radio luminosities and sizes consistent with CSS sources. However, due to the upper limits for the e-MERLIN unresolved sources, we are unable to determine if these sources are more consistent with the more compact GPS sources. 

\begin{figure}
    \centering
    \includegraphics[width=0.45\textwidth]{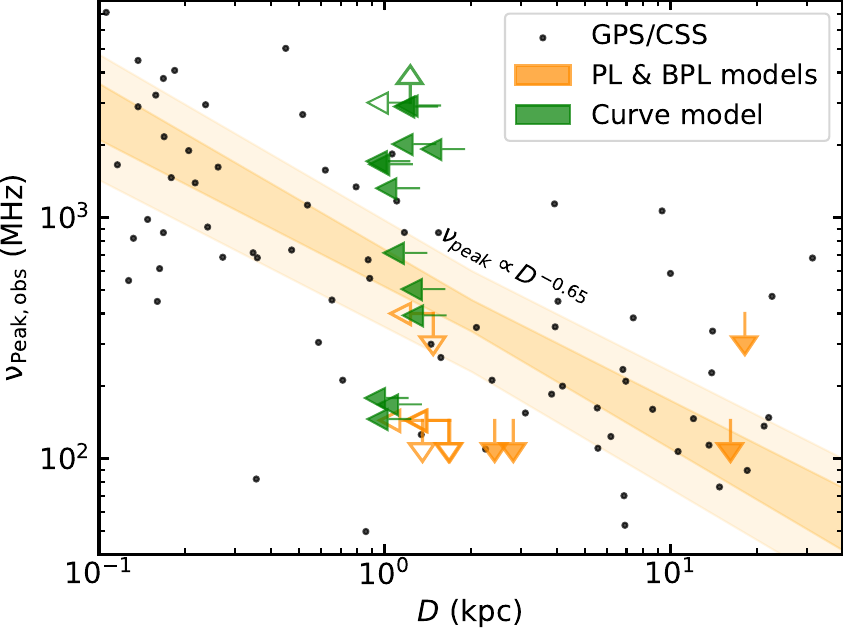}
    \caption{Peak observed frequency versus projected size obtained from e-MERLIN for the QSOs with either a curved (green), or a steep PL/BPL (orange) radio spectrum, removing the unconstrained sources (see Table~\ref{tab:radio_tab}). The peak frequency is plotted as a limit for the PL/BPL and curved sources. All sizes are displayed as upper limits. The open arrows indicate a source with an upper or lower limit on both axis. The grey dots display the GPS and CSS sources compiled in \protect\cite{jeyakumar_2016}. The darker shaded orange region displays the relationship (with scatter; Eq.~\ref{eq:css}) from \protect\cite{odea_1997} and the lighter shaded region displays the $1\sigma$ bounds. In the absence of more constraining size measurements, there is no strong evidence that these QSOs are inconsistent with the GPS/CSS population.}
    \label{fig:gps_size}
\end{figure}

Although the QSOs in our sample have similar size constraints and radio luminosities to GPS/CSS sources, we require additional information in order to robustly determine if they are consistent with jetted sources. Another parameter we can explore is the peak frequency of the spectral turnover from the radio SED fits. For GPS/CSS sources, the frequency of the spectral turnover has been found to strongly anti-correlate with the projected linear size of the radio emission, as expected for synchrotron self-absorption \citep{fanti90}. For example, \cite{odea_1997} found the simple relationship:
\begin{equation}\label{eq:css}
    \rm log\nu_{m} \simeq -0.21(\pm0.05)-0.65(\pm0.05)\times \rm log \textit{l} ,\
\end{equation}
(or $\nu_{\rm peak}$\,$\propto$\,$D^{-0.65}$) where $D$ is the linear size of the radio emission.
In Fig.~\ref{fig:gps_size} we plot the peak frequency versus projected linear size obtained from the e-MERLIN 1.4\,GHz images ($0\farcs2$; $\sim$\,kpc-scale). We also plot on the steep PL/BPL sources since these may have a radio SED that peaks beyond our frequency range (i.e., $\nu_{\rm peak}$\,$<$\,$144$\,MHz, if LoTSS data is available). None of the 15 QSOs that were best fit by the curve model are extended in e-MERLIN and therefore have size upper limits. Out of the nine QSOs best fit with a steep PL or BPL, only four have constrained e-MERLIN sizes. The other QSOs were unresolved at the resolution of e-MERLIN and therefore are plotted as upper limits for the projected linear size (see Section~\ref{sec:e-merlin} for more details on calculating the e-MERLIN sizes). For the QSOs that did not show a clear turnover within the frequency range of the radio data utilized, an upper or lower limit on the peak frequency corresponding to either the maximum or minimum frequency data point is plotted, respectively, depending on whether the overall shape of the spectrum tends to be steep or inverted. We find a few sources, both PL/BPL and curved, which are on the edge of the GPS/CSS distribution and therefore are potentially inconsistent with GPS/CSS sources, although we cannot robustly conclude this due to upper limits. Most of the other QSOs best fit by the curve model (other than the one QSO, 1103+5849, that did not display a peak in the SED and therefore also has a limit on the peak frequency) appear to be consistent with GPS/CSS-like sources within their upper limits.

To robustly conclude or rule out whether any of these sources are GPS/CSS-like we require either: 1) higher spatial resolution radio data to constrain the projected linear size, and/or 2) lower/higher frequency radio data to either constrain a turnover or, if there is still no turnover present in the spectrum, push the upper limit for the steep/inverted sources further away from the relation. For example, two of the steep QSOs do not have LoTSS data and so their upper limit is fixed at the uGMRT Band-3 data point (400\,MHz). In a future study, we will use $0\farcs05$ e-MERLIN C-band data (5\,GHz; PI: V. Fawcett; Project ID: CY18008) for the same sample of QSOs studied in this paper to improve the size constraints by a factor of four, which will help to determine whether these QSOs are GPS/CSS-like objects.

In Fig.~\ref{fig:ebv_R} we found that the peaked QSOs, which tend to be non-dusty, are mostly radio-loud and do not follow any trends with $E(B-V)$. This could suggest that these sources are similar to GPS/CSS sources, powered by small-scale radio-loud jets. On the other hand, it is not clear whether the PL/BPL sources (in particular, the steep sources which tend to be more dusty) are also powered by compact radio jets. Since these sources tend to be more radio-quiet and display a correlation with the amount of dust extinction, it could be possible that the predominant radio emission mechanism in these sources is shocks due to outflows, which may be driven by radio jets or accretion disc winds (e.g., \citealt{haidar}). This is in agreement with \cite{calistro_2024}, who found an enhancement in the [\ion{O}{iii}] outflow velocities for radio-intermediate red QSOs, but not in the radio-faint regime, suggesting that dust was the key to producing large-scale ionized outflows and radio-emitting shocks. Additionally, although larger \ion{C}{iv} outflows were found in the red QSOs compared to the blue, no differences in the \ion{C}{iv} velocity distribution were found as a function of radio-loudness. \cite{calistro_2024} concluded that the mechanism producing the enhanced radio emission in red QSOs is not directly associated with the accretion disc, but on circumnuclear narrow line region scales such as jet-driven outflows, although this does not rule out wind-driven outflows (see Section~\ref{sec:winds}). Simulations also predict that the interplay between compact radio jets and circumnuclear environment is expected to produce both outflows and shocks \citep{mukherjee,bicknell_18,young_24}. This is also in agreement with recent work that has found an intrinsic, continuous connection between the amount of dust extinction in QSOs and the radio detection fraction \citep{fawcett23,petley_2024,calistro_2024}. These results could suggest that there are not distinct mechanisms producing the radio emission in dusty and non-dusty QSOs, but instead a continuous evolution such as jets of different sizes and/or ages \citep{hardcastle}. However, this scenario is also consistent with wind-driven shocks driving the radio emission in these QSOs, which we discuss in the following section.

\subsection{Winds as the origin of the radio emission}\label{sec:winds}

Shocks from radiatively driven winds due to AGN radiation on dust (e.g., \citealt{ishibashi,ishibashi_18,leftley_19,marta_20,arakwa_22}) has been suggested by multiple authors to explain the origin of compact radio emission \citep{nims,honig,calistro,yamada}. For example, exploring a sample of ``Extremely Red Quasars'' (ERQs; \citealt{hamann}), selected to have red optical--MIR colours, \cite{hwang} similarly found that the radio spectral slopes of the ERQs were steeper compared to typical blue QSOs. They also found that the radio emission and [\ion{O}{iii}] velocity widths were well correlated, concluding that wind-driven shocks were the mostly likely origin of the steep radio spectra in the ERQs. Another interesting QSO sub-population, Broad Absorption Line QSOs (BALQSOs; \citealt{weymann}) which are selected to have broad absorption troughs blueward of the \ion{C}{iv} emission line, indicative of powerful outflows, have been found to display redder optical spectra compared to non-BALQSOs \citep{morabito,petley}. Similarly, they are more radio-detected in the radio-quiet regime compared to non-BALQSOs \citep{morabito}, with radio emission that can be explained by a wind-shock model \citep{petley_2024}. The similarities between BALQSOs and dusty QSOs could suggest that winds are also more prevalent in the dusty QSO population. 

\begin{figure}
    \centering
    \includegraphics[width=0.45\textwidth]{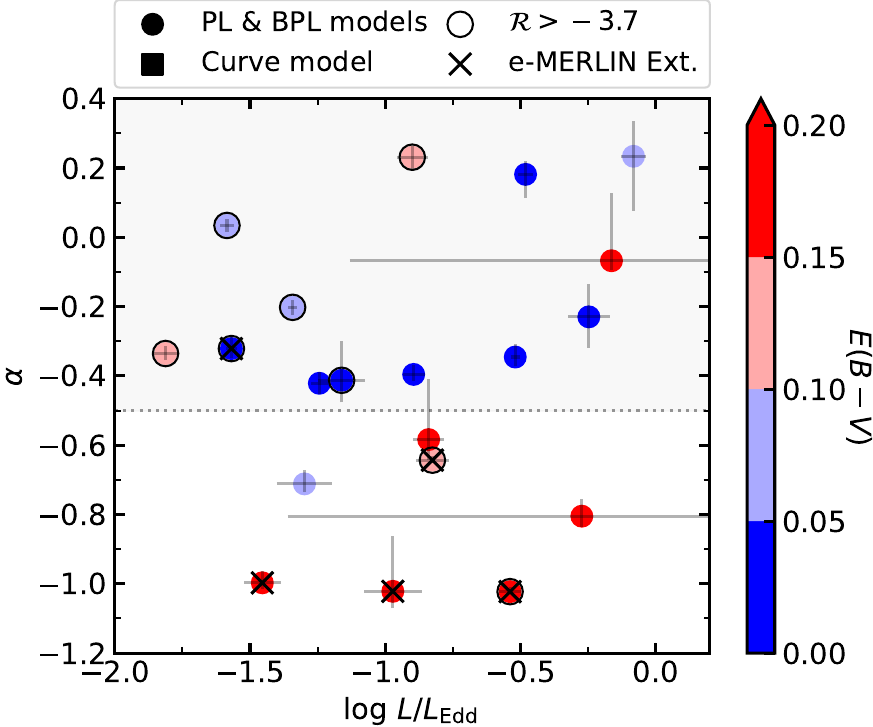}
    \vspace{2mm}
    \includegraphics[width=0.45\textwidth]{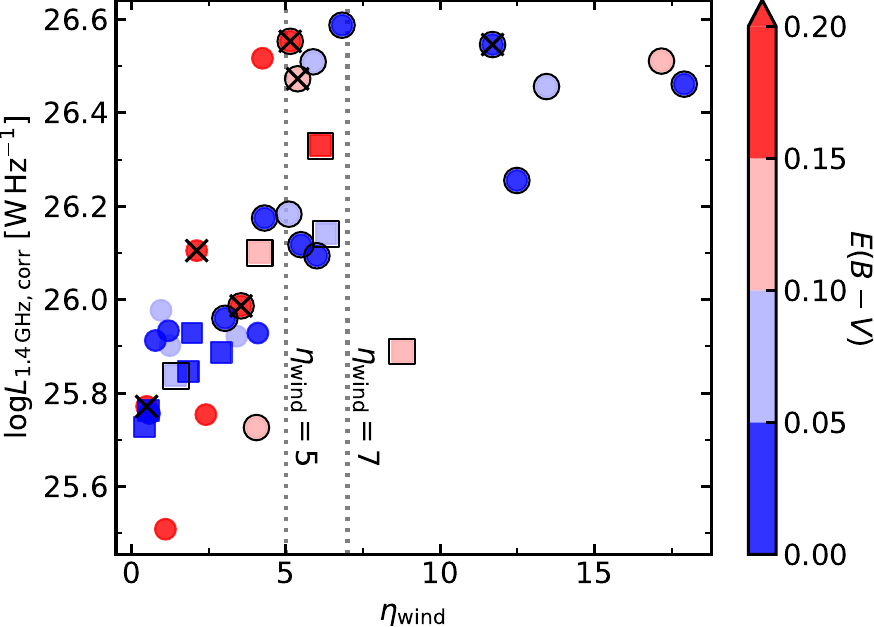}
    \caption{(Top) radio spectral slope from the PL/BPL models versus Eddington ratio, removing the unconstrained sources (see Table~\ref{tab:radio_tab}). There is no significant correlation between the steepness of the spectral slope and the Eddington ratio. It is possible that the QSOs with the steepest spectral slopes that are also close to the Eddington limit could have a strong wind component. (Bottom) 1.4\,GHz luminosity, calculated with the new measured $\alpha$ values for the sources best fit by either the PL/BPL (circles) or curve (square) models. For the curved sources, the standard $\alpha$\,$=$\,$-0.5$ is utilised for the radio luminosity. The vertical dotted lines indicates a wind efficiency of $\eta_{\rm wind}$\,$=$\,$5$ and $7$ per~cent; we find 6/38 (16$\pm$4 per~cent) of QSOs have an efficiency greater than 7 per~cent and so accretion disc winds are unlikely to account for the observed radio luminosity. In both plots the colours represent the measured dust extinction, the extreme radio loud sources are indicated by a black outline, and the e-MERLIN extended sources are indicated by the black crosses.}
    \label{fig:Redd}
\end{figure}

Exploring the 5--8.4\,GHz radio spectral slope for a sample of 25 radio-quiet Palomar-Green (PG) QSOs \citep{boroson}, \cite{laor} found a significant correlation between the Eddington ratio ($L/L_{\rm Edd}$) and the steepness of the radio spectral slope. Since a high $L/L_{\rm Edd}$ is often associated with a radiatively driven wind \citep{baskin}, they suggested that this correlation is due to wind-driven shocks in radio-quiet QSOs. However, they did not find a similar correlation between $L/L_{\rm Edd}$ and the steepness of the radio spectral slope in radio-loud QSOs, which could suggest two different mechanisms are at play for the radio-quiet and radio-loud QSOs (see also \citealt{med}). Comparing the 1.4--3\,GHz radio spectral slopes of a sample of MIR-selected red and blue QSOs, \cite{glikman22} found that the red QSOs had on average steeper radio spectral slopes ($\alpha$\,$=$\,$-0.70\pm0.05$ and $\alpha$\,$=$\,$-0.34\pm0.04$ for the red and blue QSOs, respectively). Due to the significantly higher Eddington ratios in their red sample compared to the blue, they followed the same argument as in \cite{laor}, suggesting that the steeper slopes are due to AGN-driven winds shocking the surrounding dust and gas. On the other hand, \cite{rosario} analysed the 144\,MHz--1.4\,GHz radio spectral slopes of a sample of SDSS selected red and blue QSOs and found no significant differences, with both populations peaking around $\alpha$\,$=$\,$-0.7$. In \cite{fawcett22} we found that SDSS red QSOs have on average slightly lower $L/L_{\rm Edd}$ values compared to blue QSOs, albeit with a small sample. However, other studies have found either no differences in the $L/L_{\rm Edd}$ values for red and blue QSOs \citep{calistro} or much higher $L/L_{\rm Edd}$ values for red QSOs compared to typical blue QSOs \citep{urrutia12,kim18,kim_2024_redd}.  

To test whether accretion-driven outflows could be responsible for the relationship between the steepness of the radio spectral slope and dust extinction, we calculated $L/L_{\rm Edd}$ for our sample utilizing the \cite{rakshit} catalogue. Bolometric luminosities ($L_{\rm bol}$) are provided in \cite{rakshit}, but these are inferred from the rest-frame UV-optical continuum measurements and have not been corrected for dust extinction (see e.g., \citealt{calistro,kim_2023}). Therefore we derive $L_{\rm bol}$ from the 6\,$\upmu$m luminosity by applying $L_{\rm bol}$\,$=$\,$BC_{\rm 6\upmu m}$\,$\times$\,$L_{\rm 6\upmu m}$, where $BC_{\rm 6\upmu m}$\,$=$\,8 from \cite{richards_2006}. We inferred $L_{\rm Edd}$ from the black hole mass (M$_{\rm BH}$), which we estimated using the viral relation. We adopted the M$_{\rm BH}$ calibration from \cite{Shen_2012} with the broad \ion{Mg}{ii} FWHM and $L_{\rm 3000}$ from \cite{rakshit}. We first corrected the $L_{\rm 3000}$ for dust extinction using the $E(B-V)$ values calculated in Section~\ref{sec:dust_method}. The uncertainties for $L/L_{\rm Edd}$ are calculated from the \ion{Mg}{ii} emission line fitting errors presented in \cite{rakshit}. The top panel of Fig.~\ref{fig:Redd} displays the resulting values for $L/L_{\rm Edd}$ compared to the steepness of the PL/BPL spectral slope; we find no significant correlation between the two properties ($\rho$\,$=$\,0.14, $p$\,$=$\,$0.54$), although our sample includes very radio-loud sources. This suggests that accretion-driven winds are unlikely to be the main driver behind the correlation between radio spectral steepness and dust extinction. However, \cite{temple_23} found that accretion-driven winds, measured from the blueshift of the \ion{C}{iv}$\lambda$1550 emission line, were correlated with both Eddington ratio and the black hole mass. Since our Eddington ratios are inferred from the black hole mass, future work utilising an independent measurement on Eddington ratio, such as \ion{C}{iv} distance \citep{richards_2011,rankine}\footnote{At the redshifts of our sample the \ion{C}{iv} emission line is outside of the SDSS wavelength coverage.}, is required to robustly test any correlations with $\alpha$. We find that one of the reddest QSOs (1153+5651) displays a very steep slope, is extended in e-MERLIN, and has a high Eddington ratio. For this source, it is possible that the origin of the radio emission is due to a wind-driven shock.

To further explore whether winds can explain the observed radio SEDs of our sample, we can test a popular wind-shock model presented in \cite{nims}. They first assume that a wind travelling at $0.1c$ through a typical ambient medium can reach $>1$\,kpc scales over the lifetime of the QSO. They then calculate the thermal and non-thermal emission that would be produced by the shock, including the associated synchrotron emission which they predict to be in the form:
\begin{equation}
    \nu L_{\nu} \approx 10^{-5} \eta_{\rm wind}L_{\rm bol} \bigg (\frac{L_{\rm wind}}{0.05 L_{\rm bol}}\bigg)
\end{equation}
Assuming an $L_{\rm wind}$\,$=$\,$0.05 L_{\rm bol}$, we calculate the values of $\eta_{\rm wind}$ required to produce the observed radio luminosities of our sample. This method produces an average value of $\eta_{\rm wind}$ to be 4 per~cent; this is in agreement with previous estimates based on observations of ionized outflows, which suggest an $\eta_{\rm wind}$\,$\lesssim$\,7\,per~cent \citep{liu_2013,sun}. However, six QSOs (16$\pm$4 per~cent), which are also the most radio-loud ($\mathcal{R}$\,$>$\,$-3.2$), require an $\eta_{\rm wind}$\,$>$\,7 per~cent, which is likely unfeasible (see bottom panel of Fig.~\ref{fig:Redd}). These sources therefore require an additional radio mechanism than an accretion-disc wind alone. 

An additional signature for wind-driven shocks is the detection of ionized outflows, which can be traced by the broadening or blue shifting of forbidden lines. The most popular of these lines, due to being typically bright, is the [\ion{O}{iii}]$\lambda$5007 emission line; due to the redshift range of our sample (1.0\,$<$\,$z$\,$<$\,1.55) we are unable to explore the [\ion{O}{iii}] ionized outflow properties in our sample. Previous studies have found an enhancement in the strength of [\ion{O}{iii}] outflows in reddened QSOs (e.g., \citealt{calistro, stepney}), although other studies have found no significant differences \citep{fawcett22}. To robustly determine whether enhanced ionized outflows are present in dusty QSOs we require NIR spectroscopy, which can observe [\ion{O}{iii}] out to higher redshifts; for example, the future Multi-Object Optical-to-NIR Spectrograph (MOONS; \citealt{moons}).

In conclusion, outflow-driven shocks are a likely mechanism to explain the relationship between dust extinction and radio spectral steepness for the radio-quiet/intermediate sources. A basic wind-shock model can account for the radio luminosities in the majority of these sources; however, it is not possible to completely rule out compact/low power radio jets as the outflow driving mechanism. In the following section we discuss how these results add to the overall discussion on whether dusty QSOs represent a phase in the evolution of QSOs.

\subsection{Are dusty QSOs in a younger evolutionary phase?}\label{sec:evo}
Although we cannot conclusively say what is the mechanism driving the radio emission in dusty QSOs (i.e., winds or jets), the higher fraction of PL/BPL sources (Section~\ref{sec:sed}) that are more likely to be radio-quiet/intermediate (Fig.~\ref{fig:ebv_R}) and the connection between the steepness of the PL slope and dust extinction (Fig.~\ref{fig:ebv_alpha}) suggests that outflow-driven (wind and/or jet) shocks are the most plausible origin of the radio emission in dusty QSOs. This would be consistent with the dusty ``blow-out'' phase (e.g., \citealt{hop6,glik7}), since the shocks will heat the surrounding dust, destroying it and eventually revealing an unobscured blue QSO. 

Further evidence that dusty QSOs are in a younger evolutionary phase arises from the LoTSS extended sources. Out of the seven QSOs with extended LoTSS radio emission, five are non-dusty ($E(B-V)$\,$\leq$\,$0.06$\,mag), with sizes consistent with FRII sources (Fig.~\ref{fig:radio_size}). Inspecting the LoTSS images for these sources, we find that some indeed do look like FRII objects with clear extended, symmetric bright lobes (0951+5253, 1203+4510, and 1630+3847). Others are less clear, with the presence of one bright lobe but not a second symmetric lobe (1531+4528 and 1046+3427). The remaining are only slightly extended in LoTSS, with no clear visible lobes (1304+3206 and 1428+2916). By selection, all of our sources display no extended core emission in the FIRST images (two sources do display large scale extension, see Appendix~\ref{sec:large_scale}), suggesting this extended low frequency emission may be evidence of relic radio emission. Indeed, \cite{An_2012} suggested that there is an evolutionary track in the radio luminosity--size plane, whereby compact GPS/CSS jets evolve over time into large-scale FRII-like jets (see Figure 1 therein). This is in agreement with non-dusty QSOs representing an older phase in QSO evolution compared to the dusty QSOs, with large, older jet-like structures that are only observable at the lower radio frequencies. 

Furthermore, three non-dusty QSOs compared to only one dusty QSO originally had upturned radio SEDs. After removing the LoTSS data point, all three of these QSOs display a peaked spectrum at $\sim$\,650--3000\,MHz frequencies, consistent with a young radio source \citep{kukreti,kukreti_2024}, and a steep radio spectrum at 144--400\,MHz, consistent with relic radio emission from previous AGN episodes (also known as ``restarted'' AGN; e.g., \citealt{baum_90,hancock,kharb_2016,Silpa2020,jurlin,dhanya}). This further suggests that these QSOs have already undergone a blow-out phase, with an older radio jet that has evolved (e.g., \citealt{An_2012}) and managed to escape the host-galaxy. The fact that a higher fraction of these QSOs are non-dusty supports this scenario, since we expect the dust responsible for the reddening in QSOs to get either destroyed or dispersed during a blow-out phase. Alternatively, it may be evidence that these QSOs are undergoing multiple cycles of activity on short timescales \citep[see for example][]{Nyland2020,harrison_2024}. However, we also find a higher fraction of the non-dusty QSOs that display a peaked spectrum (Section~\ref{sec:sed}), which are often associated with young sources (e.g., \citealt{kukreti,kukreti_2024}). This is in disagreement with the radio emission in non-dusty QSOs being older, although there might have been previous, older relic activity that has already faded and is no longer detectable with the low frequency radio data. A larger uGMRT sample and/or a wider statistical study is required to robustly tie down the fraction of QSOs with peaked radio SEDs and low frequency extended radio emission.

\section{Conclusions}
We have constructed 4--5 band radio SEDs from new uGMRT observations (at 400\,MHz and 650\,MHz), in combination with archival radio data, to analyse and compare the radio SEDs of 38 QSOs at $1.0$\,$<$\,$z$\,$<$\,$1.5$ ($L_{\rm 1.4\,GHz}$\,$\sim$\,$2\times10^{25-26}$\,W\,Hz$^{-1}$ and $L_{\rm 6\upmu m}$\,$\sim$\,$10^{45-46}$\,erg\,s$^{-1}$). Each SED was fit by either a power-law (PL), broken power-law (BPL), or curve model over a frequency range of 0.144--3\,GHz. With the additional constraints from our $0\farcs2$ e-MERLIN 1.4\,GHz imaging, we have explored trends between the radio spectral shape, dust obscuration, radio loudness, and morphology. From our analyses we found that:

\begin{itemize}
    \item \textbf{The majority of the QSOs have radio SEDs that are best-fit by a standard or broken power-law:} overall, 61$\pm$8 per~cent of QSOs are best fit by the PL/BPL models, with the remaining 39$\pm$7 per~cent best fit by the the curved model. Out of the PL/BPL sources, 39$\pm$9 per~cent are steep ($\alpha$\,$<$\,$-0.5$), indicating standard synchrotron emission, and 61$\pm$11 per~cent are flat, which either indicates a self-absorbed core or unresolved jet-base, both on very compact scales (Section~\ref{sec:sed}).
    \item \textbf{There is a trend between dust obscuration and the steepness of the radio spectral slope:} splitting the QSOs best fit by the PL/BPL models into two bins of dust extinction (boundary: $E(B-V)$\,$=$\,$0.1$\,mag), we found that a higher fraction of the dusty QSOs are steep (46$\pm$12 per~cent compared to 12$\pm$4 per~cent; Section~\ref{sec:sed}). Furthermore, we found a correlation between the amount of dust extinction and the steepness of the spectral slope, which is marginally statistically significant (Fig.~\ref{fig:ebv_alpha}) and confirms the result from exploring only the uGMRT Band-3--4 radio spectral slopes (Fig.~\ref{fig:uGMRT_alpha}). This is in conjunction with the previously known result that the dustier sources are also more likely to have extended 1.4\,GHz radio structures on a few kpc scales (\citetalias{rosario_21}). 
    \item \textbf{A higher fraction of the non-dusty QSOs display a peaked spectrum, which are consistent with radio-loud jetted systems:} 48$\pm$9 per~cent of the non-dusty QSOs ($E(B-V)$\,$<$\,$0.1$\,mag) are peaked compared to the dusty QSOs (23$\pm$8 per~cent). A peak in the radio spectral shape is associated with either a frustrated or young, compact jet. This is also in agreement with that fact that the majority of the curved sources are extremely radio-loud ($\mathcal{R}$\,$>$\,$-3.7$; Fig.~\ref{fig:ebv_R}). Plotting the turnover frequency versus the linear projected size obtained from the e-MERLIN imaging, we found that the majority of the QSOs best fit by the curve model are broadly consistent with GPS/CSS sources (Figs.~\ref{fig:radio_size} and \ref{fig:gps_size}). However, the majority of our sample are unresolved in the e-MERLIN imaging and, therefore, only have upper limits on the projected linear sizes. 
    \item \textbf{The radio emission in dusty QSOs is likely due to jet- or wind- driven shocks:} we found that the radio emission calculated from a wind-shock model is unable to recreate the radio luminosities observed in 39$\pm$7 per~cent of our sample, which represent the most radio-loud objects (Section~\ref{sec:winds}). One of the reddest QSOs that displays a very steep radio spectral slope also has a very high Eddington ratio, which could suggest that accretion-driven winds are responsible for the radio emission in this object (Fig.~\ref{fig:Redd}). However, for the other QSOs it is currently unclear whether jets or winds are responsible for driving the radio emission. Due to the connection between the steepness of the radio spectral slope, dust extinction, and radio loudness, with dusty objects tending to be more radio-quiet and display steeper spectral slopes, we argue that the radio emission in dusty QSOs is likely dominated by outflow-driven shocks (either launched by winds or low-powered jets) in the surrounding ISM. 
\end{itemize}

Overall, our results are consistent with previous studies which suggest an intrinsic connection between the amount of dust in a QSO and the production of radio emission \citep{fawcett23}. This connection is likely due to outflow-driven shocks in the surrounding ISM and appears to be more important in the radio-quiet/intermediate regime, where powerful jets are unlikely to be dominating (Fig.~\ref{fig:Redd}). This is consistent with dusty QSOs residing in a younger evolutionary phase compared to bluer QSOs, whereby the shocks destroy/disperse the surrounding dust/gas, eventually revealing a typical unobscured blue QSO. Furthermore, we find a higher fraction of non-dusty QSOs with extended LoTSS emission, which could be due to relic radio emission from a previous AGN episode. Therefore, this may be an indication that non-dusty QSOs are on average older than dusty QSOs, although a larger sample is needed to robustly test this (Section~\ref{sec:evo}).

Ultimately, in order to confirm whether outflow-driven shocks are responsible for the radio emission in dusty QSOs, we need to analyse the radio SEDs of a larger sample of QSOs, pushing to redder and more radio-quiet objects to robustly test the observed trends. Furthermore, future high-resolution e-MERLIN C-band (5\,GHz) imaging ($0\farcs05$; $\sim$\,500\,pc scales) will help to constrain the radio morphology of these QSOs, revealing if they display jet-like structures and comparing their sizes to GPS/CSS sources. Complementary statistical studies exploring the radio spectral slopes of red and blue QSOs (e.g., Sargent et~al. \textit{in prep}) will also aid our understanding of how the dust and radio properties in QSOs connect.

\section*{Acknowledgements}
We would like to thank the anonymous referee for their constructive comments.

VAF and CMH acknowledge funding from an United Kingdom Research and Innovation grant (code: MR/V022830/1). DMA thanks the Science Technology Facilities Council (STFC) for support from the Durham consolidated grant (ST/T000244/1). PK, JB, and SG acknowledge the support of the Department of Atomic Energy, Government of India, under the project 12-R\&D-TFR-5.02-0700. JP acknowledges support from the CAS–NWO programme for radio astronomy with project number 629.001.024, which is financed by the NWO. JP acknowledges support from the ERC Starting Grant ClusterWeb 804208. SS acknowledges financial support from Millenium Nucleus NCN23\_002 (TITANs) and Comit\'{e} Mixto ESO-Chile. CS acknowledges STFC for studentship support (ST/Y509346/1).

We thank the staff of the GMRT that made these observations possible. GMRT is run by the National Centre for Radio Astrophysics of the Tata Institute of Fundamental Research.

\section*{Data Availability}
The uGMRT data underlying this article are available via the GMRT online archive facility: \url{https://naps.ncra.tifr.res.in/goa/data/search}. 
The uGMRT Band-3 and Band-4 radio images utilized in this paper can be found online at \url{https://doi.org/10.25405/data.ncl.27854934.v2}. The archival radio data are publicly available online (see Section~\ref{sec:archival}).

An electronic table containing the extracted uGMRT fluxes, uGMRT image details, SED classification and best fitting parameters, and the SDSS identifier for the red and blue QSO samples can be found in the online Supplementary material.

\appendix

\section{GMRT observation details}\label{sec:gmrt_obs}
\begin{table*}
\centering
\caption{Table displaying the uGMRT observation details for our rQSO and cQSO samples. The columns from left to right display the: (1) SDSS name, (2) whether the QSO is part of the red (rQSO) or blue (cQSO) sample, (3)-(10) the integrated flux, RMS, beam size, and flagging fraction for Band-3 and Band-4, and (11) the phase calibrator used in the observations. An electronic table containing the uGMRT Band-3 and Band-4 extracted fluxes can be found online. \newline \textsuperscript{\S} No Band-3 data.}
\begin{tabular}{cc|cccc|cccc|c}
\hline
\hline
Name & Sample & $F_{\rm B3}$ & RMS$_{\rm B3}$ & Beam size (B3)  & Flag (B3) & $F_{\rm B4}$ & RMS$_{\rm B4}$ & Beam size (B4) & Flag (B4) & Phase cal.  \\
& & [mJy] & [mJy] & [arcsec] & [\%] & [mJy] & [mJy] & [arcsec] & [\%] & [IAU name] \\
\hline
0823+5609 & rQSO & 4.6$\pm$0.4 & 0.27 & $8.7\times5.7$ & 27.3 & 5.3$\pm$0.4 & 0.27 & $5.4\times3.8$ & 7.7 & 0944+520 \\
0828+2731 & rQSO & 5.0$\pm$0.3 & 0.13 & $8.1\times5.3$ & 34.0 & 3.6$\pm$0.1 & 0.13 & $4.3\times3.7$ & 14.8 & 0746+258 \\
0946+2548 & rQSO & 26.6$\pm$0.5 & 0.31 & $7.6\times5.5$ & 25.6 & 15.0$\pm$0.8 & 0.31 & $4.4\times3.8$ & 9.5 & 1022+306 \\
0951+5253 & rQSO & 19.0$\pm$1.7 & 0.13 & $9.5\times5.5$ & 29.7 & 11.6$\pm$0.4 & 0.13 & $5.4\times3.6$ & 9.8 & 0944+520 \\
1007+2853 & rQSO & 33.8$\pm$0.6 & 0.23 & $7.9\times5.5$ & 25.4 & 19.4$\pm$1& 0.23 & $4.5\times3.6$ & 9.6 & 1022+306 \\
1057+3119 & rQSO & 15.0$\pm$0.3 & 0.21 & $8.9\times5.7$ & 23.3 & 14.4$\pm$0.4 & 0.21 & $4.9\times3.5$ & 8.1 & 1022+306 \\
1122+3124 & rQSO & 26.0$\pm$0.4 & 0.18 & $7.8\times5.5$ & 23.3 & 18.4$\pm$0.3 & 0.18 & $4.6\times3.7$ & 7.9 & 1215+348 \\
1140+4416 & rQSO & 3.6$\pm$0.2 & 0.16 & $8.9\times5.6$ & 23.0 & 5.0$\pm$0.1 & 0.16 & $5.1\times3.6$ & 8.1 & 1146+539 \\
1153+5651 & rQSO & 97.9$\pm$1.1 & 0.18 & $10.3\times5.7$ & 24.2 & 56.8$\pm$1.6 & 0.18 & $5.9\times3.7$ & 10.0 & 1146+539 \\
1159+2151 & rQSO & 13.8$\pm$0.2 & 0.14 & $8.0\times5.6$ & 24.3 & 9.7$\pm$0.2 & 0.14 &$ 4.7\times3.6$ & 8.1 & 1215+348 \\
1202+6317 & rQSO & 2.4$\pm$0.5 & 0.39 & $11.4\times5.7$ & 22.1 & 2.7$\pm$0.1 & 0.39 & $6.4\times3.6$ & 7.8 & 1146+539 \\
1211+2221 & rQSO & 2.2$\pm$0.2 & 0.16 & $8.2\times5.7$ & 24.1 & 2.8$\pm$0.1 & 0.16 & $4.8\times3.6$ & 8.1 & 1215+348 \\
1251+4317 & rQSO & 16.4$\pm$0.2 & 0.17 & $8.2\times5.7$ & 22.2 & 13.5$\pm$0.8 & 0.17 & $5.3\times3.6$ & 7.8 & 1215+348 \\
1315+2017\textsuperscript{\S} & rQSO & & & &  & 3.4$\pm$0.4 & 0.0 & $4.9\times3.5$ & 14.0 & 1352+314 \\
1323+3948 & rQSO & 0.8$\pm$0.1 & 0.12 & $9.2\times5.5$ & 26.7 & 2.4$\pm$0.1 & 0.12 & $5.1\times3.5$ & 13.4 & 1352+314 \\
1342+4326 & rQSO & 28.0$\pm$0.4 & 0.24 & $9.6\times5.5$ & 26.6 & 26.1$\pm$0.3 & 0.24 & $5.4\times3.5$ & 13.2 & 1352+314 \\
1410+4016 & rQSO & 10.3$\pm$0.3 & 0.14 & $9.9\times5.5$ & 27.9 & 8.2$\pm$0.1 & 0.14 & $5.5\times3.6$ & 13.4 & 1352+314 \\
1531+4528 & rQSO & 15.2$\pm$0.8 & 0.21 & $12.0\times5.4$ & 27.8 & 10.2$\pm$0.6 & 0.21 & $6.2\times3.6$ & 13.4 & 1500+478 \\
1535+2434 & rQSO & 9.9$\pm$0.3 & 0.2 & $10.8\times5.8$ & 23.0 & 9.8$\pm$0.2 & 0.2 & $5.5\times3.7$ & 7.8 & 1609+266 \\
\hline
0748+2200 & cQSO & 7.8$\pm$0.4 & 0.24 & $7.7\times5.3$ & 34.1 & 8.7$\pm$0.1 & 0.24 & $4.2\times3.8$ & 13.9 & 0746+258 \\
1003+2727 & cQSO & 25.9$\pm$0.3 & 0.16 & $7.8\times5.5$ & 24.2 & 16.3$\pm$0.1 & 0.16 & $4.4\times3.7$ & 9.0 & 1022+306 \\
1019+2817 & cQSO & 16.6$\pm$0.4 & 0.17 & $8.0\times5.5$ & 24.0 & 17.8$\pm$0.2 & 0.17 & $4.6\times3.6$ & 8.2 & 1022+306 \\
1038+4155 & cQSO & 2.3$\pm$0.1 & 0.12 & $8.8\times5.5$ & 23.8 & 3.2$\pm$0& 0.12 & $5.0\times3.6$ & 9.2 & 1022+306 \\
1042+4834 & cQSO & 13.2$\pm$0.2 & 0.13 & $10.2\times5.5$ & 28.6 & 12.6$\pm$0.1 & 0.13 & $5.5\times3.6$ & 9.7 & 0944+520 \\
1046+3427 & cQSO & 18.6$\pm$0.6 & 0.17 & $8.6\times5.5$ & 25.2 & 29.1$\pm$0.3 & 0.17 & $4.9\times3.6$ & 7.8 & 1022+306 \\
1057+3315 & cQSO & 3.1$\pm$0.1 & 0.13 & $8.9\times5.6$ & 24.1 & 2.2$\pm$0.1 & 0.13 & $4.9\times3.6$ & 8.1 & 1022+306 \\
1103+5849 & cQSO & 4.6$\pm$0.1 & 0.18 & $9.9\times5.7$ & 22.0 & 5.7$\pm$0.1 & 0.18 & $5.8\times3.7$ & 8.1 & 1146+539 \\
1203+4510 & cQSO & 20.9$\pm$1.6 & 0.23 & $8.8\times5.7$ & 24.7 & 32.3$\pm$0.8 & 0.23 & $5.1\times3.6$ & 10.5 & 1215+348 \\
1222+3723 & cQSO & 21.4$\pm$0.4 & 0.16 & $8.6\times5.6$ & 22.9 & 15.1$\pm$0.2 & 0.16 & $5.0\times3.6$ & 7.9 & 1215+348 \\
1304+3206 & cQSO & 6.4$\pm$0.3 & 0.13 & $8.5\times5.5$ & 27.7 & 6.3$\pm$0.2 & 0.13 & $4.9\times3.5$ & 13.5 & 1352+314 \\
1410+2217 & cQSO & 34.8$\pm$0.7 & 0.19 & $9.1\times5.8$ & 29.3 & 29.6$\pm$1.6 & 0.19 & $5.0\times3.6$ & 16.4 & 1352+314 \\
1428+2916 & cQSO & 10.9$\pm$1.2 & 0.42 & $10.2\times5.7$ & 26.9 & 18.3$\pm$1.4 & 0.42 & $5.3\times3.6$ & 13.2 & 1352+314 \\
1432+2925 & cQSO & 24.8$\pm$0.5 & 0.15 & $10.1\times5.6$ & 28.0 & 22.8$\pm$0.3 & 0.15 & $5.3\times3.6$ & 13.6 & 1352+314 \\
1530+2310 & cQSO & 7.8$\pm$0.4 & 0.24 & $10.9\times5.8$ & 23.0 & 5.9$\pm$0.1 & 0.24 & $5.5\times3.7$ & 7.9 & 1609+266 \\
1554+2859 & cQSO & 8.9$\pm$0.1 & 0.14 & $11.6\times5.7$ & 24.4 & 7.5$\pm$0.1 & 0.14 & $5.7\times3.7$ & 7.8 & 1609+266 \\
1602+4530 & cQSO & 12.2$\pm$0.2 & 0.16 & $14.3\times5.3$ & 28.5 & 8.2$\pm$0.1 & 0.16 & $7.0\times3.6$ & 14.3 & 1500+478 \\
1630+3847 & cQSO & 2.4$\pm$0.3 & 0.21 & $13.4\times5.4$ & 23.2 & 4.4$\pm$0.3 & 0.21 & $6.5\times3.6$ & 8.4 & 1609+266 \\
1657+2045 & cQSO & 13.2$\pm$0.4 & 0.18 & $15.5\times5.6$ & 24.2 & 10.0$\pm$0.2 & 0.18 & $7.0\times3.7$ & 7.9 & 1609+266 \\
\hline
\hline
\end{tabular}
\label{tab:ugmrt_obs}
\end{table*}

Table~\ref{tab:ugmrt_obs} displays the uGMRT Band-3 and Band-4 imaging details and extracted flux values.

\section{Radio variability}\label{sec:variability}

A potential bias introduced when combining multiple archival radio surveys is variability. It has been long established that QSOs vary across multiple frequencies, including radio, on day-to-year timescales \citep{wagner2,baravainis}. Therefore, utilizing different radio data that span a large timescale could result in skewed radio SEDs that do not capture the true nature of the source.

\begin{figure}
    \centering
    \includegraphics[width=0.45\textwidth]{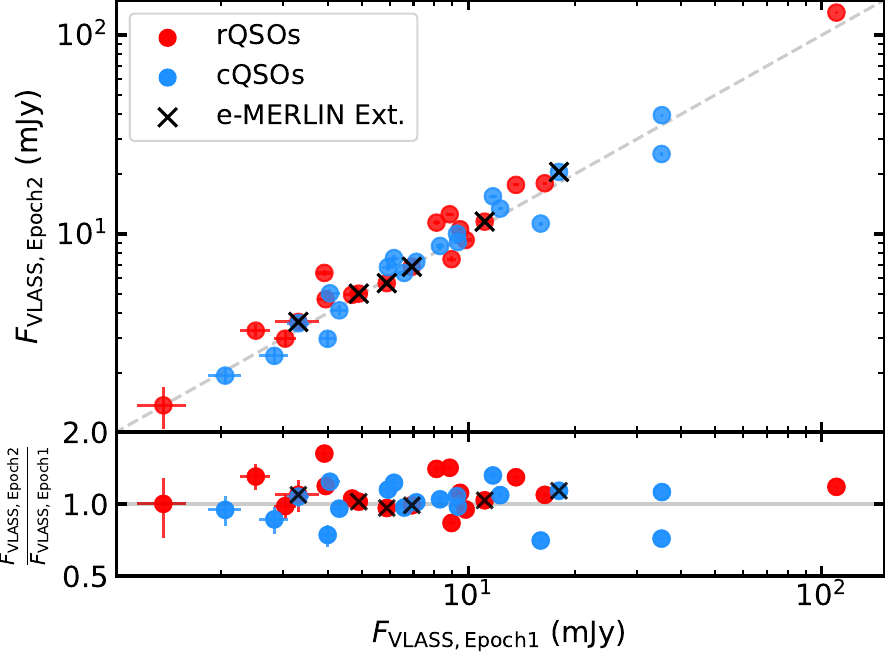}
    \caption{Flux densities from VLASS Epoch 2 versus Epoch 1 for rQSOs (red) and cQSOs (blue). The dashed line in the top panel indicates the 1:1 relation. The e-MERLIN extended sources are indicated by the black crosses. The bottom panel displays the fractional difference between Epoch 2 and Epoch 1; little variability is seen between the two epochs, with a maximum offset between the VLASS flux densities of $\sim1.6\times$.}
    \label{fig:vlass_variability}
\end{figure}

Although it is not possible to have a complete understanding of the variability of these sources due to the wide range of timescales, we can compare the VLASS Epoch 1 and 2 data to gain some insight into short-term variability. The VLASS Epoch 1 and 2 catalogues contain two epochs of data that were taken $\sim$\,$2$ years apart. Matching to both catalogues ($0\farcs5$ matching radius for both) resulted in all 38 QSOs detected. Fig.~\ref{fig:vlass_variability} displays a comparison between the VLASS Epoch 1 and 2 integrated flux densities for the red and blue QSOs. Overall, there is little variability, with a maximum offset between the VLASS flux densities of both the red and blue QSOs of $\sim$\,$1.6\times$. Similarly, Sargent et~al. (\textit{in prep}) explored the variability of the in-band VLASS spectral slopes versus the FIRST--VLASS spectral slopes of a larger sample of SDSS red and blue QSOs and found no significant differences. Therefore, it is unlikely that radio variability will strongly affect the interpretation of our SED fitting overall; however, the SED fits for some individual sources may be impacted by variability. Interestingly, the source that have a VLASS flux much higher than the fitted radio SED model (1251+4317; see Section~\ref{sec:sed}) does not show signs of significant variability based on the VLASS Epoch 1 to Epoch 2 integrated fluxes. 

Another indication that there is no strong radio variability in our sources come from inspecting the lower resolution archival radio data, displayed on SEDs (e.g., Fig.~\ref{fig:example_seds}), but not used in the fitting (see Table~\ref{tab:radio}). We find that despite the $\sim$\,25 year time span of the various data, the majority of the data points from these lower resolution surveys are consistent with the model fitted to the uGMRT, FIRST, LoTSS, and VLASS data, and also cover the same frequency range. The only exceptions are for the upturned sources, where the lower resolution, lower frequency radio data picks up the relic radio emission, as expected (e.g., QSO 1046+3427, which displays a much higher WENNS $54''$ data point). For these sources, the lower frequency LoTSS point is removed and the SED refitted.

\section{Large scale radio emission}\label{sec:large_scale}
\begin{figure}
    \centering
    \includegraphics[width=0.45\textwidth]{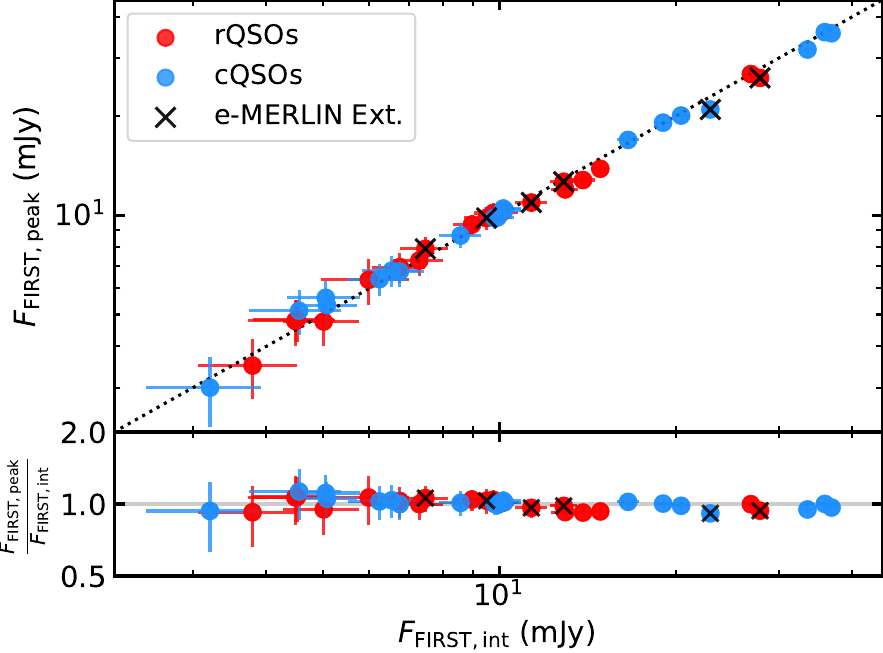}
    \caption{FIRST peak flux versus FIRST integrated flux for the rQSOs (red) and cQSOs (blue). The e-MERLIN extended sources are indicated by the black crosses. The majority of sources display very similar integrated and peak fluxes which is expected for unresolved sources.}
    \label{fig:peak_int}
\end{figure}

By selection, the QSOs in our sample were chosen to have no visually extended radio emission at 1.4 GHz in the FIRST images (see Section~\ref{sec:sample}). To confirm this and to justify our use of integrated fluxes rather than peak fluxes in our SED fitting, in Fig.~\ref{fig:peak_int} we plot the peak versus integrated flux density from FIRST. We find a very tight correlation which is expected for unresolved sources.

\begin{figure*}
    \centering
    \includegraphics[width=0.21\textwidth]{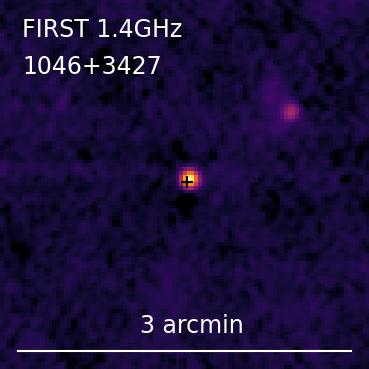}
    \includegraphics[width=0.21\textwidth]{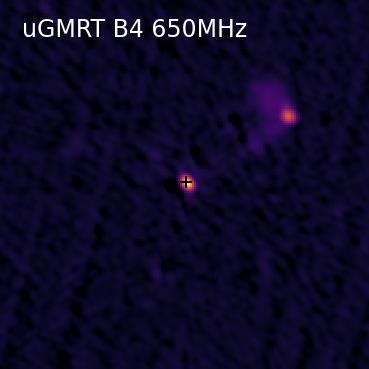}
    \includegraphics[width=0.21\textwidth]{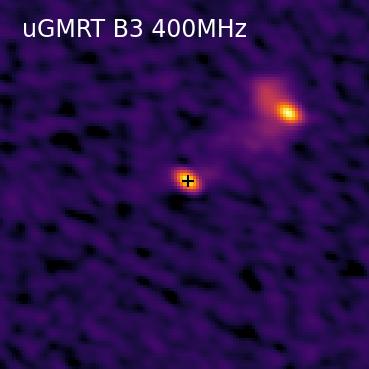}
    \includegraphics[width=0.21\textwidth]{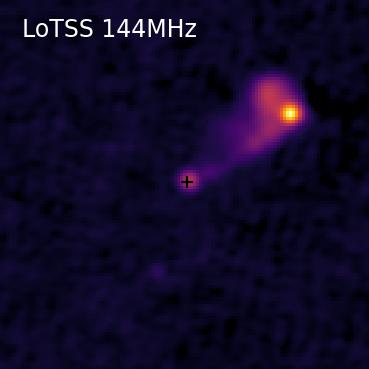}
    \vspace{3mm}
    
    \includegraphics[width=0.21\textwidth]{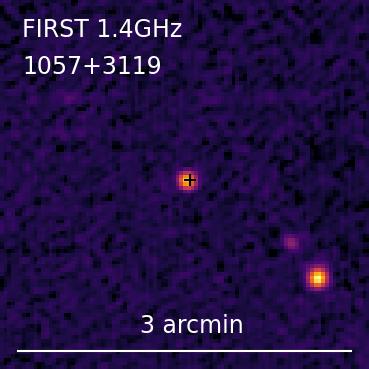}
    \includegraphics[width=0.21\textwidth]{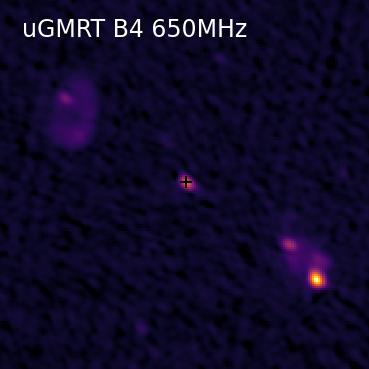}
    \includegraphics[width=0.21\textwidth]{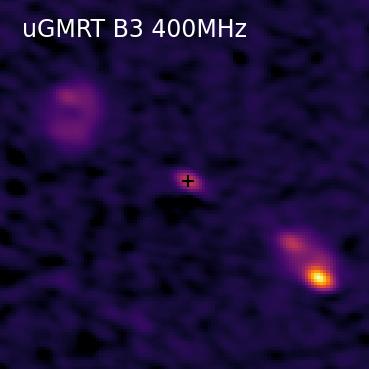}
    \includegraphics[width=0.21\textwidth]{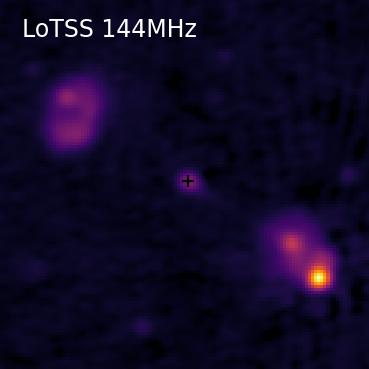}
    \caption{(Left to right) FIRST (1.4\,GHz), uGMRT B4 (650\,MHz) and B3 (400\,MHz), and LoTSS (144\,MHz) $3'$ thumbnails of the cQSO 1046+3427 (top) and rQSO 1057+3119 (bottom), which both display large scale extended radio emission that gets increasingly brighter at lower frequencies. Both QSOs were classified as unresolved in FIRST due to the lack of a clear connection between the extended radio emission and the core (indicated by the black cross which shows the location of the optical QSO) and also the lack of a symmetrical second lobe. The cQSO 1046+3426 originally had an upturned radio SED which is due to the LoTSS flux measurement incorporating the extended radio emission, which was then refit by a flat spectrum after removing the LoTSS data point. The rQSO 1057+3119 has a flat radio SED, due to the non-association of the LoTSS flux from the radio lobes.}
    \label{fig:extended_radio}
\end{figure*}

However, from inspecting the radio images of our sample we found seven sources with extended radio emission in LoTSS that appear compact in FIRST (see Section~\ref{sec:evo}). To understand whether this is relic radio emission, which is not be present in the higher frequency data, or missed extended emission we visually inspected $3'$ cutouts for all the QSOs in all the bands utilized in the SED fitting. We found two sources (1046+3427 and 1057+3119) that displayed extended radio emission in all bands that appear as multiple entries in the FIRST and VLASS catalogues (Fig.~\ref{fig:extended_radio}). The extended LoTSS radio emission in QSO 1046+3427 was treated as associated in the \cite{lotss_opt} catalogue, and therefore originally had an upturned radio SED. The LoTSS radio emission for QSO 1057+3119 on the other hand appears as three entries in the catalogue, and therefore the radio SED is flat due to the core emission. The other five QSOs with extended LoTSS emission either show extension in multiple bands, suggesting diffuse structures that increase in brightness with decreasing frequency, or display extension in only LoTSS and faintly in uGMRT Band-3, which could be evidence of relic emission.


\section{Comparison to the radio spectral index quadrant}\label{sec:appendix_quad}
\begin{figure}
    \centering
    \includegraphics[width=0.45\textwidth]{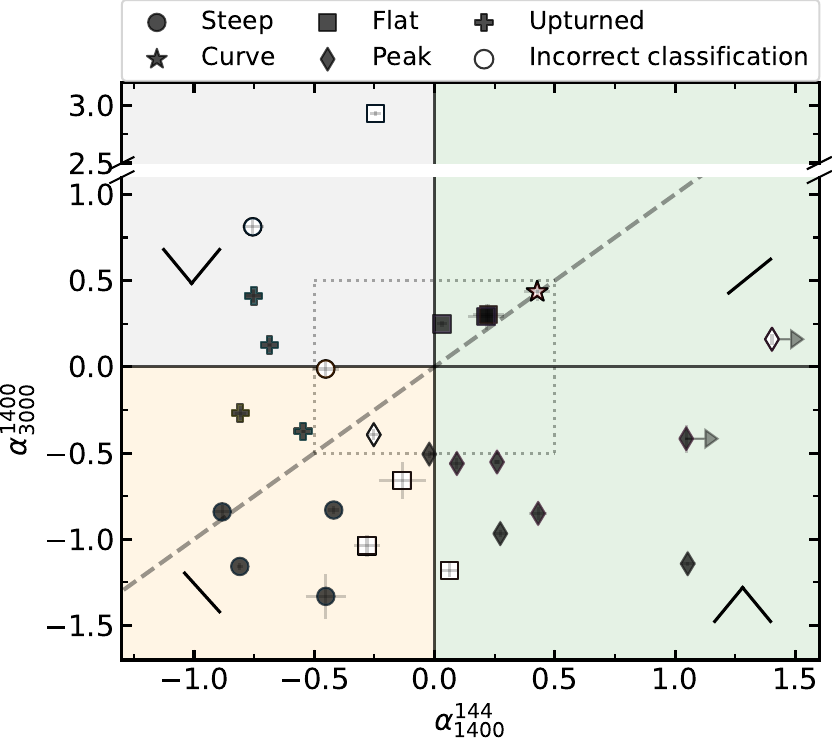}
    \caption{Comparing the classifications from the radio SED fitting with the simple quadrant diagram, where the bottom left (yellow) is steep, top left (grey) is upturned, the top right (green) is inverted, the bottom right (green) is peaked, and the grey dotted region is flat. The open symbols indicate the sources that would be misclassified by utilising the radio spectral quadrant alone.}
    \label{fig:alpha_quad}
\end{figure}

The majority of radio spectral studies are limited to two or three data points, due to the large publicly available radio catalogues. A popular approach to classify the shape of the radio spectrum, rather than just obtaining a two-point spectral index, is to use the radio spectral index quadrant plot (e.g., \citealt{mahony,Patil_2022,sinha,kukreti}; Fig.~\ref{fig:alpha_quad}). Typically, a higher frequency radio spectral slope ($\alpha_{\rm high}$) is plotted against a lower frequency radio spectral slope ($\alpha_{\rm low}$), with each quadrant representing a different global spectral shape. For example, both a steep $\alpha_{\rm high}$ and $\alpha_{\rm low}$ would give a globally steep spectral index, and likewise, if both $\alpha_{\rm high}$ and $\alpha_{\rm low}$ were inverted, then the global spectral index would be inverted. The top left quadrant (inverted $\alpha_{\rm high}$ and steep $\alpha_{\rm low}$) would produce an upturned spectrum and the bottom right quadrant (steep $\alpha_{\rm high}$ and inverted $\alpha_{\rm low}$) would produce a peaked spectrum. A region around $\alpha_{\rm high}$\,$=$\,0 and $\alpha_{\rm low}$\,$=$\,0 is usually defined to represent flat spectrum sources. To assess the reliability of this method we can compare our robust spectral classifications from the SED modelling to what we would have concluded with only three data points (e.g., LoTSS, FIRST, and VLASS).

Fig.~\ref{fig:alpha_quad} displays the radio spectral index quadrant plot for our sample, with each marker representing a different spectral classification from our SED modelling. Unfortunately, due to the lack of LoTSS coverage we can only compare the classifications of 27 QSOs in our sample. Defining our flat region to be within $|\alpha_{\rm high}|$\,$<$\,0.5 and $|\alpha_{\rm low}|$\,$<$\,0.5, we find the $\sim$18/27 of the QSOs fall in the correct quadrant (not including one borderline source). Inspecting the other nine QSO classifications we find six sources (1057+3315, 1103+5849, 1222+3723, 1323+3948, 1554+2859, and 1602+4530) are on the borderline of the classifications. The other three sources (1251+4317, 1304+3206, and 1531+4528) have one data point that is unusually higher than the other data points (either VLASS or FIRST), which may be due to variability. In conclusion, for the majority of sources the radio spectral quadrant provides an effective method for determining the global shape of the radio SED. For a small fraction of unusual or borderline cases, additional data points are required to fully describe the radio SED.

\section{Radio SEDs and Images}\label{sec:appendix_sed}
\begin{figure*}
    \centering
    \subfloat{\includegraphics[width=0.79\textwidth]{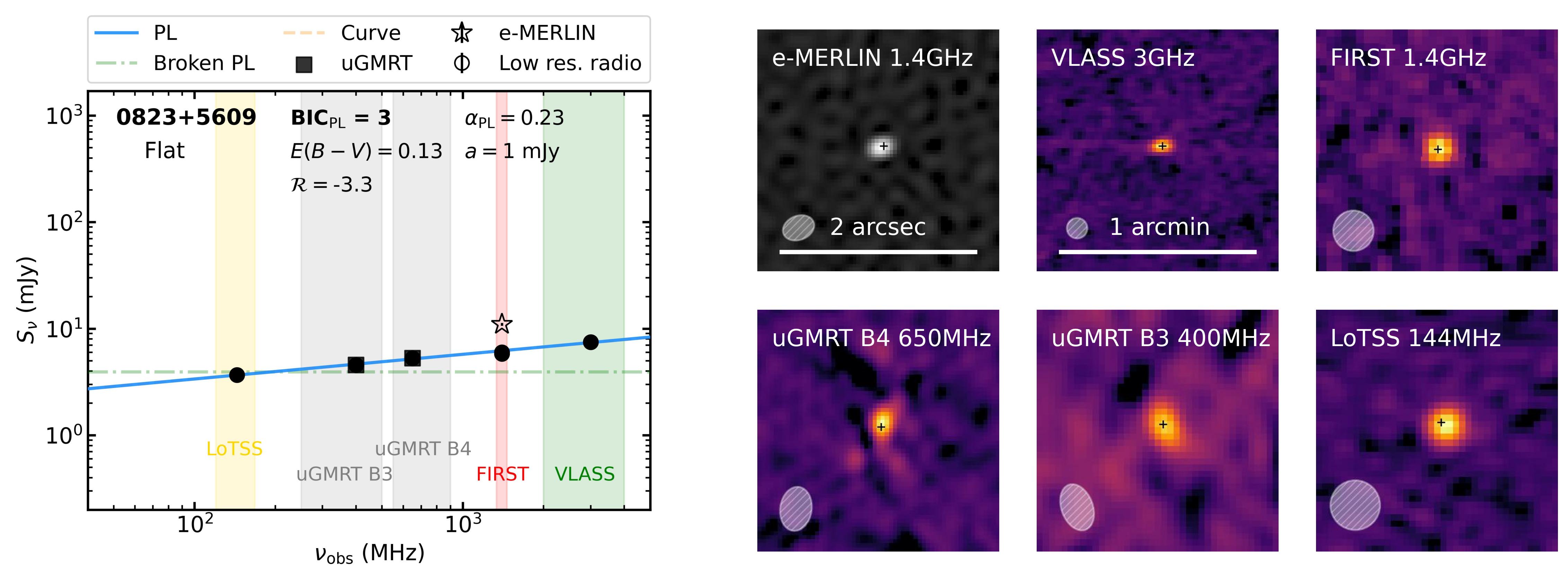}}\\
    \subfloat{\includegraphics[width=0.79\textwidth]{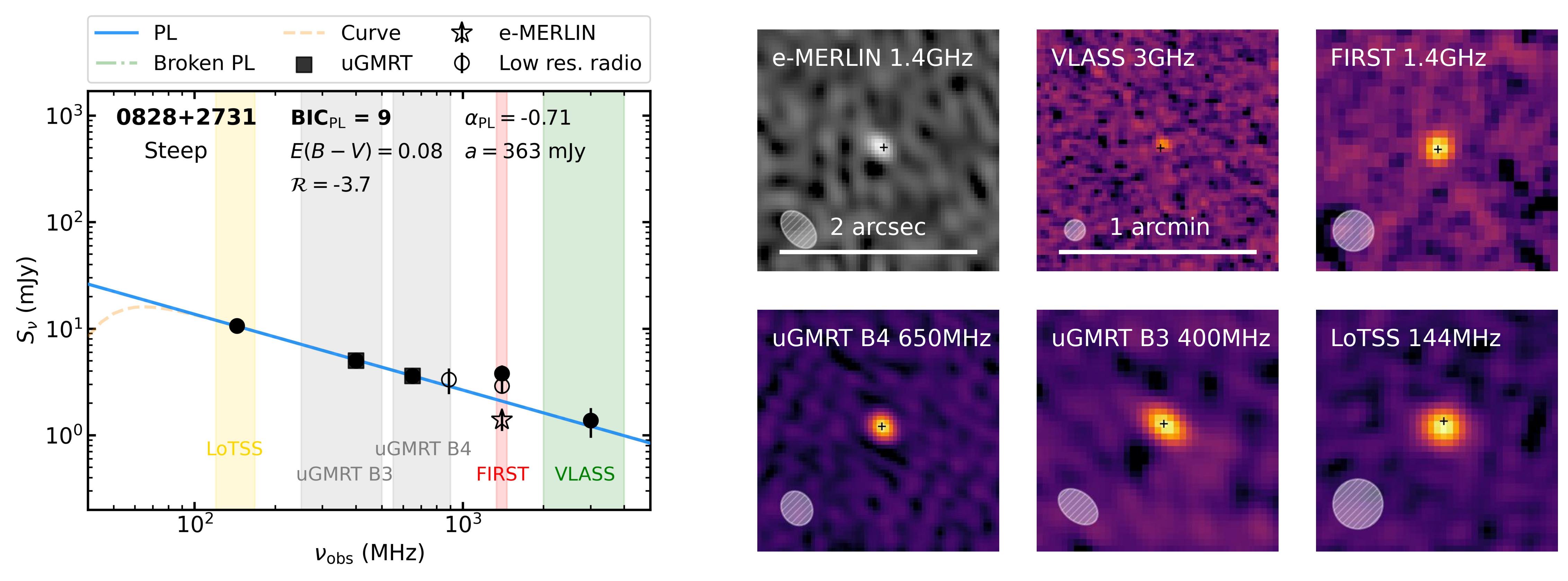}}\\
    \subfloat{\includegraphics[width=0.79\textwidth]{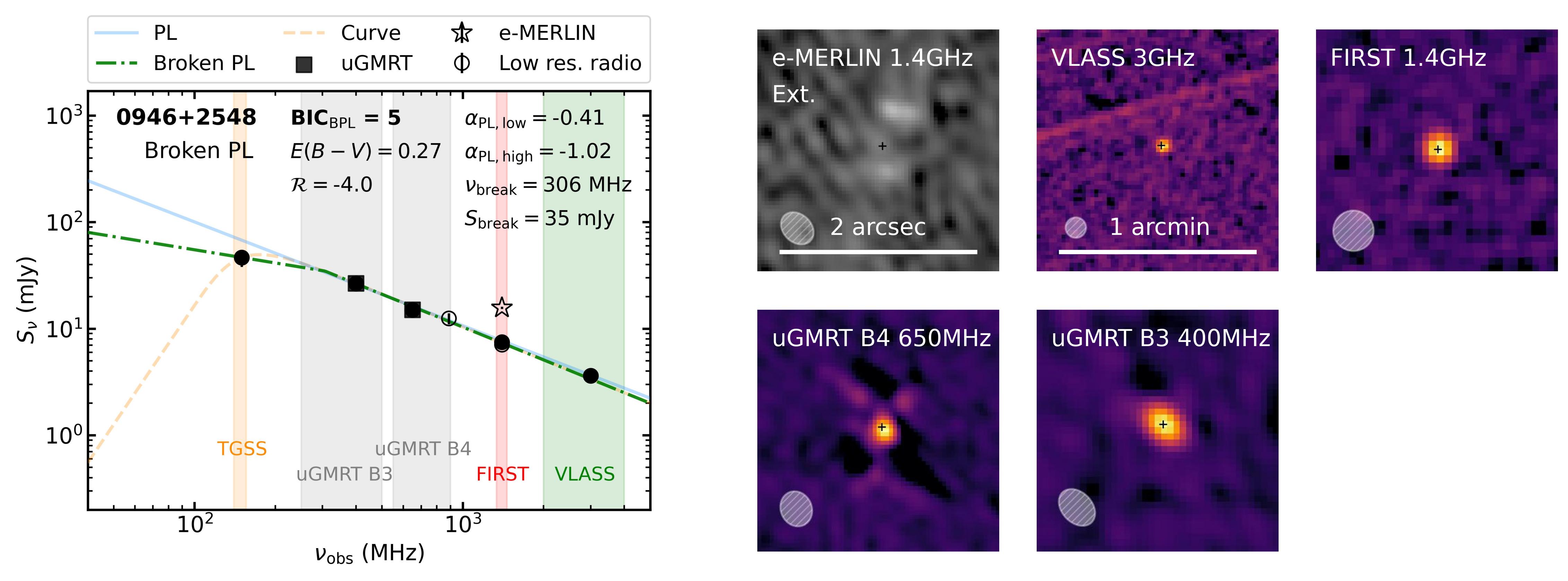}}\\
    \subfloat{\includegraphics[width=0.79\textwidth]{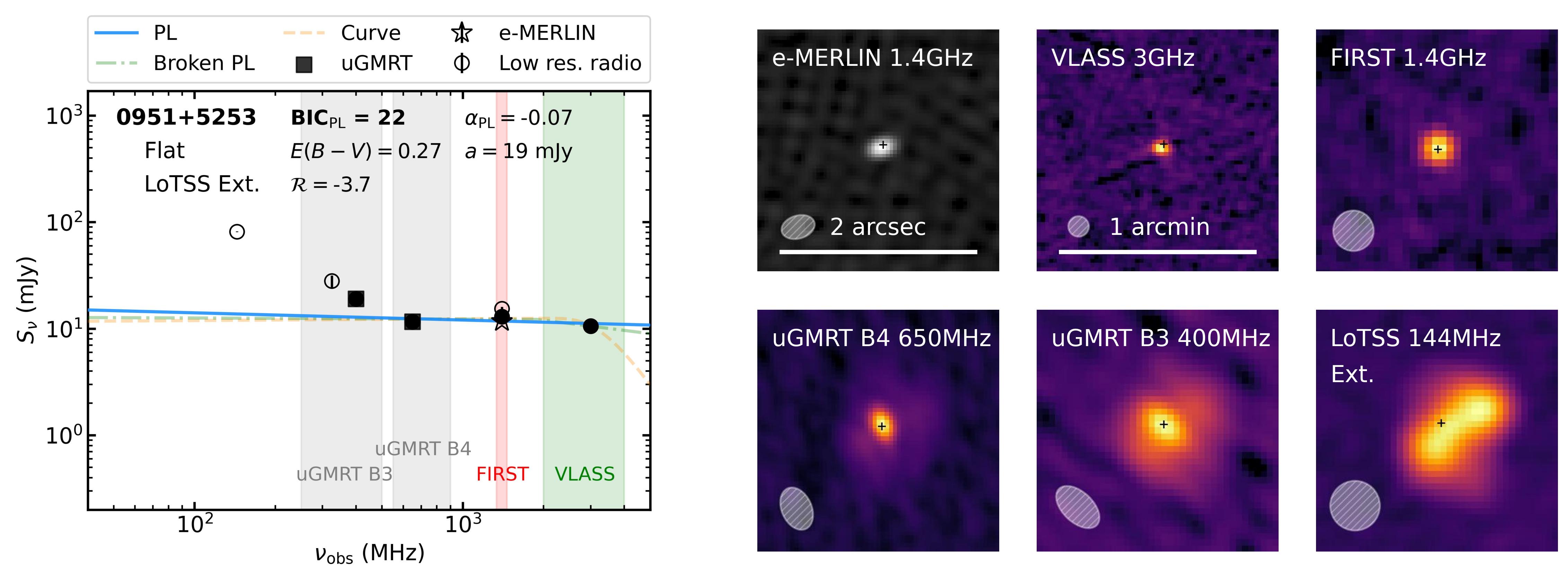}}
    \caption{The radio SEDs (left) and radio images (right thumbnails) for the red QSO sample. The best-fit power-law, broken power-law and curved model is displayed on the SED as a solid blue, dotted-dashed green, and dashed orange line, respectively. The various radio data utilized in the fitting are shown as solid black points and the band width shown as the shaded regions. Our uGMRT observations are shown as the squares and the e-MERLIN flux is shown as the star. Additional archival low resolution radio emission not used in the SED fitting is displayed by the open circles. For some sources the LoTSS data point is also shown as an open marker, indicating that these sources were originally classified as “upturned” and so the SEDs were refitted excluding the LoTSS data point. In the thumbnails, the QSO Gaia optical position is shown as the black plus.}
    \label{fig:SED_rQSO}
\end{figure*}

\begin{figure*}
    \ContinuedFloat
    \centering
    \subfloat{\includegraphics[width=0.79\textwidth]{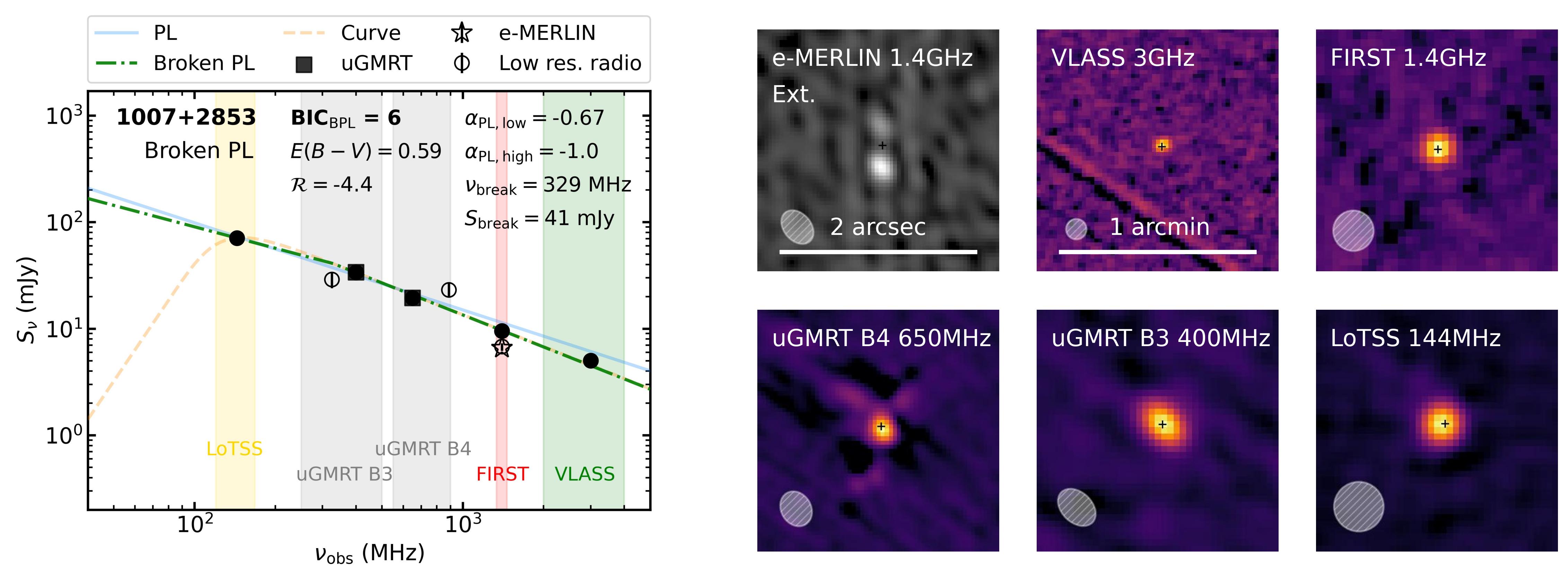}}\\
    \subfloat{\includegraphics[width=0.79\textwidth]{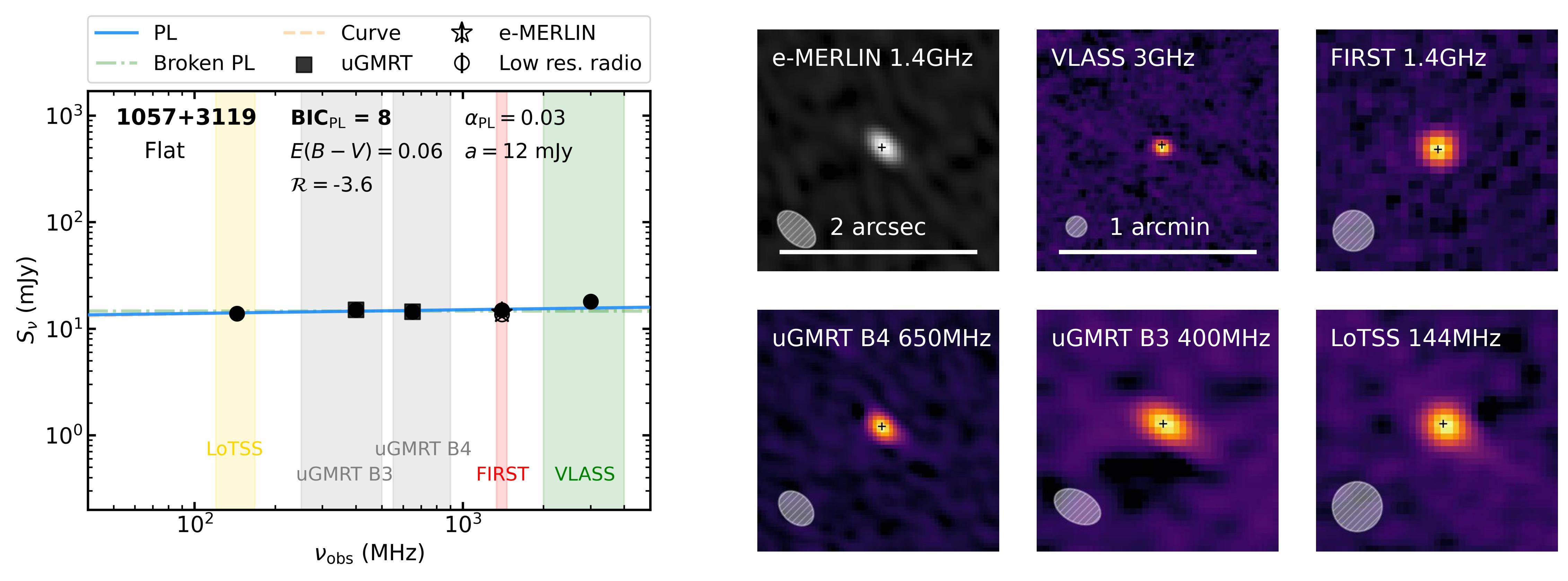}}\\
    \subfloat{\includegraphics[width=0.79\textwidth]{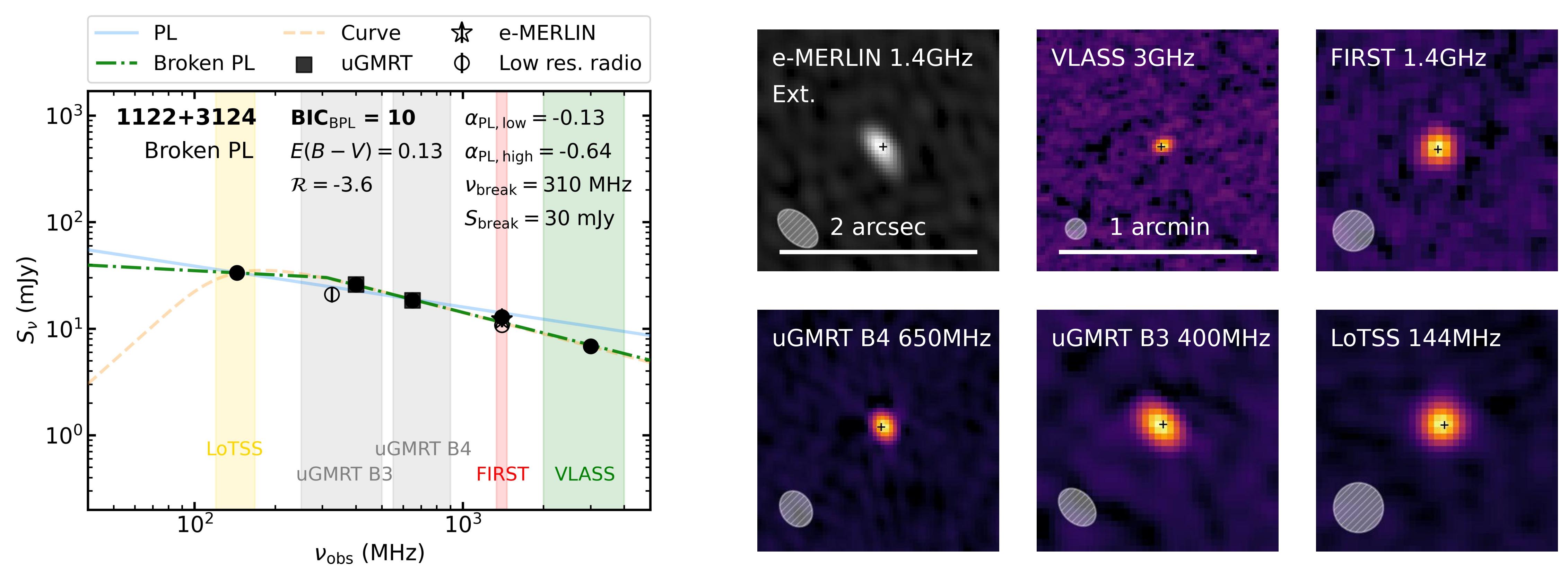}}\\
    \subfloat{\includegraphics[width=0.79\textwidth]{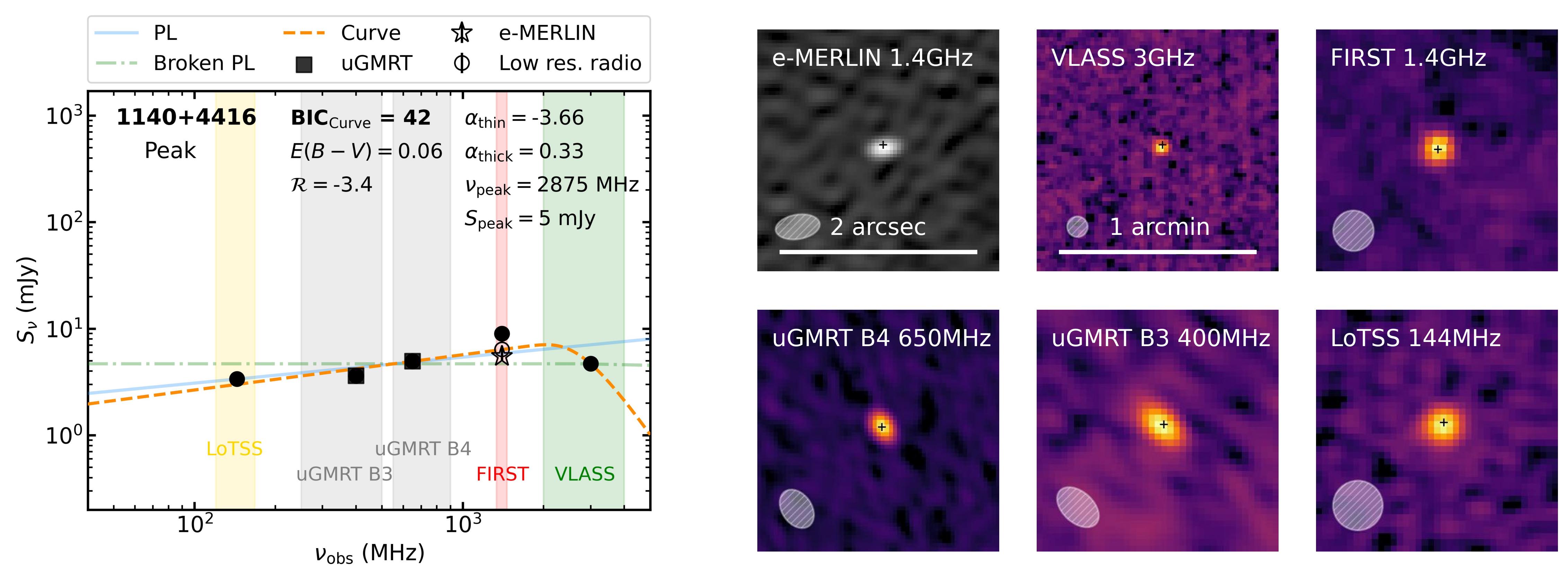}}
    \caption{Continued.}
\end{figure*}

\begin{figure*}
    \ContinuedFloat
    \centering
    \subfloat{\includegraphics[width=0.79\textwidth]{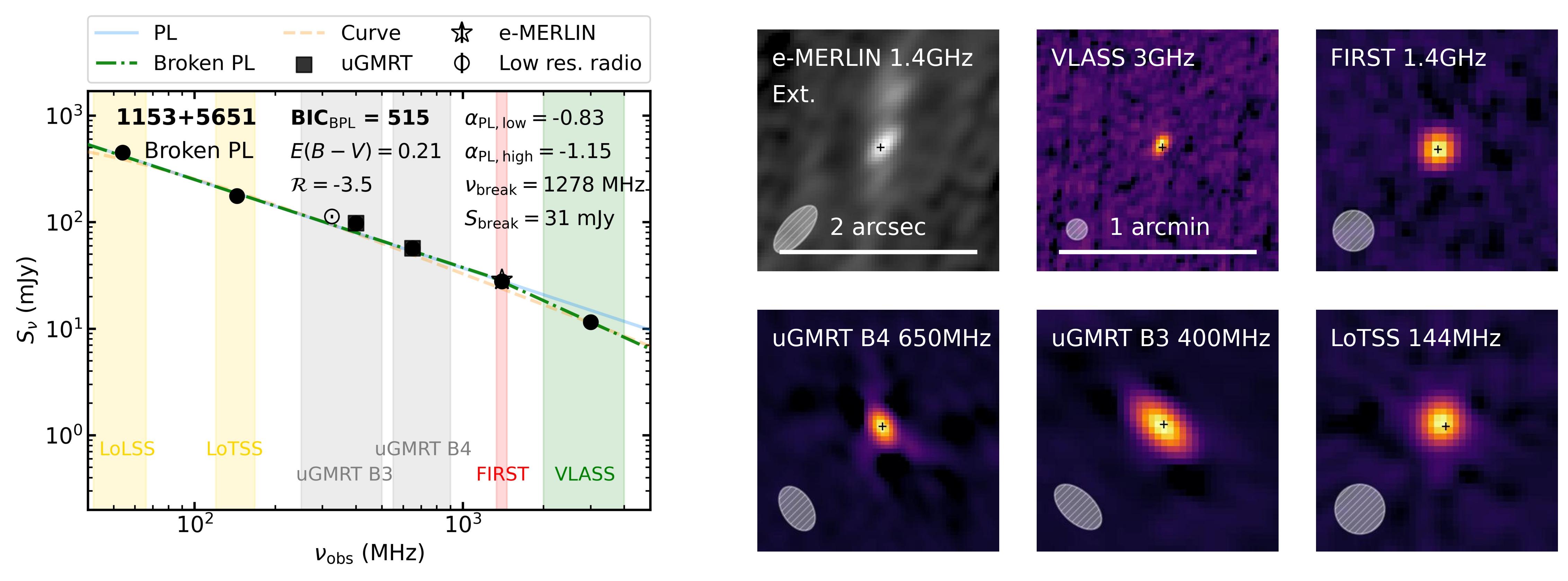}}\\
    \subfloat{\includegraphics[width=0.79\textwidth]{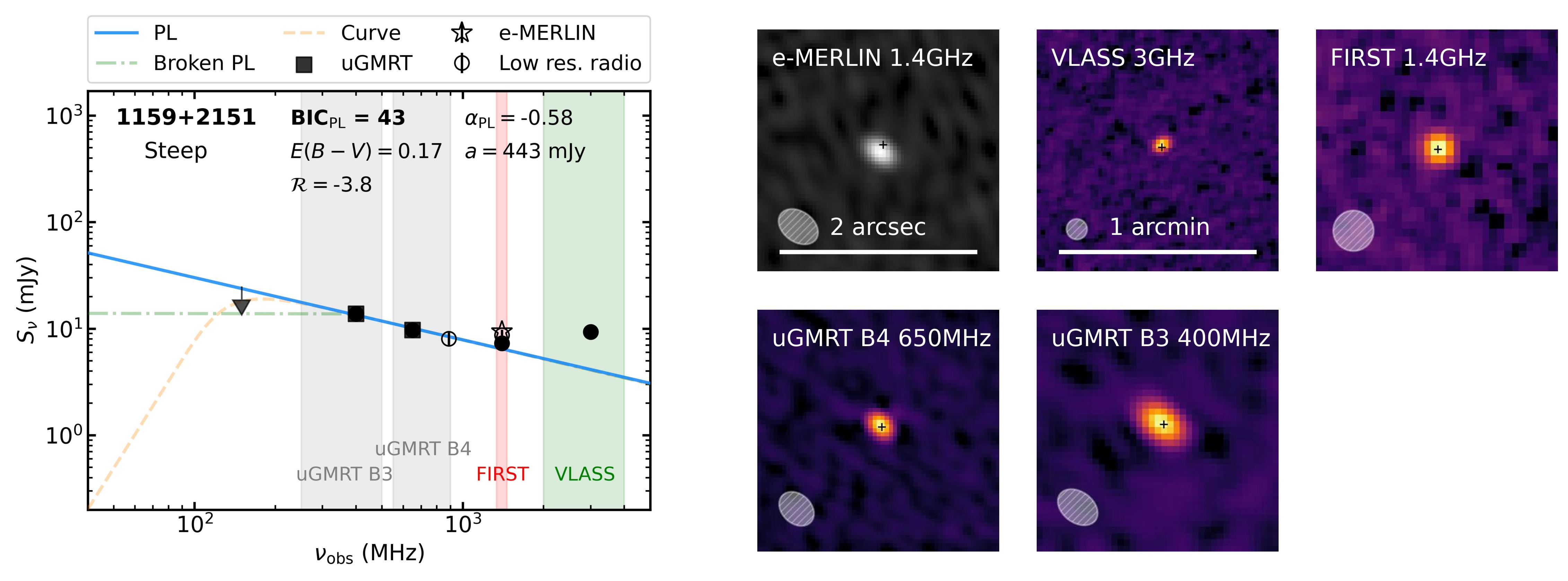}}\\
    \subfloat{\includegraphics[width=0.79\textwidth]{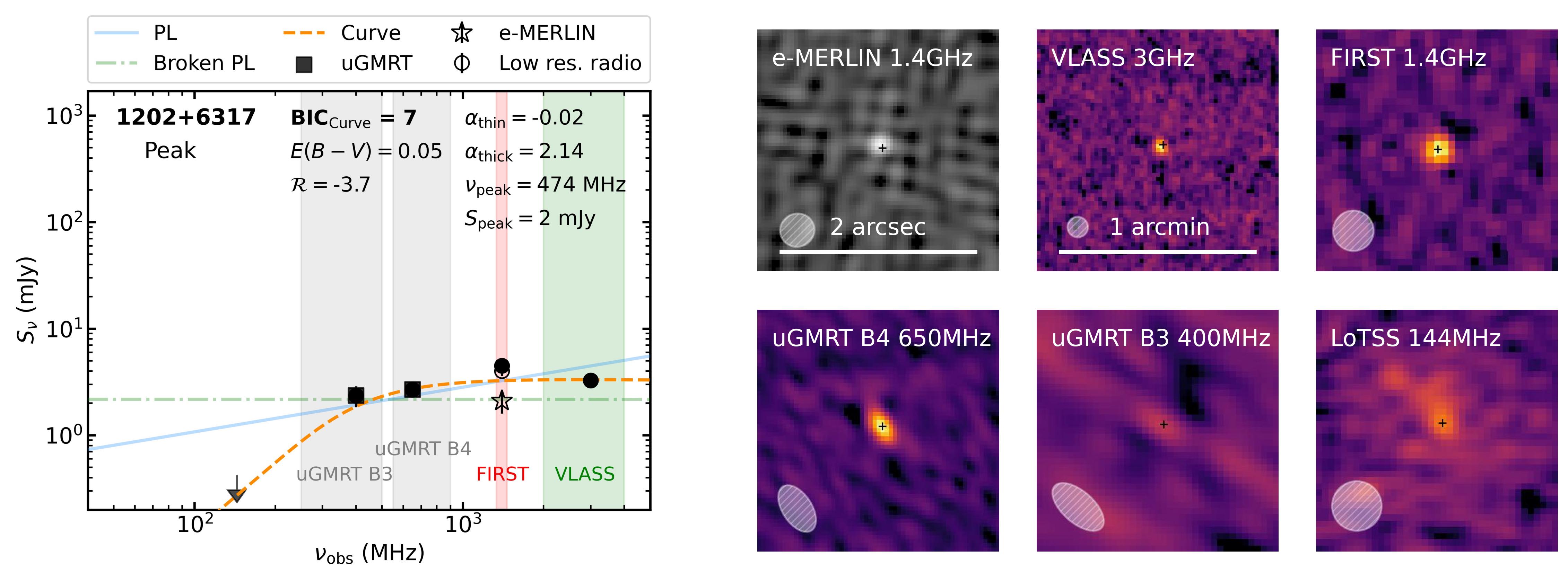}}\\
    \subfloat{\includegraphics[width=0.79\textwidth]{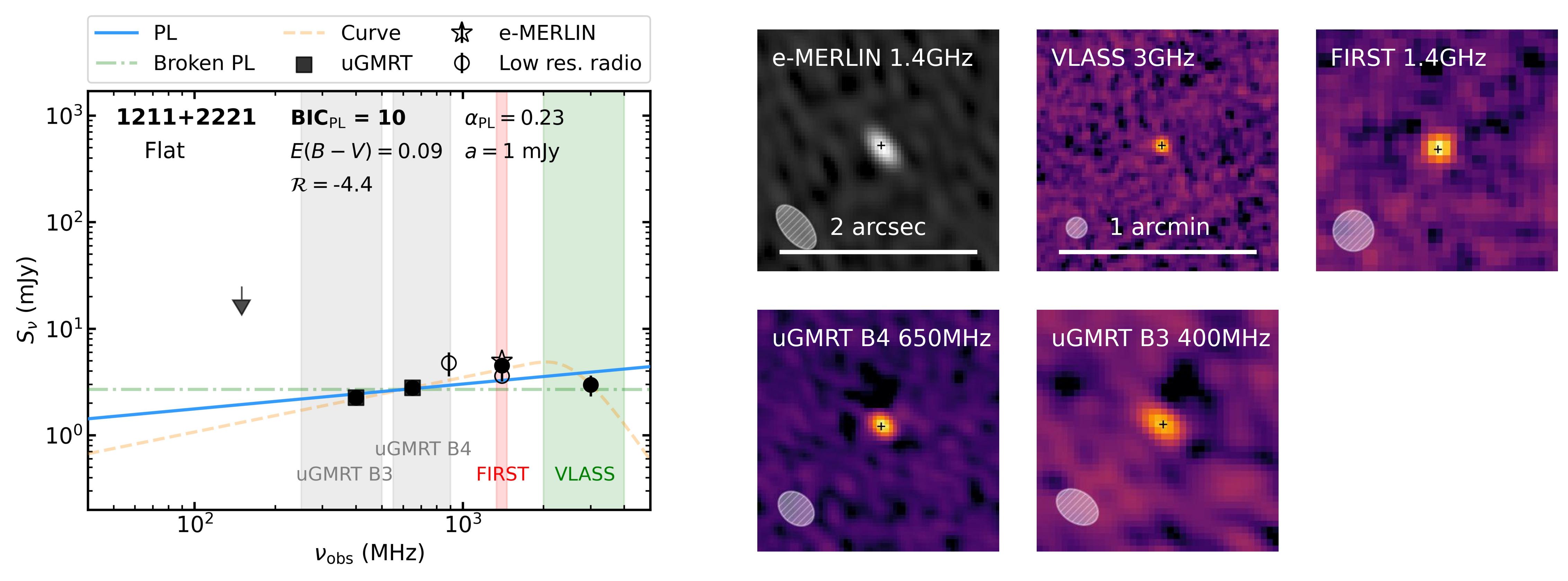}}
    \caption{Continued.}
\end{figure*}

\begin{figure*}
    \ContinuedFloat
    \centering
    \subfloat{\includegraphics[width=0.79\textwidth]{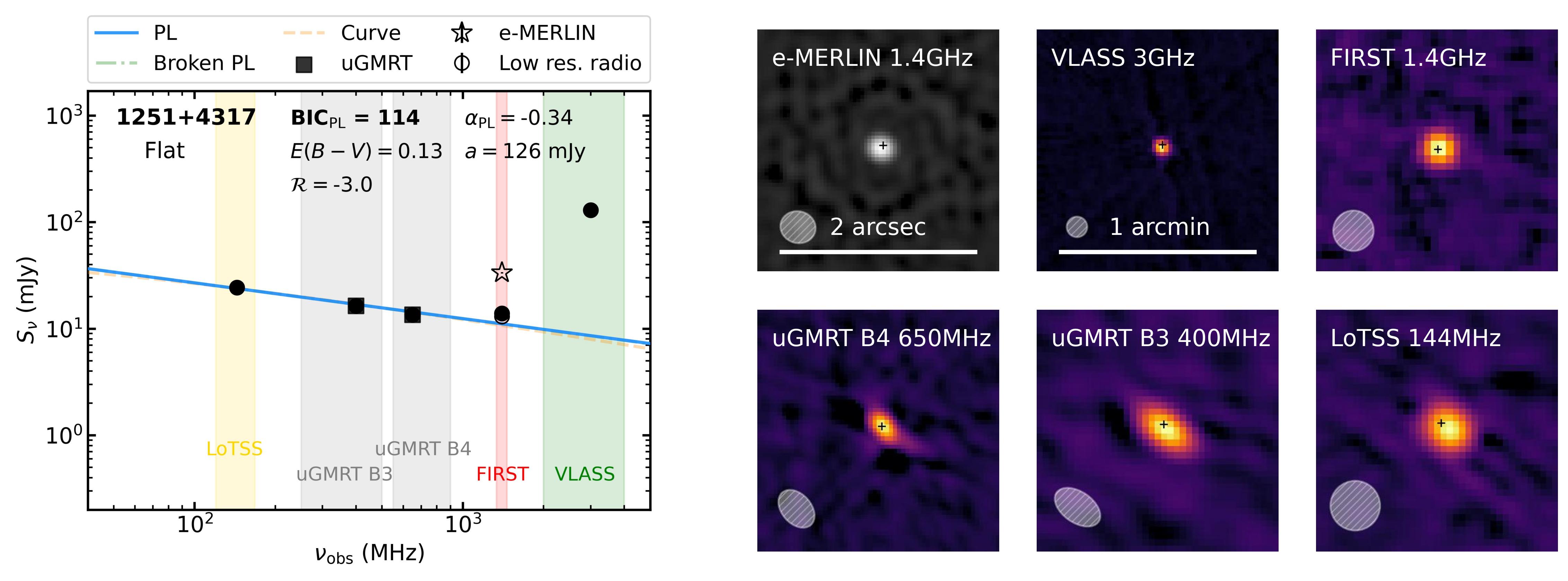}}\\
    \subfloat{\includegraphics[width=0.79\textwidth]{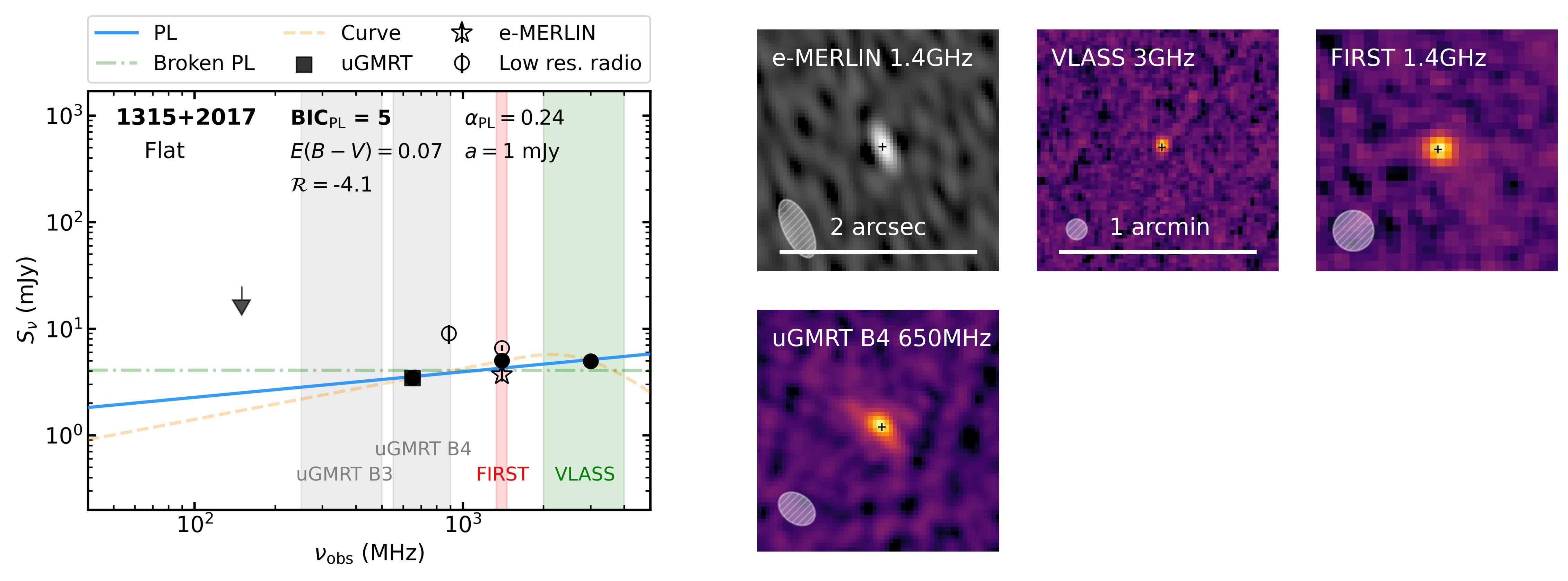}}\\
    \subfloat{\includegraphics[width=0.79\textwidth]{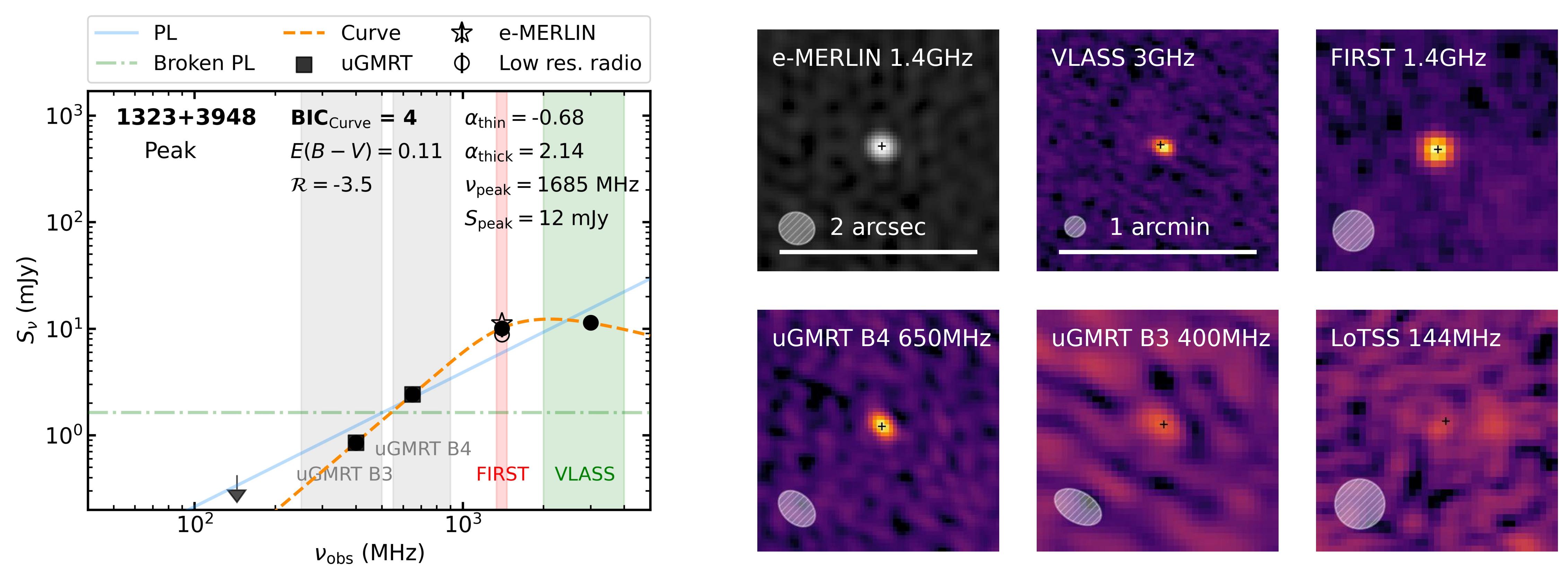}}\\
    \subfloat{\includegraphics[width=0.79\textwidth]{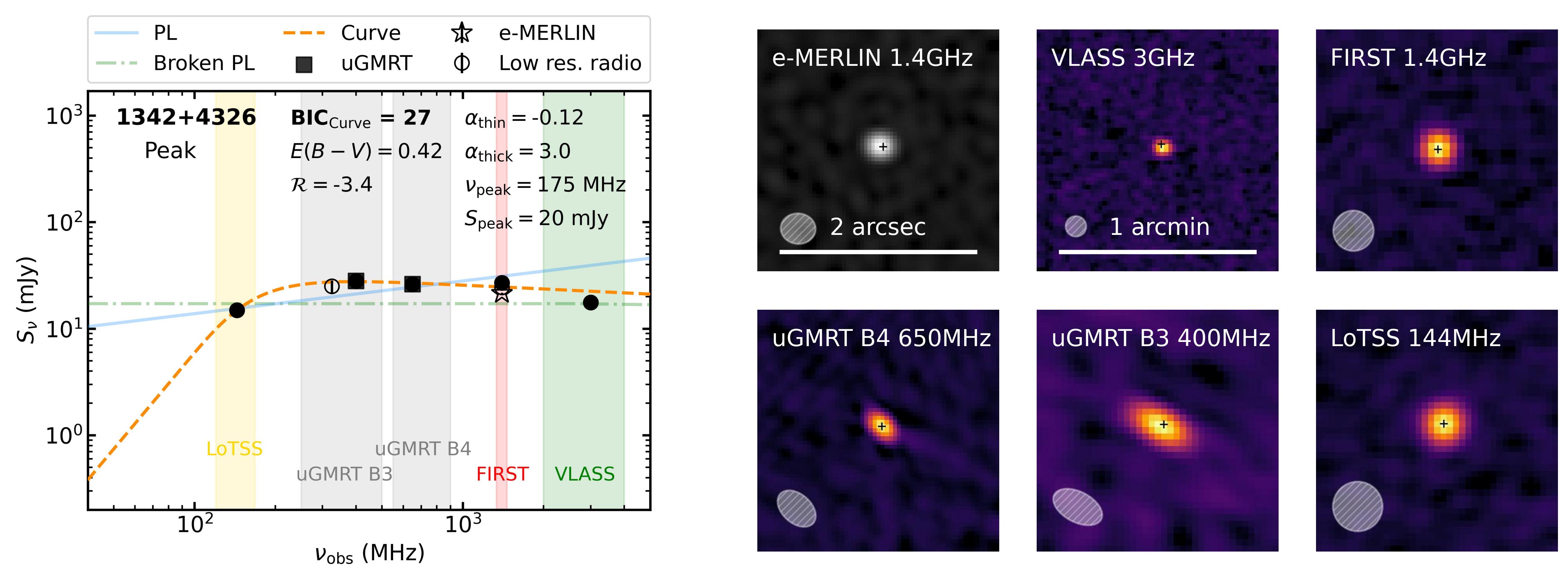}}
    \caption{Continued.}
\end{figure*}

\begin{figure*}
    \ContinuedFloat
    \centering
    \subfloat{\includegraphics[width=0.79\textwidth]{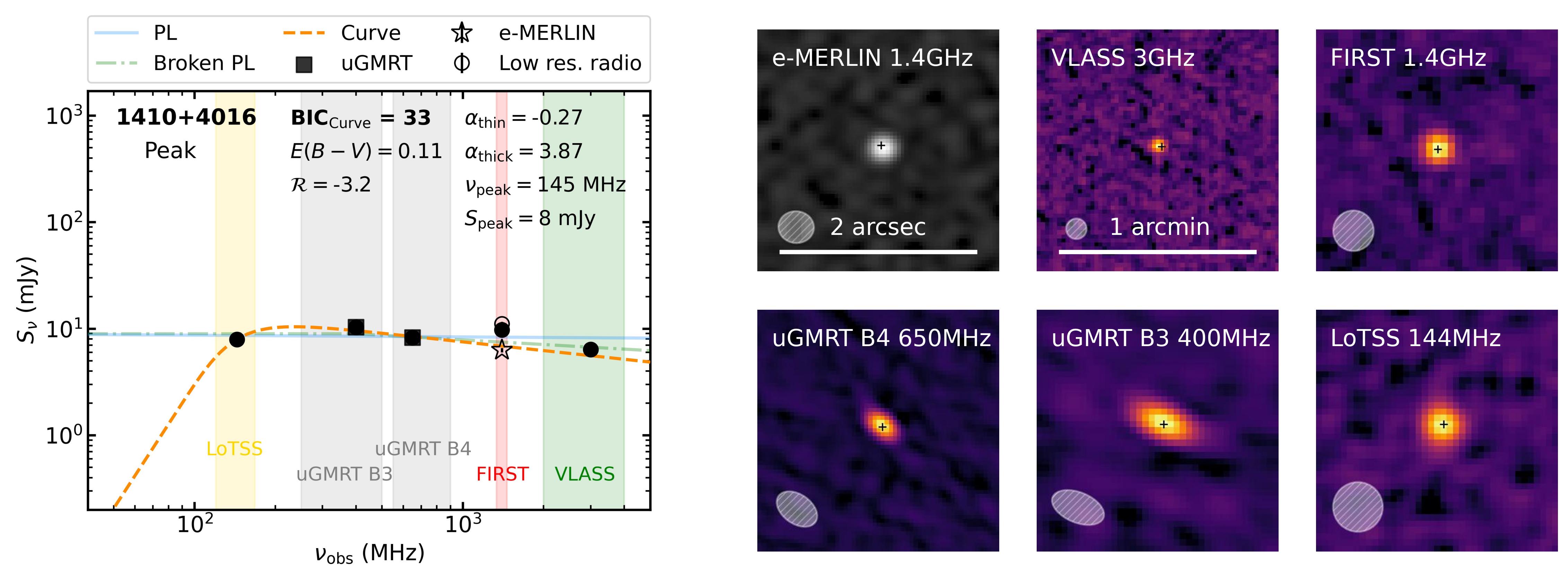}}\\
    \subfloat{\includegraphics[width=0.79\textwidth]{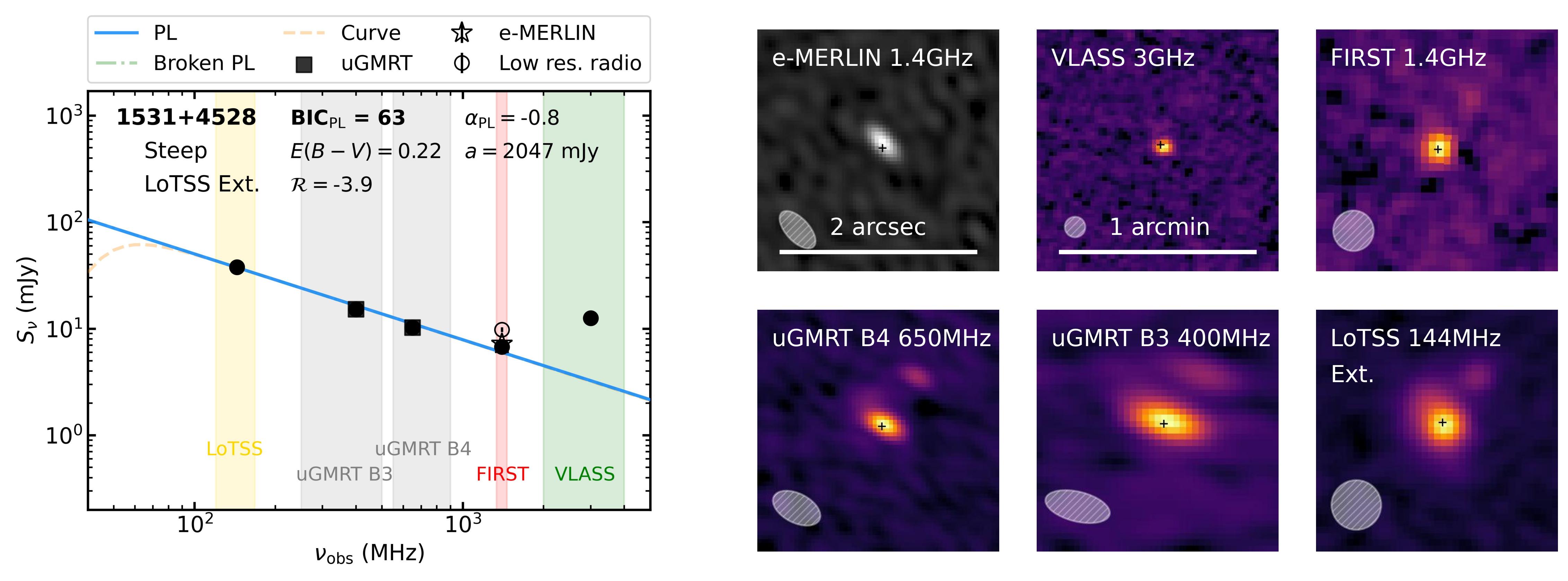}}\\
    \subfloat{\includegraphics[width=0.79\textwidth]{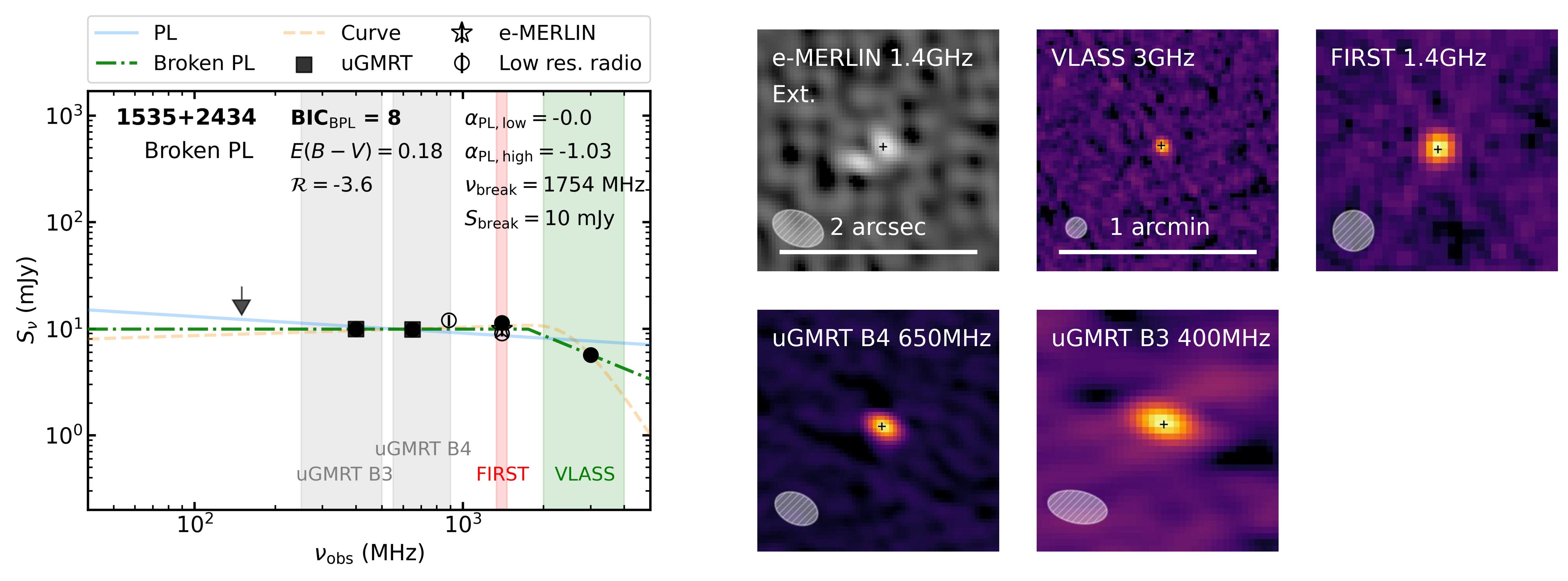}}
    \caption{Continued.}
\end{figure*}

\begin{figure*}
    \centering
    \subfloat{\includegraphics[width=0.79\textwidth]{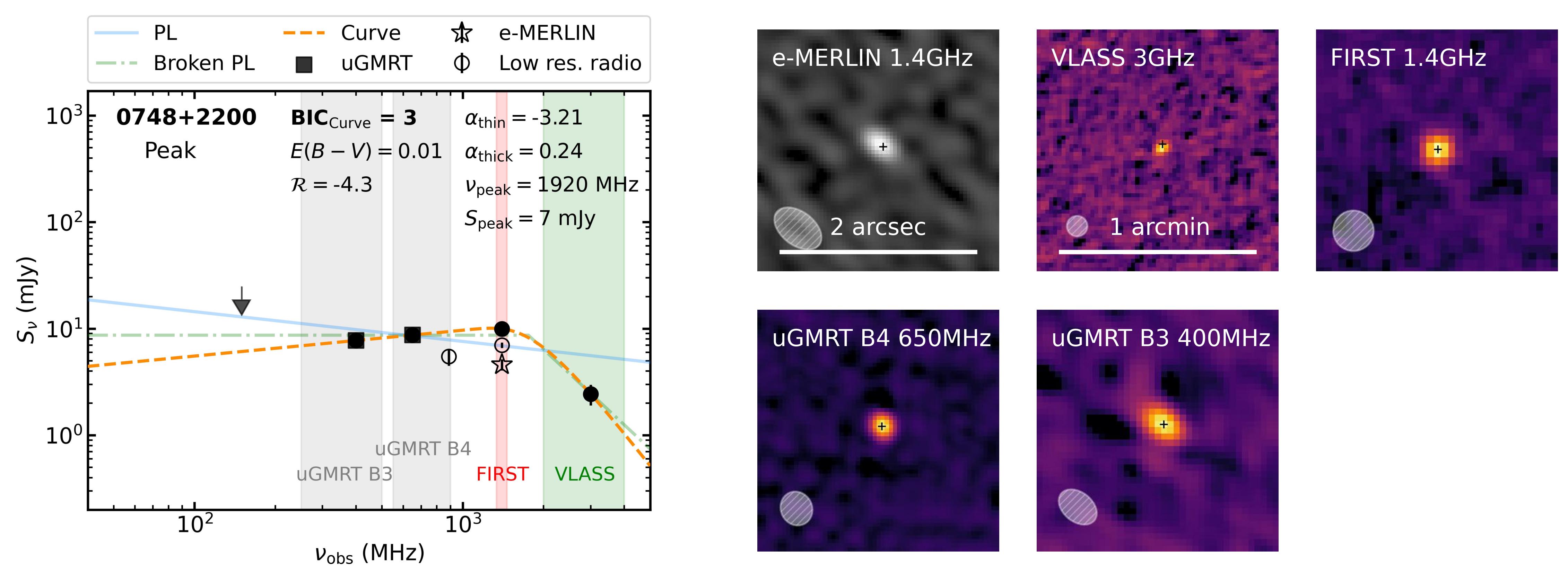}}\\
    \subfloat{\includegraphics[width=0.79\textwidth]{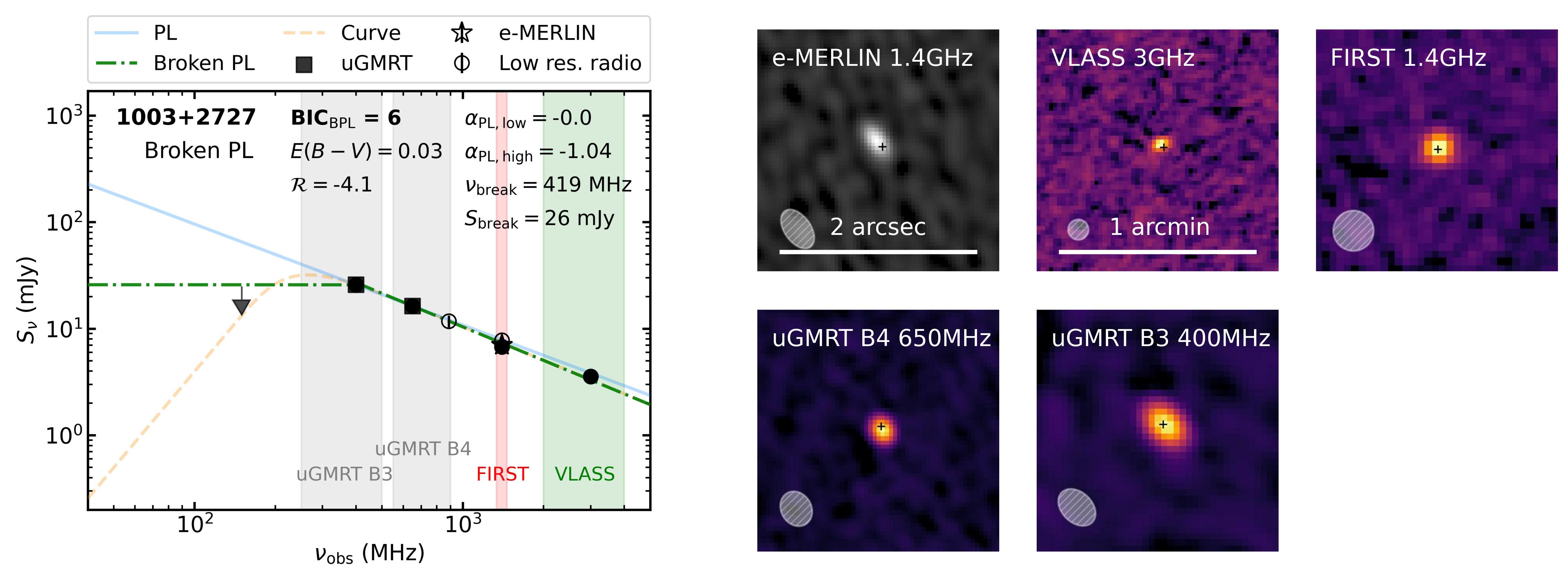}}\\
    \subfloat{\includegraphics[width=0.79\textwidth]{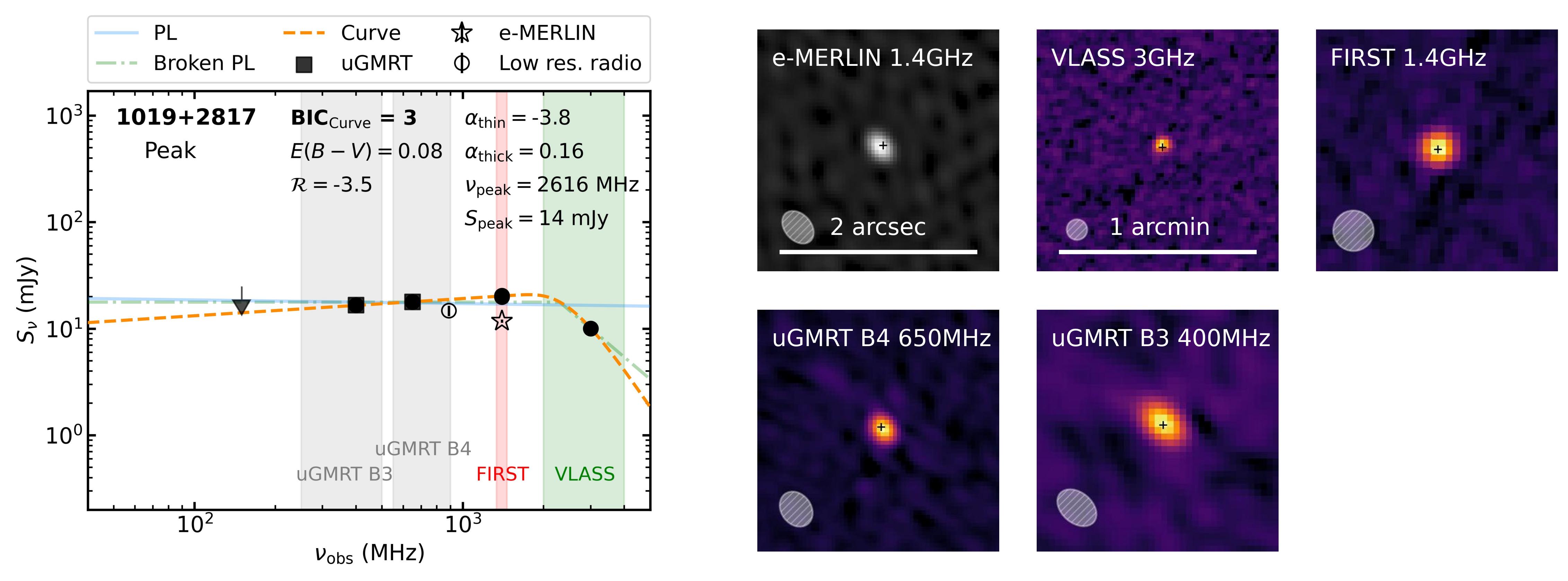}}\\
    \subfloat{\includegraphics[width=0.79\textwidth]{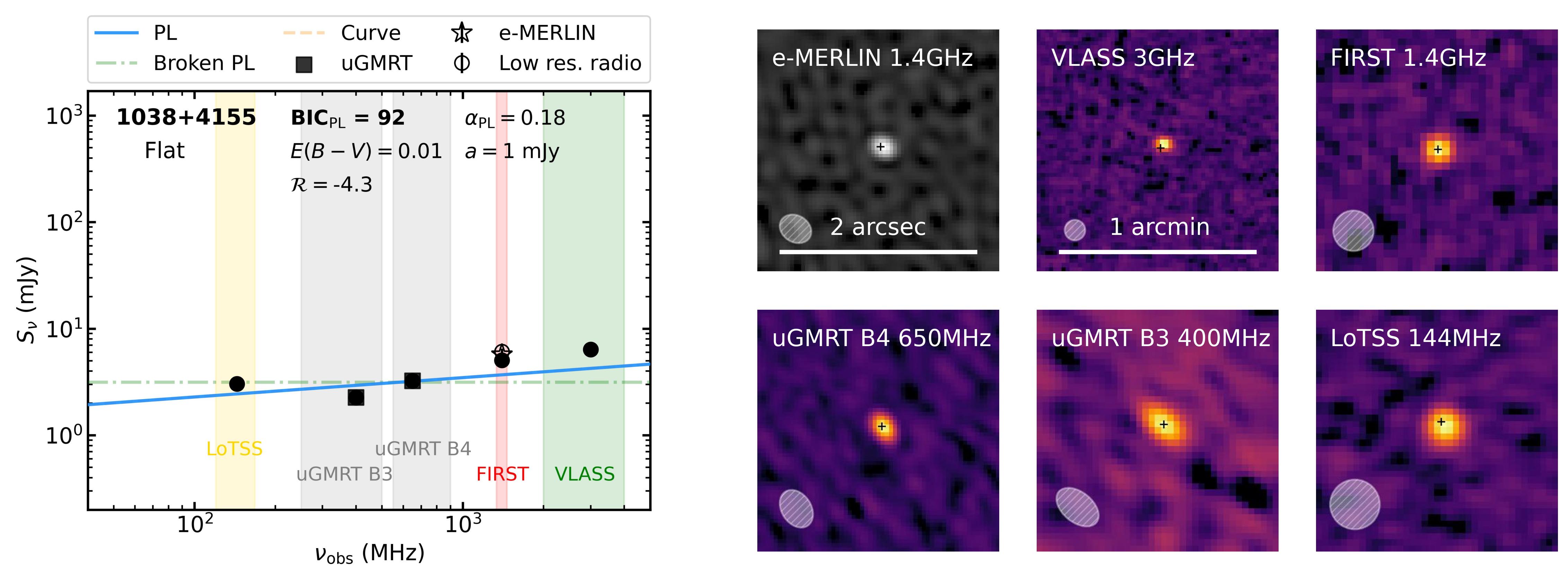}}
    \caption{The radio SEDs (left) and radio images (right thumbnails) for the blue QSO sample. The best-fit power-law, broken power-law and curved model is displayed on the SED as a solid blue, dotted-dashed green, and dashed orange line, respectively. The various radio data utilized in the fitting are shown as solid black points and the band width shown as the shaded regions. Our uGMRT observations are shown as the squares and the e-MERLIN flux is shown as the star. Additional archival low resolution radio emission not used in the SED fitting is displayed by the open circles. For some sources the LoTSS data point is also shown as an open marker, indicating that these sources were originally classified as “upturned” and so the SEDs were refitted excluding the LoTSS data point. In the thumbnails, the QSO Gaia optical position is shown as the black plus.}
    \label{fig:SED_cQSO}
\end{figure*}

\begin{figure*}
    \ContinuedFloat
    \centering
    \subfloat{\includegraphics[width=0.79\textwidth]{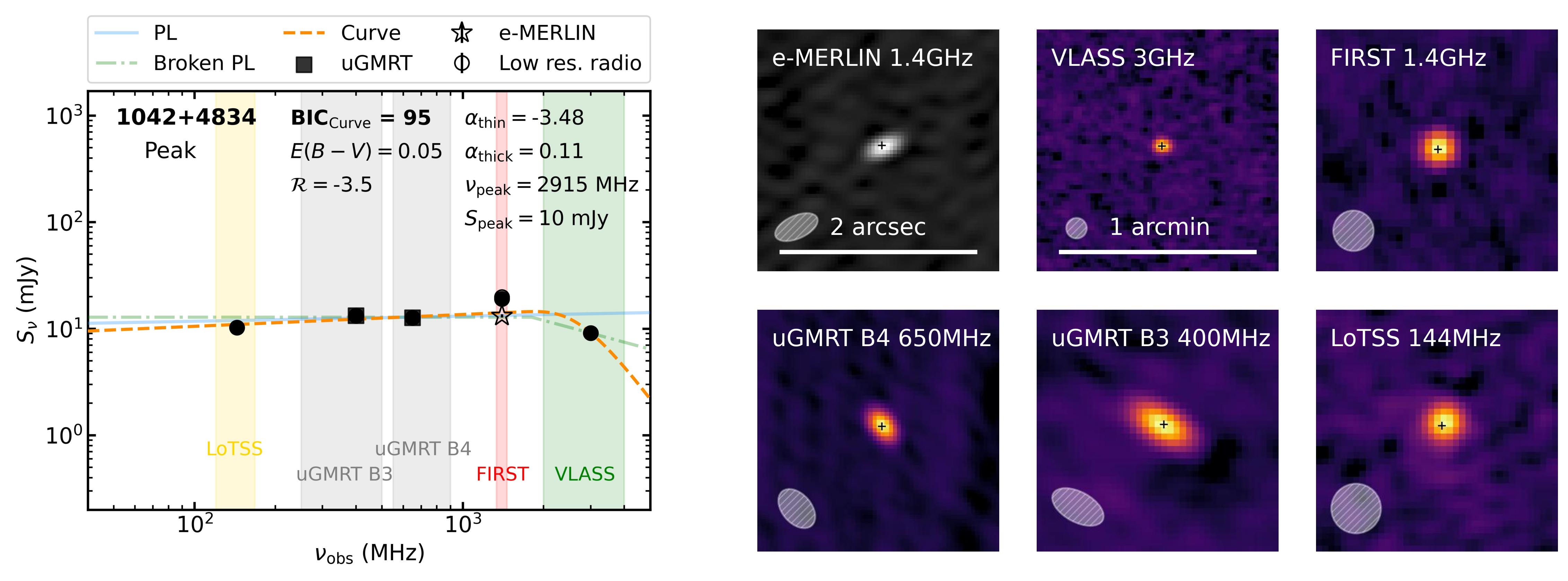}}\\
    \subfloat{\includegraphics[width=0.79\textwidth]{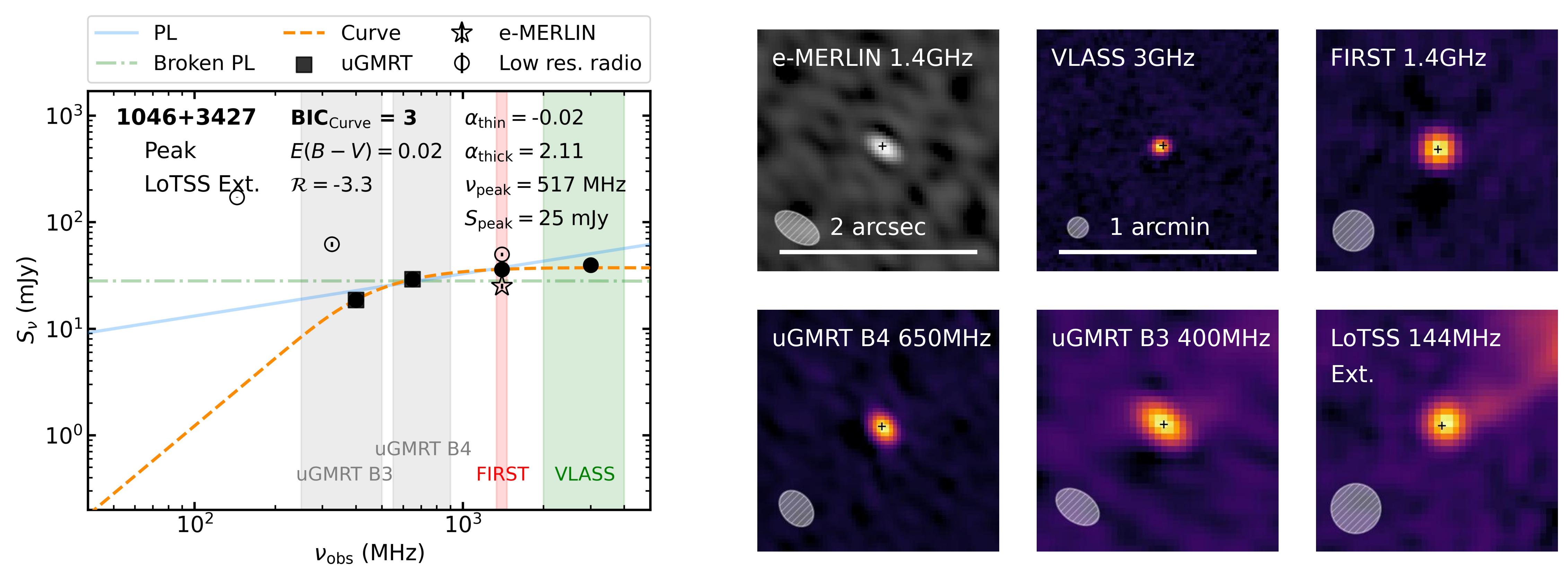}}\\
    \subfloat{\includegraphics[width=0.79\textwidth]{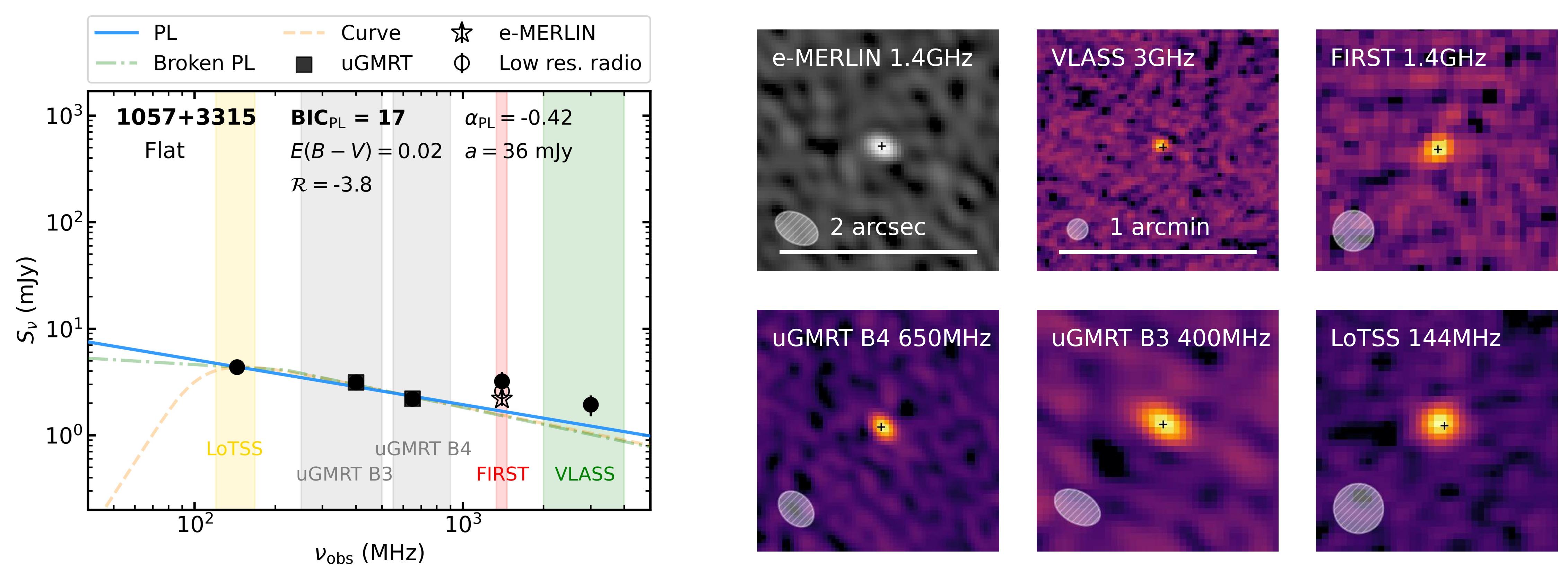}}\\
    \subfloat{\includegraphics[width=0.79\textwidth]{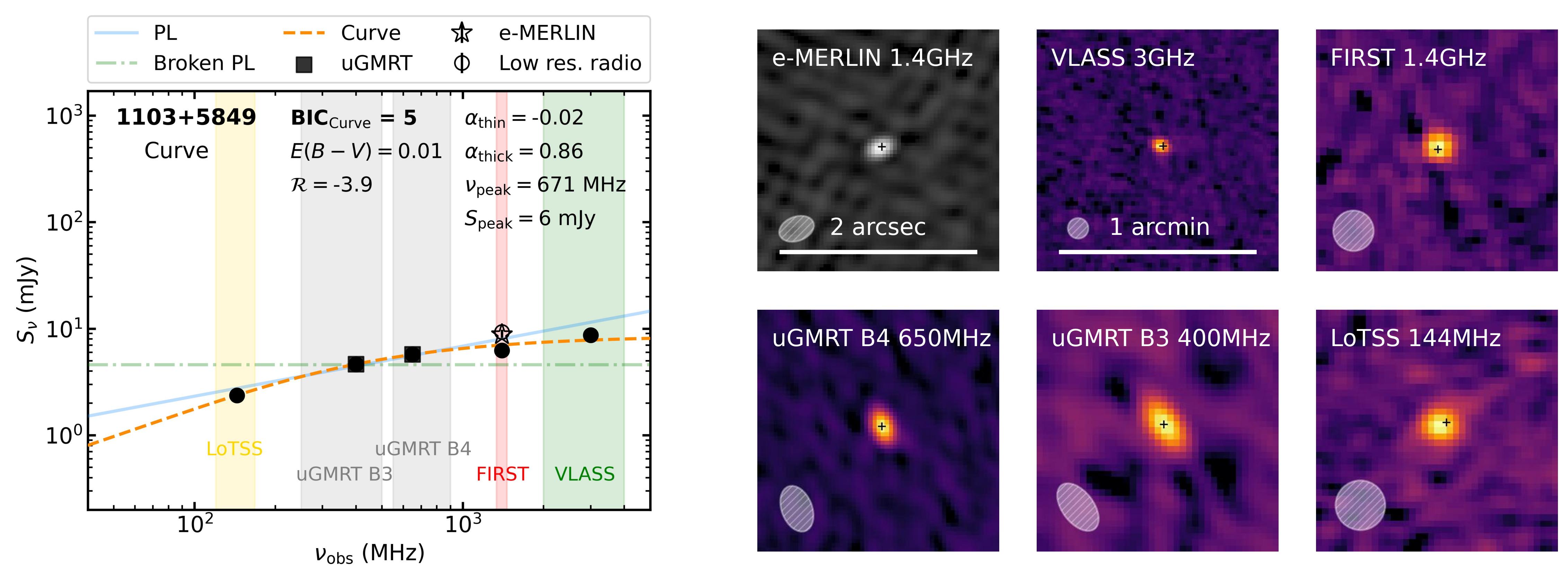}}
    \caption{Continued.}
\end{figure*}

\begin{figure*}
    \ContinuedFloat
    \centering
    \subfloat{\includegraphics[width=0.79\textwidth]{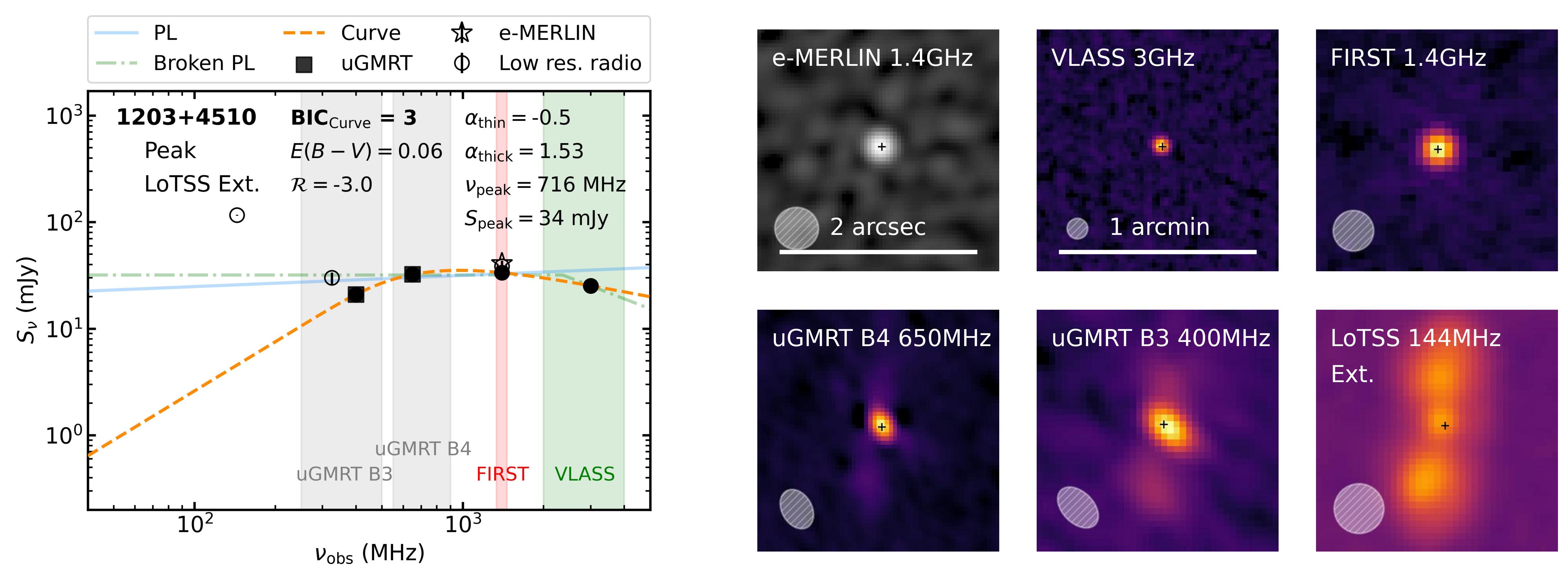}}\\
    \subfloat{\includegraphics[width=0.79\textwidth]{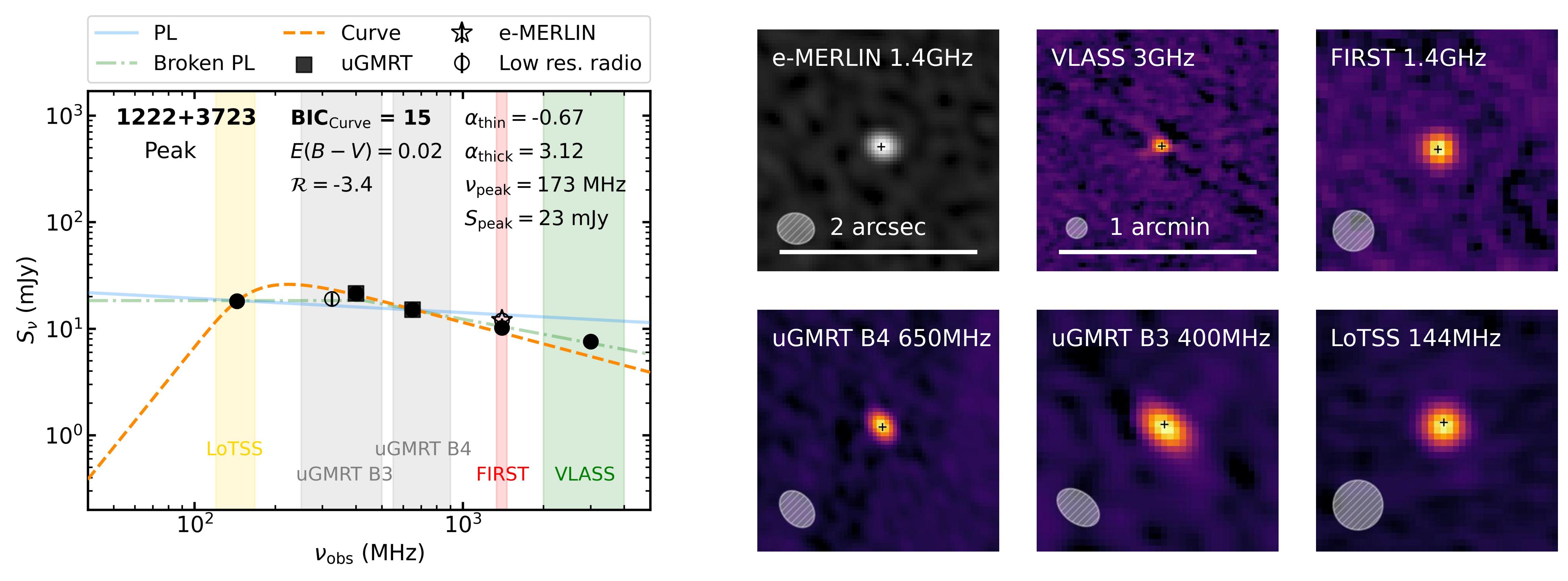}}\\
    \subfloat{\includegraphics[width=0.79\textwidth]{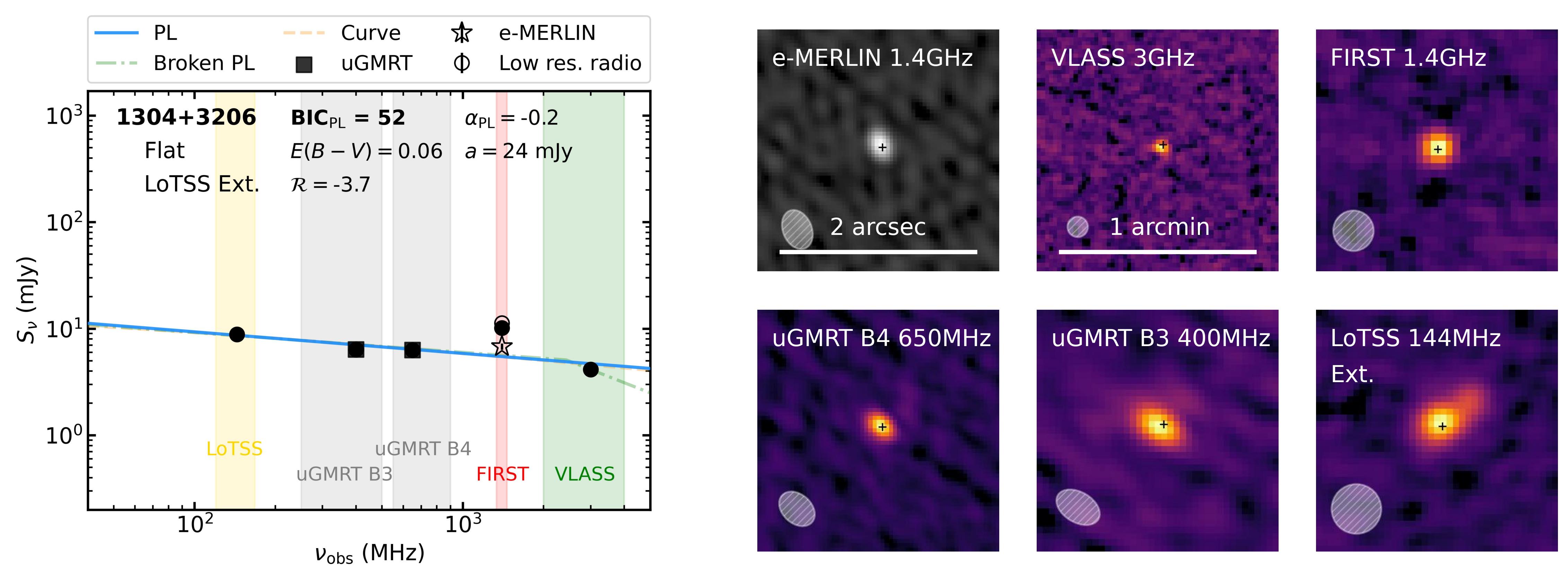}}\\
    \subfloat{\includegraphics[width=0.79\textwidth]{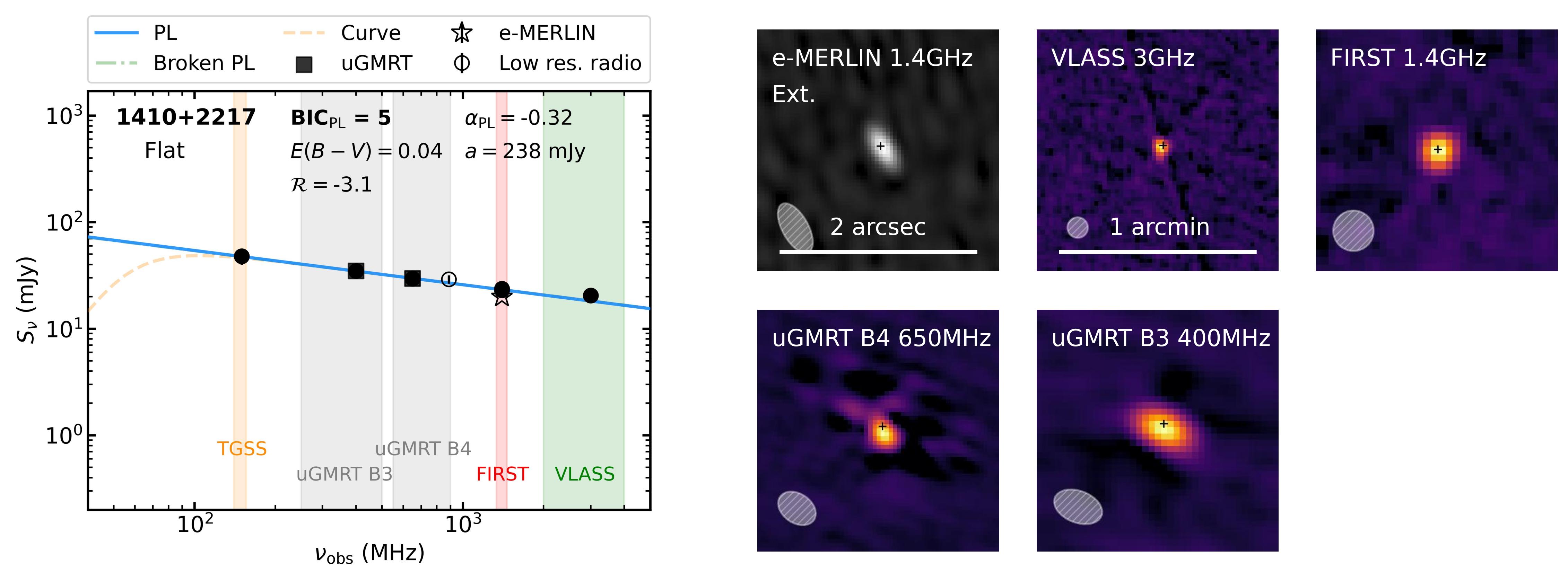}}
    \caption{Continued.}
\end{figure*}

\begin{figure*}
    \ContinuedFloat
    \centering
    \subfloat{\includegraphics[width=0.79\textwidth]{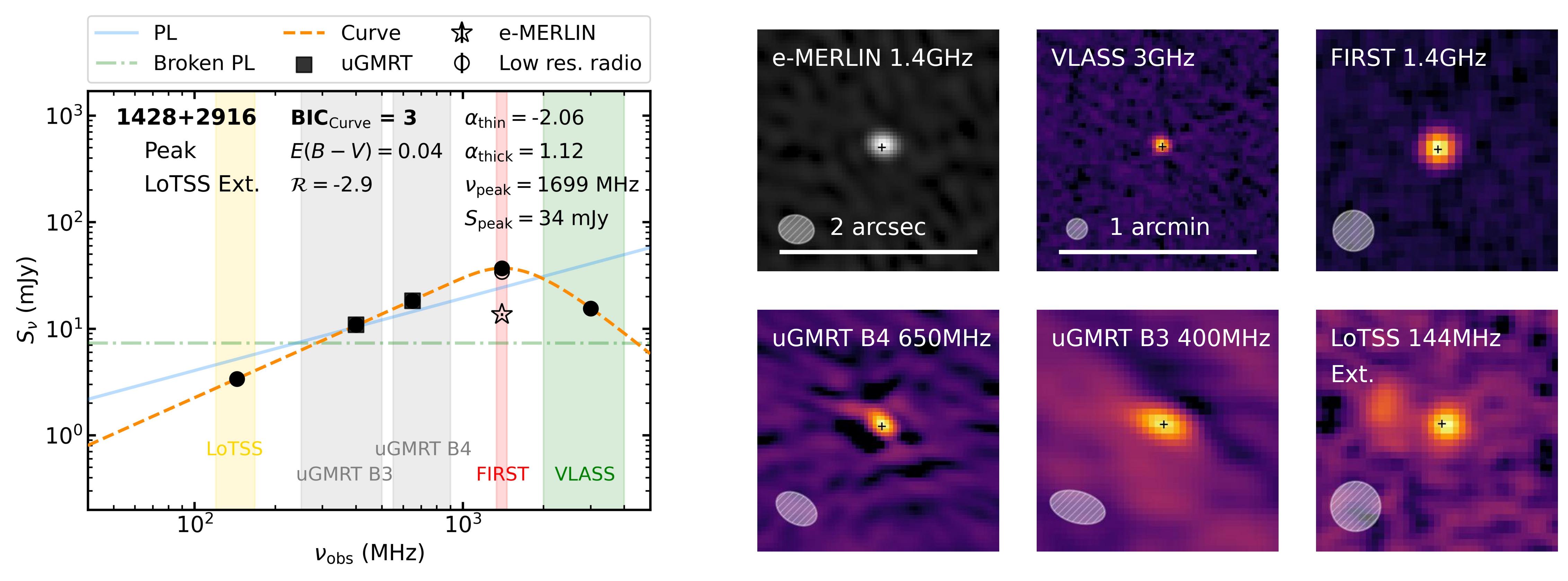}}\\
    \subfloat{\includegraphics[width=0.79\textwidth]{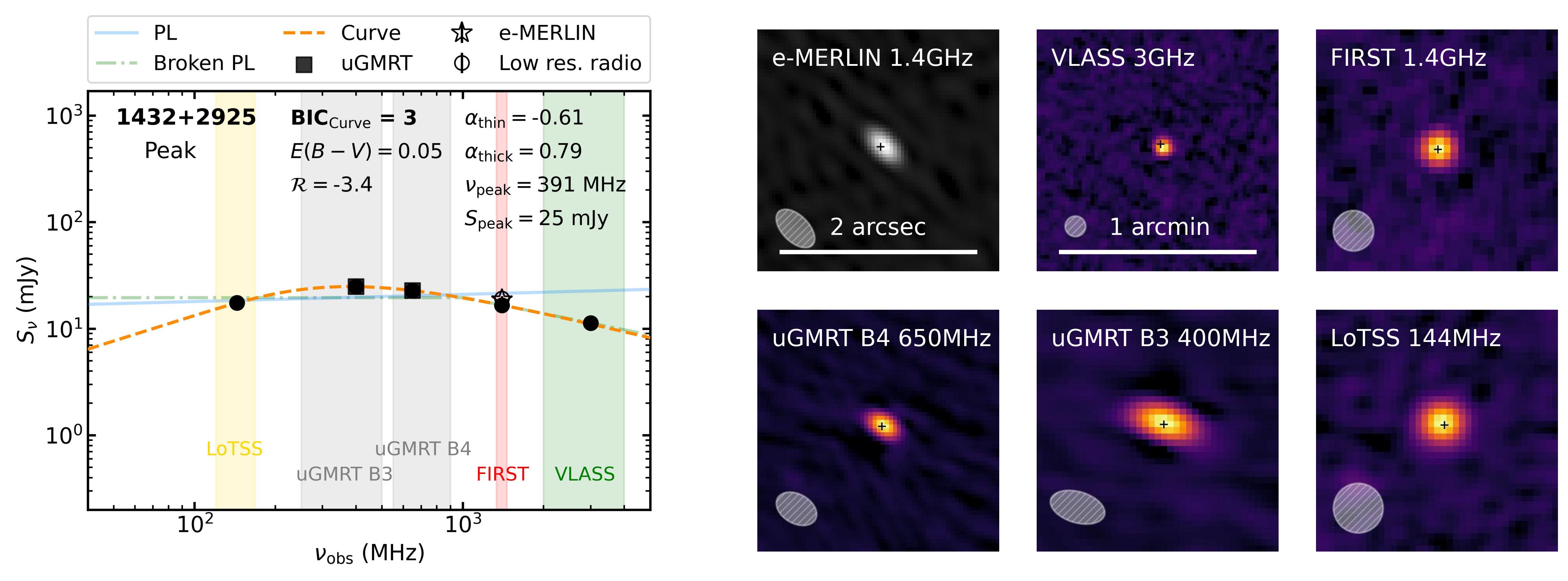}}\\
    \subfloat{\includegraphics[width=0.79\textwidth]{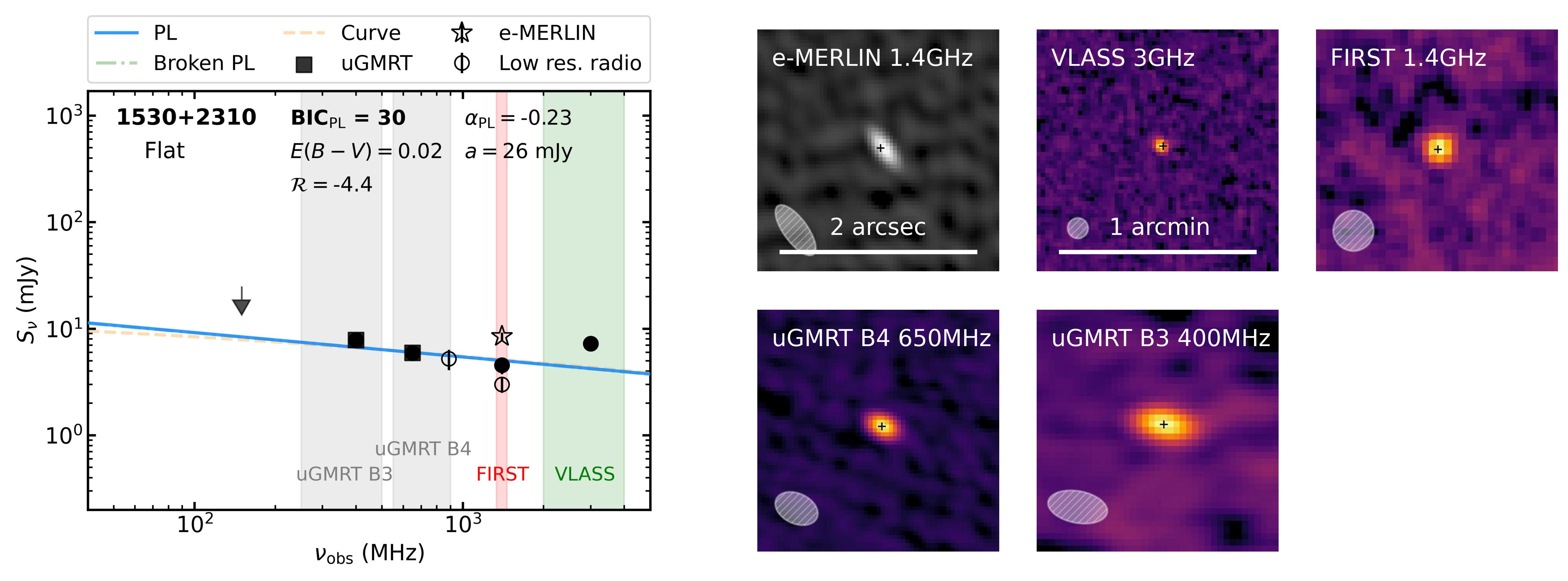}}\\
    \subfloat{\includegraphics[width=0.79\textwidth]{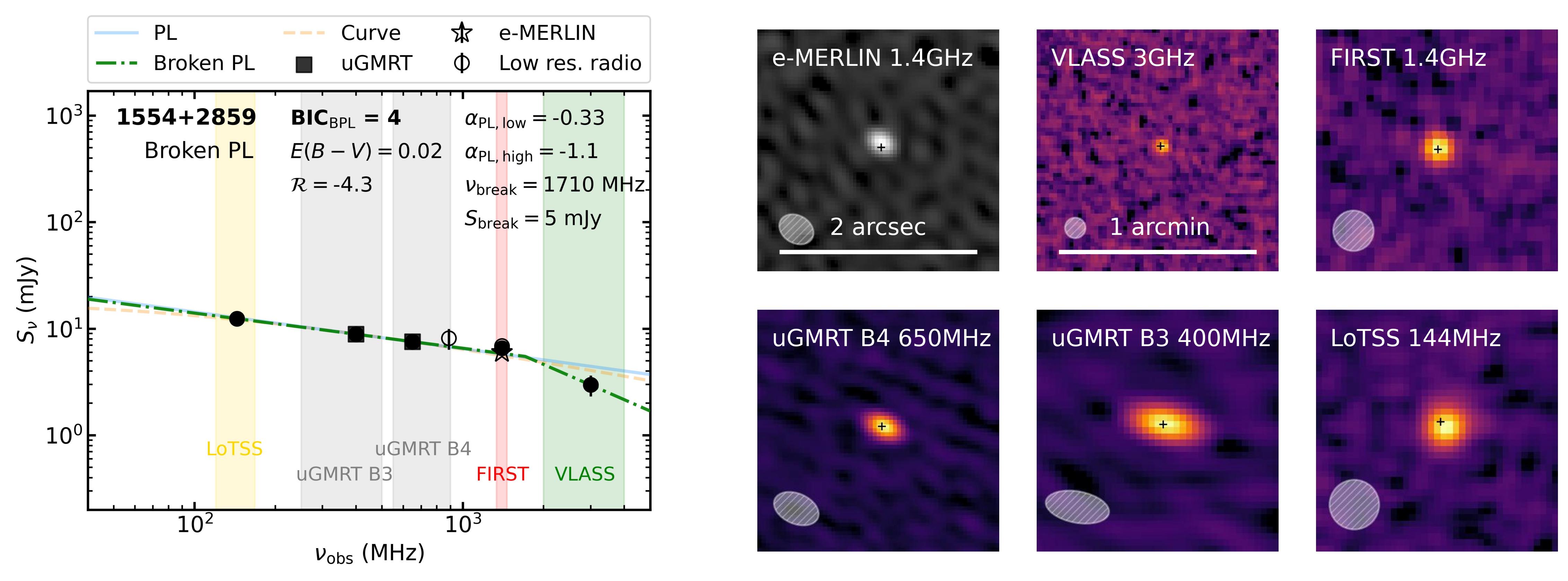}}
    \caption{Continued.}
\end{figure*}

\begin{figure*}
    \ContinuedFloat
    \centering
    \subfloat{\includegraphics[width=0.79\textwidth]{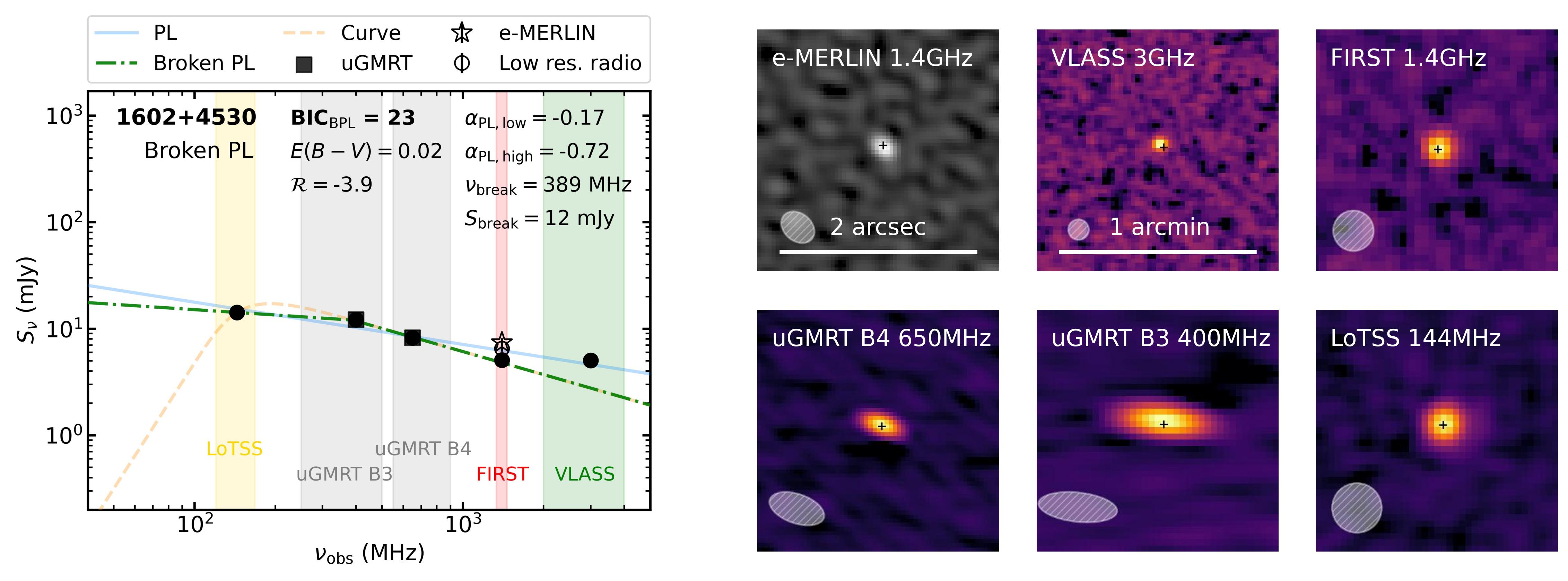}}\\
    \subfloat{\includegraphics[width=0.79\textwidth]{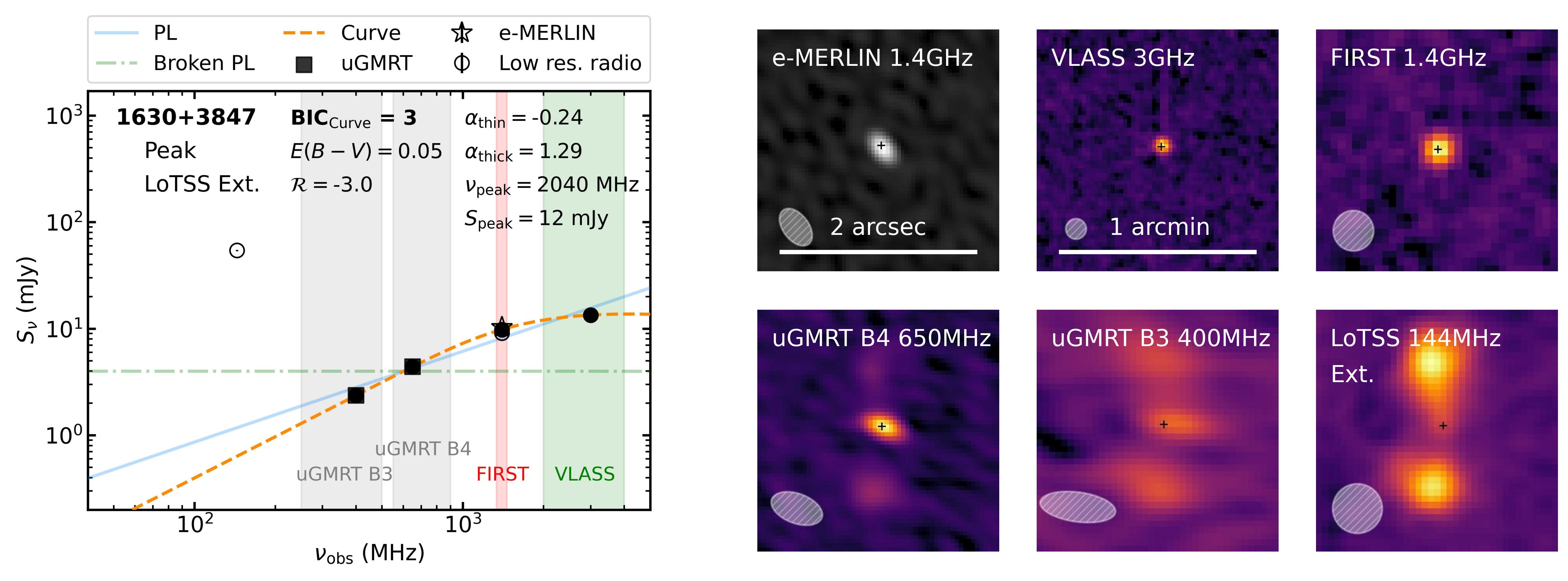}}\\
    \subfloat{\includegraphics[width=0.79\textwidth]{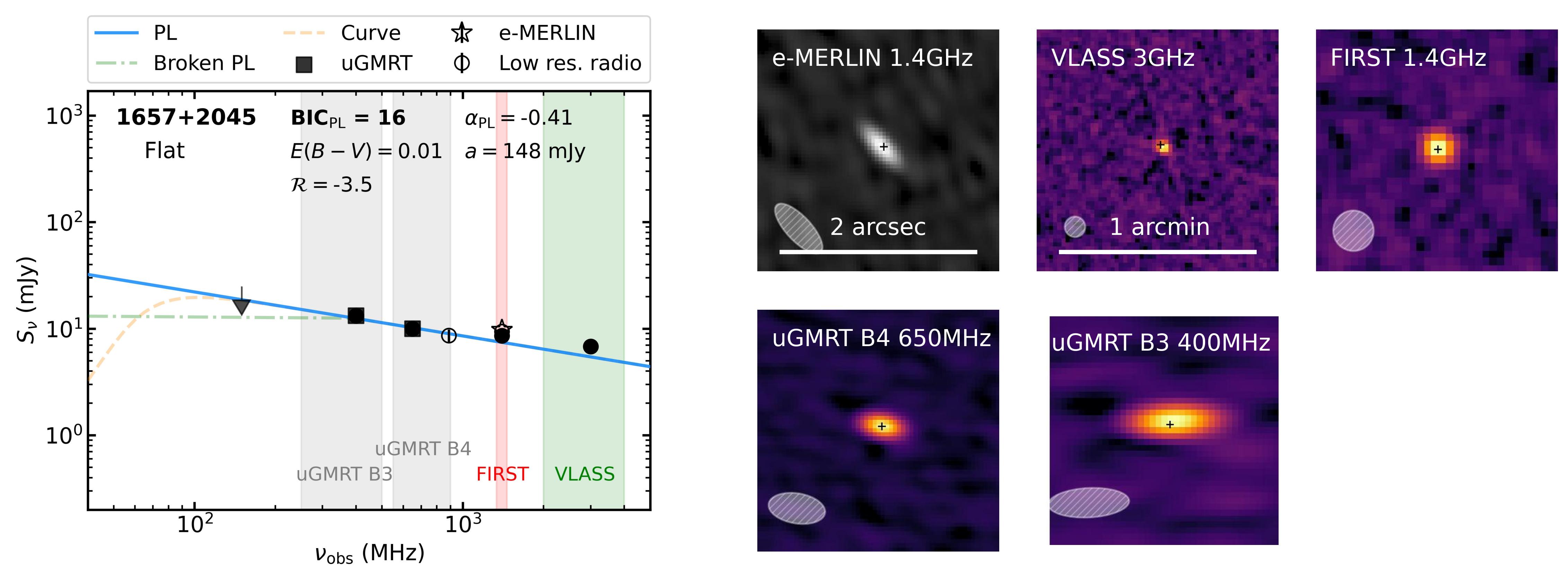}}
    \caption{Continued.}
\end{figure*}



\bibliographystyle{mnras}
\bibliography{bib} 

\begin{thebibliography}{}
\makeatletter
\relax
\def\mn@urlcharsother{\let\do\@makeother \do\$\do\&\do\#\do\^\do\_\do\%\do\~}
\def\mn@doi{\begingroup\mn@urlcharsother \@ifnextchar [ {\mn@doi@} {\mn@doi@[]}}
\def\mn@doi@[#1]#2{\def\@tempa{#1}\ifx\@tempa\@empty \href {http://dx.doi.org/#2} {doi:#2}\else \href {http://dx.doi.org/#2} {#1}\fi \endgroup}
\def\mn@eprint#1#2{\mn@eprint@#1:#2::\@nil}
\def\mn@eprint@arXiv#1{\href {http://arxiv.org/abs/#1} {{\tt arXiv:#1}}}
\def\mn@eprint@dblp#1{\href {http://dblp.uni-trier.de/rec/bibtex/#1.xml} {dblp:#1}}
\def\mn@eprint@#1:#2:#3:#4\@nil{\def\@tempa {#1}\def\@tempb {#2}\def\@tempc {#3}\ifx \@tempc \@empty \let \@tempc \@tempb \let \@tempb \@tempa \fi \ifx \@tempb \@empty \def\@tempb {arXiv}\fi \@ifundefined {mn@eprint@\@tempb}{\@tempb:\@tempc}{\expandafter \expandafter \csname mn@eprint@\@tempb\endcsname \expandafter{\@tempc}}}

\bibitem[\protect\citeauthoryear{Alexander et~al.,}{Alexander et~al.}{2003}]{alex2003}
Alexander D.~M.,  et~al., 2003, \mn@doi [\aj] {10.1086/346088}, 125, 383

\bibitem[\protect\citeauthoryear{An \& Baan}{An \& Baan}{2012}]{An_2012}
An T.,  Baan W.~A.,  2012, \mn@doi [\apj] {10.1088/0004-637x/760/1/77}, 760, 77

\bibitem[\protect\citeauthoryear{{Ananthakrishnan}}{{Ananthakrishnan}}{1995}]{gmrt}
{Ananthakrishnan} S.,  1995, JA\&A, \href {https://ui.adsabs.harvard.edu/abs/1995JApAS..16..427A} {16, 427}

\bibitem[\protect\citeauthoryear{Andonie et~al.,}{Andonie et~al.}{2022}]{andonie}
Andonie C.,  et~al., 2022, \mn@doi [\mnras] {10.1093/mnras/stac2800}, 517, 2577

\bibitem[\protect\citeauthoryear{{Arakawa}, {Fabian}, {Ferland}  \& {Ishibashi}}{{Arakawa} et~al.}{2022}]{arakwa_22}
{Arakawa} N.,  {Fabian} A.~C.,  {Ferland} G.~J.,   {Ishibashi} W.,  2022, \mn@doi [\mnras] {10.1093/mnras/stac3044}, \href {https://ui.adsabs.harvard.edu/abs/2022MNRAS.517.5069A} {517, 5069}

\bibitem[\protect\citeauthoryear{{Banerji}, {McMahon}, {Hewett}, {Alaghband-Zadeh}, {Gonzalez-Solares}, {Venemans}  \& {Hawthorn}}{{Banerji} et~al.}{2012}]{ban12}
{Banerji} M.,  {McMahon} R.~G.,  {Hewett} P.~C.,  {Alaghband-Zadeh} S.,  {Gonzalez-Solares} E.,  {Venemans} B.~P.,   {Hawthorn} M.~J.,  2012, \mn@doi [\mnras] {10.1111/j.1365-2966.2012.22099.x}, \href {https://ui.adsabs.harvard.edu/\#abs/2012MNRAS.427.2275B} {427, 2275}

\bibitem[\protect\citeauthoryear{{Barvainis}, {Leh{\'a}r}, {Birkinshaw}, {Falcke}  \& {Blundell}}{{Barvainis} et~al.}{2005}]{baravainis}
{Barvainis} R.,  {Leh{\'a}r} J.,  {Birkinshaw} M.,  {Falcke} H.,   {Blundell} K.~M.,  2005, \mn@doi [\apj] {10.1086/425859}, \href {https://ui.adsabs.harvard.edu/abs/2005ApJ...618..108B} {618, 108}

\bibitem[\protect\citeauthoryear{Baskin \& Laor}{Baskin \& Laor}{2005}]{baskin}
Baskin A.,  Laor A.,  2005, \mn@doi [MNRAS] {10.1111/j.1365-2966.2005.08841.x}, 358, 1043

\bibitem[\protect\citeauthoryear{{Baum}, {O'Dea}, {Murphy}  \& {de Bruyn}}{{Baum} et~al.}{1990}]{baum_90}
{Baum} S.~A.,  {O'Dea} C.~P.,  {Murphy} D.~W.,   {de Bruyn} A.~G.,  1990, \aap, \href {https://ui.adsabs.harvard.edu/abs/1990A&A...232...19B} {232, 19}

\bibitem[\protect\citeauthoryear{{Becker}, {White}  \& {Helfand}}{{Becker} et~al.}{1995}]{becker}
{Becker} R.~H.,  {White} R.~L.,   {Helfand} D.~J.,  1995, \mn@doi [\apj] {10.1086/176166}, \href {https://ui.adsabs.harvard.edu/abs/1995ApJ...450..559B} {450, 559}

\bibitem[\protect\citeauthoryear{Bicknell, Dopita  \& O{\textquotesingle}Dea}{Bicknell et~al.}{1997}]{Bicknell_1997}
Bicknell G.~V.,  Dopita M.~A.,   O{\textquotesingle}Dea C. P.~O.,  1997, \mn@doi [\apj] {10.1086/304400}, 485, 112

\bibitem[\protect\citeauthoryear{Bicknell, Mukherjee, Wagner, Sutherland  \& Nesvadba}{Bicknell et~al.}{2018}]{bicknell_18}
Bicknell G.~V.,  Mukherjee D.,  Wagner A.~Y.,  Sutherland R.~S.,   Nesvadba N. P.~H.,  2018, \mn@doi [\mnras] {10.1093/mnras/sty070}, 475, 3493

\bibitem[\protect\citeauthoryear{{Blandford} \& {K{\"o}nigl}}{{Blandford} \& {K{\"o}nigl}}{1979}]{blandford_79}
{Blandford} R.~D.,  {K{\"o}nigl} A.,  1979, \mn@doi [\apj] {10.1086/157262}, \href {https://ui.adsabs.harvard.edu/abs/1979ApJ...232...34B} {232, 34}

\bibitem[\protect\citeauthoryear{{Boroson} \& {Green}}{{Boroson} \& {Green}}{1992}]{boroson}
{Boroson} T.~A.,  {Green} R.~F.,  1992, \mn@doi [\apjs] {10.1086/191661}, \href {https://ui.adsabs.harvard.edu/abs/1992ApJS...80..109B} {80, 109}

\bibitem[\protect\citeauthoryear{{Calistro Rivera} et~al.,}{{Calistro Rivera} et~al.}{2017}]{calistro_17}
{Calistro Rivera} G.,  et~al., 2017, \mn@doi [\mnras] {10.1093/mnras/stx1040}, \href {https://ui.adsabs.harvard.edu/abs/2017MNRAS.469.3468C} {469, 3468}

\bibitem[\protect\citeauthoryear{{Calistro Rivera} et~al.,}{{Calistro Rivera} et~al.}{2021}]{calistro}
{Calistro Rivera} G.,  et~al., 2021, \mn@doi [\aap] {10.1051/0004-6361/202040214}, \href {https://ui.adsabs.harvard.edu/abs/2021A&A...649A.102C} {649, A102}

\bibitem[\protect\citeauthoryear{{Calistro Rivera} et~al.,}{{Calistro Rivera} et~al.}{2024}]{calistro_2024}
{Calistro Rivera} G.,  et~al., 2024, \mn@doi [arXiv e-prints] {10.48550/arXiv.2312.10177}, \href {https://ui.adsabs.harvard.edu/abs/2023arXiv231210177C} {p. arXiv:2312.10177}

\bibitem[\protect\citeauthoryear{{Callingham} et~al.,}{{Callingham} et~al.}{2017}]{callingham}
{Callingham} J.~R.,  et~al., 2017, \mn@doi [\apj] {10.3847/1538-4357/836/2/174}, \href {https://ui.adsabs.harvard.edu/abs/2017ApJ...836..174C} {836, 174}

\bibitem[\protect\citeauthoryear{{Cameron}}{{Cameron}}{2011}]{cam}
{Cameron} E.,  2011, \mn@doi [\pasa] {10.1071/AS10046}, \href {https://ui.adsabs.harvard.edu/\#abs/2011PASA...28..128C} {28, 128}

\bibitem[\protect\citeauthoryear{{Carvalho}}{{Carvalho}}{1985}]{carvalho}
{Carvalho} J.~C.,  1985, \mn@doi [\mnras] {10.1093/mnras/215.3.463}, \href {https://ui.adsabs.harvard.edu/abs/1985MNRAS.215..463C} {215, 463}

\bibitem[\protect\citeauthoryear{{Chen} et~al.,}{{Chen} et~al.}{2024}]{chen_24}
{Chen} S.,  et~al., 2024, arXiv e-prints, \href {https://ui.adsabs.harvard.edu/abs/2024arXiv240815934C} {p. arXiv:2408.15934}

\bibitem[\protect\citeauthoryear{{Cirasuolo} et~al.,}{{Cirasuolo} et~al.}{2020}]{moons}
{Cirasuolo} M.,  et~al., 2020, \mn@doi [The Messenger] {10.18727/0722-6691/5195}, \href {https://ui.adsabs.harvard.edu/abs/2020Msngr.180...10C} {180, 10}

\bibitem[\protect\citeauthoryear{{Condon}, {Cotton}, {Greisen}, {Yin}, {Perley}, {Taylor}  \& {Broderick}}{{Condon} et~al.}{1998}]{NVSS}
{Condon} J.~J.,  {Cotton} W.~D.,  {Greisen} E.~W.,  {Yin} Q.~F.,  {Perley} R.~A.,  {Taylor} G.~B.,   {Broderick} J.~J.,  1998, \mn@doi [\aj] {10.1086/300337}, \href {https://ui.adsabs.harvard.edu/abs/1998AJ....115.1693C} {115, 1693}

\bibitem[\protect\citeauthoryear{{Condon}, {Kellermann}, {Kimball}, {Ivezi{\'c}}  \& {Perley}}{{Condon} et~al.}{2013}]{condon}
{Condon} J.~J.,  {Kellermann} K.~I.,  {Kimball} A.~E.,  {Ivezi{\'c}} {\v{Z}}.,   {Perley} R.~A.,  2013, \mn@doi [ApJ] {10.1088/0004-637X/768/1/37}, \href {https://ui.adsabs.harvard.edu/abs/2013ApJ...768...37C} {768, 37}

\bibitem[\protect\citeauthoryear{{DESI Collaboration} et~al.,}{{DESI Collaboration} et~al.}{2016a}]{desi}
{DESI Collaboration} et~al., 2016a, arXiv e-prints, \href {https://ui.adsabs.harvard.edu/abs/2016arXiv161100036D} {p. arXiv:1611.00036}

\bibitem[\protect\citeauthoryear{{DESI Collaboration} et~al.,}{{DESI Collaboration} et~al.}{2016b}]{desiII}
{DESI Collaboration} et~al., 2016b, arXiv e-prints, \href {https://ui.adsabs.harvard.edu/abs/2016arXiv161100037D} {p. arXiv:1611.00037}

\bibitem[\protect\citeauthoryear{{Duffy} \& {Blundell}}{{Duffy} \& {Blundell}}{2012}]{duffy}
{Duffy} P.,  {Blundell} K.~M.,  2012, \mn@doi [\mnras] {10.1111/j.1365-2966.2011.20239.x}, \href {https://ui.adsabs.harvard.edu/abs/2012MNRAS.421..108D} {421, 108}

\bibitem[\protect\citeauthoryear{{Eckart}, {Witzel}, {Biermann}, {Johnston}, {Simon}, {Schalinski}  \& {Kuhr}}{{Eckart} et~al.}{1986}]{eckart}
{Eckart} A.,  {Witzel} A.,  {Biermann} P.,  {Johnston} K.~J.,  {Simon} R.,  {Schalinski} C.,   {Kuhr} H.,  1986, \aap, \href {https://ui.adsabs.harvard.edu/abs/1986A&A...168...17E} {168, 17}

\bibitem[\protect\citeauthoryear{Fanaroff \& Riley}{Fanaroff \& Riley}{1974}]{fr}
Fanaroff B.~L.,  Riley J.~M.,  1974, \mn@doi [\mnras] {10.1093/mnras/167.1.31P}, 167, 31P

\bibitem[\protect\citeauthoryear{{Fanti}, {Fanti}, {Schilizzi}, {Spencer}, {Nan Rendong}, {Parma}, {van Breugel}  \& {Venturi}}{{Fanti} et~al.}{1990}]{fanti90}
{Fanti} R.,  {Fanti} C.,  {Schilizzi} R.~T.,  {Spencer} R.~E.,  {Nan Rendong} {Parma} P.,  {van Breugel} W.~J.~M.,   {Venturi} T.,  1990, \aap, \href {https://ui.adsabs.harvard.edu/abs/1990A&A...231..333F} {231, 333}

\bibitem[\protect\citeauthoryear{{Fanti}, {Fanti}, {Dallacasa}, {Schilizzi}, {Spencer}  \& {Stanghellini}}{{Fanti} et~al.}{1995}]{fanti}
{Fanti} C.,  {Fanti} R.,  {Dallacasa} D.,  {Schilizzi} R.~T.,  {Spencer} R.~E.,   {Stanghellini} C.,  1995, \aap, \href {https://ui.adsabs.harvard.edu/abs/1995A&A...302..317F} {302, 317}

\bibitem[\protect\citeauthoryear{{Faucher-Gigu{\`e}re} \& {Quataert}}{{Faucher-Gigu{\`e}re} \& {Quataert}}{2012}]{faucher}
{Faucher-Gigu{\`e}re} C.-A.,  {Quataert} E.,  2012, \mn@doi [\mnras] {10.1111/j.1365-2966.2012.21512.x}, \href {https://ui.adsabs.harvard.edu/abs/2012MNRAS.425..605F} {425, 605}

\bibitem[\protect\citeauthoryear{{Fawcett}, {Alexander}, {Rosario}, {Klindt}, {Fotopoulou}, {Lusso}, {Morabito}  \& {Calistro Rivera}}{{Fawcett} et~al.}{2020}]{fawcett20}
{Fawcett} V.~A.,  {Alexander} D.~M.,  {Rosario} D.~J.,  {Klindt} L.,  {Fotopoulou} S.,  {Lusso} E.,  {Morabito} L.~K.,   {Calistro Rivera} G.,  2020, \mn@doi [\mnras] {10.1093/mnras/staa954}, \href {https://ui.adsabs.harvard.edu/abs/2020MNRAS.494.4802F} {494, 4802}

\bibitem[\protect\citeauthoryear{{Fawcett}, {Alexander}, {Rosario}  \& {Klindt}}{{Fawcett} et~al.}{2021}]{fawcett21}
{Fawcett} V.~A.,  {Alexander} D.~M.,  {Rosario} D.~J.,   {Klindt} L.,  2021, \mn@doi [Galaxies] {10.3390/galaxies9040107}, \href {https://ui.adsabs.harvard.edu/abs/2021Galax...9..107F} {9, 107}

\bibitem[\protect\citeauthoryear{{Fawcett}, Alexander, Rosario, Klindt, Lusso, Morabito  \& Calistro Rivera}{{Fawcett} et~al.}{2022}]{fawcett22}
{Fawcett} V.~A.,  Alexander D.~M.,  Rosario D.~J.,  Klindt L.,  Lusso E.,  Morabito L.~K.,   Calistro Rivera G.,  2022, \mn@doi [\mnras] {10.1093/mnras/stac945}, 513, 1254

\bibitem[\protect\citeauthoryear{{Fawcett} et~al.,}{{Fawcett} et~al.}{2023}]{fawcett23}
{Fawcett} V.~A.,  et~al., 2023, \mn@doi [\mnras] {10.1093/mnras/stad2603}, \href {https://ui.adsabs.harvard.edu/abs/2023MNRAS.525.5575F} {525, 5575}

\bibitem[\protect\citeauthoryear{{Foreman-Mackey}, {Hogg}, {Lang}  \& {Goodman}}{{Foreman-Mackey} et~al.}{2013}]{emcee}
{Foreman-Mackey} D.,  {Hogg} D.~W.,  {Lang} D.,   {Goodman} J.,  2013, \mn@doi [\pasp] {10.1086/670067}, \href {https://ui.adsabs.harvard.edu/abs/2013PASP..125..306F} {125, 306}

\bibitem[\protect\citeauthoryear{{Gaia Collaboration} et~al.,}{{Gaia Collaboration} et~al.}{2023}]{gaia_dr3}
{Gaia Collaboration} et~al., 2023, \mn@doi [\aap] {10.1051/0004-6361/202243940}, \href {https://ui.adsabs.harvard.edu/abs/2023A&A...674A...1G} {674, A1}

\bibitem[\protect\citeauthoryear{{Georgakakis}, Clements, Bendo, Rowan-Robinson, Nandra  \& Brotherton}{{Georgakakis} et~al.}{2009}]{georg}
{Georgakakis} A.,  Clements D.~L.,  Bendo G.,  Rowan-Robinson M.,  Nandra K.,   Brotherton M.~S.,  2009, \mn@doi [\mnras] {10.1111/j.1365-2966.2008.14344.x}, 394, 533

\bibitem[\protect\citeauthoryear{{Georgakakis}, {Grossi}, {Afonso}  \& {Hopkins}}{{Georgakakis} et~al.}{2012}]{georg12}
{Georgakakis} A.,  {Grossi} M.,  {Afonso} J.,   {Hopkins} A.~M.,  2012, \mn@doi [\mnras] {10.1111/j.1365-2966.2012.20446.x}, \href {https://ui.adsabs.harvard.edu/abs/2012MNRAS.421.2223G} {421, 2223}

\bibitem[\protect\citeauthoryear{{Girdhar} et~al.,}{{Girdhar} et~al.}{2022}]{girdhar}
{Girdhar} A.,  et~al., 2022, \mn@doi [\mnras] {10.1093/mnras/stac073}, \href {https://ui.adsabs.harvard.edu/abs/2022MNRAS.512.1608G} {512, 1608}

\bibitem[\protect\citeauthoryear{{Glikman}, {Helfand}, {White}, {Becker}, {Gregg}  \& {Lacy}}{{Glikman} et~al.}{2007}]{glik7}
{Glikman} E.,  {Helfand} D.~J.,  {White} R.~L.,  {Becker} R.~H.,  {Gregg} M.~D.,   {Lacy} M.,  2007, \mn@doi [\apj] {10.1086/521073}, \href {https://ui.adsabs.harvard.edu/\#abs/2007ApJ...667..673G} {667, 673}

\bibitem[\protect\citeauthoryear{{Glikman} et~al.,}{{Glikman} et~al.}{2012}]{glik12}
{Glikman} E.,  et~al., 2012, \mn@doi [\apj] {10.1088/0004-637X/757/1/51}, \href {http://adsabs.harvard.edu/abs/2012ApJ...757...51G} {757, 51}

\bibitem[\protect\citeauthoryear{{Glikman} et~al.,}{{Glikman} et~al.}{2022}]{glikman22}
{Glikman} E.,  et~al., 2022, \mn@doi [\apj] {10.3847/1538-4357/ac6bee}, \href {https://ui.adsabs.harvard.edu/abs/2022ApJ...934..119G} {934, 119}

\bibitem[\protect\citeauthoryear{{Gordon} et~al.,}{{Gordon} et~al.}{2020}]{vlass}
{Gordon} Y.~A.,  et~al., 2020, \mn@doi [RNAAS] {10.3847/2515-5172/abbe23}, \href {https://ui.adsabs.harvard.edu/abs/2020RNAAS...4..175G} {4, 175}

\bibitem[\protect\citeauthoryear{Gordon et~al.,}{Gordon et~al.}{2021}]{Gordon_2021}
Gordon Y.~A.,  et~al., 2021, \mn@doi [\apjs] {10.3847/1538-4365/ac05c0}, 255, 30

\bibitem[\protect\citeauthoryear{{Gupta} et~al.,}{{Gupta} et~al.}{2017}]{ugmrt}
{Gupta} Y.,  et~al., 2017, \mn@doi [Current Science] {10.18520/cs/v113/i04/707-714}, \href {https://ui.adsabs.harvard.edu/abs/2017CSci..113..707G} {113, 707}

\bibitem[\protect\citeauthoryear{{Haidar} et~al.,}{{Haidar} et~al.}{2024}]{haidar}
{Haidar} H.,  et~al., 2024, \mn@doi [\mnras] {10.1093/mnras/stae1596}, \href {https://ui.adsabs.harvard.edu/abs/2024MNRAS.532.4645H} {532, 4645}

\bibitem[\protect\citeauthoryear{{Hale} et~al.,}{{Hale} et~al.}{2021}]{Hale_2021}
{Hale} C.~L.,  et~al., 2021, \mn@doi [\pasa] {10.1017/pasa.2021.47}, \href {https://ui.adsabs.harvard.edu/abs/2021PASA...38...58H} {38, e058}

\bibitem[\protect\citeauthoryear{{Hamann} et~al.,}{{Hamann} et~al.}{2017}]{hamann}
{Hamann} F.,  et~al., 2017, \mn@doi [\mnras] {10.1093/mnras/stw2387}, \href {https://ui.adsabs.harvard.edu/abs/2017MNRAS.464.3431H} {464, 3431}

\bibitem[\protect\citeauthoryear{{Hancock}, {Sadler}, {Mahony}  \& {Ricci}}{{Hancock} et~al.}{2010}]{hancock}
{Hancock} P.~J.,  {Sadler} E.~M.,  {Mahony} E.~K.,   {Ricci} R.,  2010, \mn@doi [\mnras] {10.1111/j.1365-2966.2010.17199.x}, \href {https://ui.adsabs.harvard.edu/abs/2010MNRAS.408.1187H} {408, 1187}

\bibitem[\protect\citeauthoryear{Hardcastle \& Croston}{Hardcastle \& Croston}{2020}]{hardcastle}
Hardcastle M.,  Croston J.,  2020, \mn@doi [\nar] {https://doi.org/10.1016/j.newar.2020.101539}, 88, 101539

\bibitem[\protect\citeauthoryear{{Hardcastle} et~al.,}{{Hardcastle} et~al.}{2023}]{lotss_opt}
{Hardcastle} M.~J.,  et~al., 2023, \mn@doi [\aap] {10.1051/0004-6361/202347333}, \href {https://ui.adsabs.harvard.edu/abs/2023A&A...678A.151H} {678, A151}

\bibitem[\protect\citeauthoryear{{Harrison} \& {Ramos Almeida}}{{Harrison} \& {Ramos Almeida}}{2024}]{harrison_2024}
{Harrison} C.~M.,  {Ramos Almeida} C.,  2024, arXiv e-prints, \href {https://ui.adsabs.harvard.edu/abs/2024arXiv240408050H} {p. arXiv:2404.08050}

\bibitem[\protect\citeauthoryear{{Hayashi}, {Doi}  \& {Nagai}}{{Hayashi} et~al.}{2024}]{hayashi}
{Hayashi} T.~J.,  {Doi} A.,   {Nagai} H.,  2024, \mn@doi [arXiv e-prints] {10.48550/arXiv.2404.07726}, \href {https://ui.adsabs.harvard.edu/abs/2024arXiv240407726H} {p. arXiv:2404.07726}

\bibitem[\protect\citeauthoryear{{Helfand}, {White}  \& {Becker}}{{Helfand} et~al.}{2015}]{helfand}
{Helfand} D.~J.,  {White} R.~L.,   {Becker} R.~H.,  2015, \mn@doi [\apj] {10.1088/0004-637X/801/1/26}, \href {https://ui.adsabs.harvard.edu/abs/2015ApJ...801...26H} {801, 26}

\bibitem[\protect\citeauthoryear{{H{\"o}nig}}{{H{\"o}nig}}{2019}]{honig}
{H{\"o}nig} S.~F.,  2019, \mn@doi [\apj] {10.3847/1538-4357/ab4591}, \href {https://ui.adsabs.harvard.edu/abs/2019ApJ...884..171H} {884, 171}

\bibitem[\protect\citeauthoryear{{Hopkins}, {Hernquist}, {Cox}, {Di Matteo}, {Robertson}  \& {Springel}}{{Hopkins} et~al.}{2006}]{hop6}
{Hopkins} P.~F.,  {Hernquist} L.,  {Cox} T.~J.,  {Di Matteo} T.,  {Robertson} B.,   {Springel} V.,  2006, \mn@doi [\apjs] {10.1086/499298}, \href {https://ui.adsabs.harvard.edu/\#abs/2006ApJS..163....1H} {163, 1}

\bibitem[\protect\citeauthoryear{{Hopkins}, {Hernquist}, {Cox}  \& {Kere{\v{s}}}}{{Hopkins} et~al.}{2008}]{hop}
{Hopkins} P.~F.,  {Hernquist} L.,  {Cox} T.~J.,   {Kere{\v{s}}} D.,  2008, \mn@doi [\apjs] {10.1086/524362}, \href {https://ui.adsabs.harvard.edu/\#abs/2008ApJS..175..356H} {175, 356}

\bibitem[\protect\citeauthoryear{{Hotan} et~al.,}{{Hotan} et~al.}{2021}]{askap2}
{Hotan} A.~W.,  et~al., 2021, \mn@doi [\pasa] {10.1017/pasa.2021.1}, \href {https://ui.adsabs.harvard.edu/abs/2021PASA...38....9H} {38, e009}

\bibitem[\protect\citeauthoryear{Hwang, Zakamska, Alexandroff, Hamann, Greene, Perrotta  \& Richards}{Hwang et~al.}{2018}]{hwang}
Hwang H.-C.,  Zakamska N.~L.,  Alexandroff R.~M.,  Hamann F.,  Greene J.~E.,  Perrotta S.,   Richards G.~T.,  2018, \mn@doi [\mnras] {10.1093/mnras/sty742}, 477, 830

\bibitem[\protect\citeauthoryear{{Intema}, {Jagannathan}, {Mooley}  \& {Frail}}{{Intema} et~al.}{2017}]{tgss}
{Intema} H.~T.,  {Jagannathan} P.,  {Mooley} K.~P.,   {Frail} D.~A.,  2017, \mn@doi [A\&A] {10.1051/0004-6361/201628536}, \href {https://ui.adsabs.harvard.edu/abs/2017A&A...598A..78I} {598, A78}

\bibitem[\protect\citeauthoryear{{Ishibashi} \& {Fabian}}{{Ishibashi} \& {Fabian}}{2015}]{ishibashi}
{Ishibashi} W.,  {Fabian} A.~C.,  2015, \mn@doi [\mnras] {10.1093/mnras/stv944}, \href {https://ui.adsabs.harvard.edu/abs/2015MNRAS.451...93I} {451, 93}

\bibitem[\protect\citeauthoryear{{Ishibashi}, {Fabian}  \& {Maiolino}}{{Ishibashi} et~al.}{2018}]{ishibashi_18}
{Ishibashi} W.,  {Fabian} A.~C.,   {Maiolino} R.,  2018, \mn@doi [\mnras] {10.1093/mnras/sty236}, \href {https://ui.adsabs.harvard.edu/abs/2018MNRAS.476..512I} {476, 512}

\bibitem[\protect\citeauthoryear{{Jarvis} et~al.,}{{Jarvis} et~al.}{2019}]{jarvis}
{Jarvis} M.~E.,  et~al., 2019, \mn@doi [\mnras] {10.1093/mnras/stz556}, \href {https://ui.adsabs.harvard.edu/abs/2019MNRAS.485.2710J} {485, 2710}

\bibitem[\protect\citeauthoryear{{Jeyakumar}}{{Jeyakumar}}{2016}]{jeyakumar_2016}
{Jeyakumar} S.,  2016, \mn@doi [\mnras] {10.1093/mnras/stw181}, \href {https://ui.adsabs.harvard.edu/abs/2016MNRAS.458.3786J} {458, 3786}

\bibitem[\protect\citeauthoryear{{Johnston} et~al.,}{{Johnston} et~al.}{2007}]{askap1}
{Johnston} S.,  et~al., 2007, \mn@doi [\pasa] {10.1071/AS07033}, \href {https://ui.adsabs.harvard.edu/abs/2007PASA...24..174J} {24, 174}

\bibitem[\protect\citeauthoryear{{Jurlin}, {Morganti}, {Sweijen}, {Morabito}, {Brienza}, {Barthel}  \& {Miley}}{{Jurlin} et~al.}{2024}]{jurlin}
{Jurlin} N.,  {Morganti} R.,  {Sweijen} F.,  {Morabito} L.~K.,  {Brienza} M.,  {Barthel} P.,   {Miley} G.~K.,  2024, \mn@doi [\aap] {10.1051/0004-6361/202245821}, \href {https://ui.adsabs.harvard.edu/abs/2024A&A...682A.118J} {682, A118}

\bibitem[\protect\citeauthoryear{{Kale} \& {Ishwara-Chandra}}{{Kale} \& {Ishwara-Chandra}}{2021}]{capture}
{Kale} R.,  {Ishwara-Chandra} C.~H.,  2021, \mn@doi [Experimental Astronomy] {10.1007/s10686-020-09677-6}, \href {https://ui.adsabs.harvard.edu/abs/2021ExA....51...95K} {51, 95}

\bibitem[\protect\citeauthoryear{{Kellermann}, {Sramek}, {Schmidt}, {Shaffer}  \& {Green}}{{Kellermann} et~al.}{1989}]{kellermann}
{Kellermann} K.~I.,  {Sramek} R.,  {Schmidt} M.,  {Shaffer} D.~B.,   {Green} R.,  1989, \mn@doi [\aj] {10.1086/115207}, \href {https://ui.adsabs.harvard.edu/abs/1989AJ.....98.1195K} {98, 1195}

\bibitem[\protect\citeauthoryear{{Kerrison}, {Allison}, {Moss}, {Sadler}  \& {Rees}}{{Kerrison} et~al.}{2024}]{kerrison}
{Kerrison} E.~F.,  {Allison} J.~R.,  {Moss} V.~A.,  {Sadler} E.~M.,   {Rees} G.~A.,  2024, \mn@doi [arXiv e-prints] {10.48550/arXiv.2407.16201}, \href {https://ui.adsabs.harvard.edu/abs/2024arXiv240716201K} {p. arXiv:2407.16201}

\bibitem[\protect\citeauthoryear{{Kharb}, {Srivastava}, {Singh}, {Gallimore}, {Ishwara-Chandra}  \& {Ananda}}{{Kharb} et~al.}{2016}]{kharb_2016}
{Kharb} P.,  {Srivastava} S.,  {Singh} V.,  {Gallimore} J.~F.,  {Ishwara-Chandra} C.~H.,   {Ananda} H.,  2016, \mn@doi [\mnras] {10.1093/mnras/stw699}, \href {https://ui.adsabs.harvard.edu/abs/2016MNRAS.459.1310K} {459, 1310}

\bibitem[\protect\citeauthoryear{{Kharb}, {Subramanian}, {Das}, {Vaddi}  \& {Paragi}}{{Kharb} et~al.}{2021}]{Kharb2021}
{Kharb} P.,  {Subramanian} S.,  {Das} M.,  {Vaddi} S.,   {Paragi} Z.,  2021, \mn@doi [\apj] {10.3847/1538-4357/ac0c82}, \href {https://ui.adsabs.harvard.edu/abs/2021ApJ...919..108K} {919, 108}

\bibitem[\protect\citeauthoryear{{Kim} \& {Im}}{{Kim} \& {Im}}{2018}]{kim18}
{Kim} D.,  {Im} M.,  2018, \mn@doi [\aap] {10.1051/0004-6361/201731963}, \href {https://ui.adsabs.harvard.edu/\#abs/2018A&A...610A..31K} {610, A31}

\bibitem[\protect\citeauthoryear{{Kim}, {Im}, {Kim}, {Kim}, {Shin}, {Shim}  \& {Song}}{{Kim} et~al.}{2023}]{kim_2023}
{Kim} D.,  {Im} M.,  {Kim} M.,  {Kim} Y.,  {Shin} S.,  {Shim} H.,   {Song} H.,  2023, \mn@doi [\apj] {10.3847/1538-4357/aceb5e}, \href {https://ui.adsabs.harvard.edu/abs/2023ApJ...954..156K} {954, 156}

\bibitem[\protect\citeauthoryear{{Kim}, {Kim}, {Im}, {Glikman}, {Kim}, {Urrutia}  \& {Lim}}{{Kim} et~al.}{2024a}]{kim_2024_redd}
{Kim} D.,  {Kim} Y.,  {Im} M.,  {Glikman} E.,  {Kim} M.,  {Urrutia} T.,   {Lim} G.,  2024a, \mn@doi [arXiv e-prints] {10.48550/arXiv.2408.03324}, \href {https://ui.adsabs.harvard.edu/abs/2024arXiv240803324K} {p. arXiv:2408.03324}

\bibitem[\protect\citeauthoryear{{Kim}, {Kim}, {Im}  \& {Kim}}{{Kim} et~al.}{2024b}]{kim_2024_dust}
{Kim} Y.,  {Kim} D.,  {Im} M.,   {Kim} M.,  2024b, \mn@doi [\apj] {10.3847/1538-4357/ad5d5d}, \href {https://ui.adsabs.harvard.edu/abs/2024ApJ...972..171K} {972, 171}

\bibitem[\protect\citeauthoryear{Kimball, Kellermann, Condon, Ivezi{\'{c}}  \& Perley}{Kimball et~al.}{2011}]{Kimball_2011}
Kimball A.~E.,  Kellermann K.~I.,  Condon J.~J.,  Ivezi{\'{c}} {\v{Z}}.,   Perley R.~A.,  2011, \mn@doi [\apj] {10.1088/2041-8205/739/1/l29}, 739, L29

\bibitem[\protect\citeauthoryear{Klindt, Alexander, Rosario, Lusso  \& Fotopoulou}{Klindt et~al.}{2019}]{klindt}
Klindt L.,  Alexander D.~M.,  Rosario D.~J.,  Lusso E.,   Fotopoulou S.,  2019, \mn@doi [\mnras] {10.1093/mnras/stz1771}, 488, 3109

\bibitem[\protect\citeauthoryear{{Kukreti} \& {Morganti}}{{Kukreti} \& {Morganti}}{2024}]{kukreti_2024}
{Kukreti} P.,  {Morganti} R.,  2024, \mn@doi [arXiv e-prints] {10.48550/arXiv.2407.06265}, \href {https://ui.adsabs.harvard.edu/abs/2024arXiv240706265K} {p. arXiv:2407.06265}

\bibitem[\protect\citeauthoryear{{Kukreti}, {Morganti}, {Tadhunter}  \& {Santoro}}{{Kukreti} et~al.}{2023}]{kukreti}
{Kukreti} P.,  {Morganti} R.,  {Tadhunter} C.,   {Santoro} F.,  2023, \mn@doi [\aap] {10.1051/0004-6361/202245691}, \href {https://ui.adsabs.harvard.edu/abs/2023A&A...674A.198K} {674, A198}

\bibitem[\protect\citeauthoryear{{Lacy} et~al.,}{{Lacy} et~al.}{2020}]{lacy_vlass}
{Lacy} M.,  et~al., 2020, \mn@doi [\pasp] {10.1088/1538-3873/ab63eb}, \href {https://ui.adsabs.harvard.edu/abs/2020PASP..132c5001L} {132, 035001}

\bibitem[\protect\citeauthoryear{{L{\"a}hteenm{\"a}ki} \& {Valtaoja}}{{L{\"a}hteenm{\"a}ki} \& {Valtaoja}}{1999}]{lahteenm}
{L{\"a}hteenm{\"a}ki} A.,  {Valtaoja} E.,  1999, \mn@doi [\apj] {10.1086/307587}, \href {https://ui.adsabs.harvard.edu/abs/1999ApJ...521..493L} {521, 493}

\bibitem[\protect\citeauthoryear{{Laor}, {Baldi}  \& {Behar}}{{Laor} et~al.}{2019}]{laor}
{Laor} A.,  {Baldi} R.~D.,   {Behar} E.,  2019, \mn@doi [\mnras] {10.1093/mnras/sty3098}, \href {https://ui.adsabs.harvard.edu/abs/2019MNRAS.482.5513L} {482, 5513}

\bibitem[\protect\citeauthoryear{{Leftley}, {H{\"o}nig}, {Asmus}, {Tristram}, {Gandhi}, {Kishimoto}, {Venanzi}  \& {Williamson}}{{Leftley} et~al.}{2019}]{leftley_19}
{Leftley} J.~H.,  {H{\"o}nig} S.~F.,  {Asmus} D.,  {Tristram} K. R.~W.,  {Gandhi} P.,  {Kishimoto} M.,  {Venanzi} M.,   {Williamson} D.~J.,  2019, \mn@doi [\apj] {10.3847/1538-4357/ab4a0b}, \href {https://ui.adsabs.harvard.edu/abs/2019ApJ...886...55L} {886, 55}

\bibitem[\protect\citeauthoryear{{Liu}, {Zakamska}, {Greene}, {Nesvadba}  \& {Liu}}{{Liu} et~al.}{2013}]{liu_2013}
{Liu} G.,  {Zakamska} N.~L.,  {Greene} J.~E.,  {Nesvadba} N. P.~H.,   {Liu} X.,  2013, \mn@doi [\mnras] {10.1093/mnras/stt1755}, \href {https://ui.adsabs.harvard.edu/abs/2013MNRAS.436.2576L} {436, 2576}

\bibitem[\protect\citeauthoryear{{Macfarlane} et~al.,}{{Macfarlane} et~al.}{2021}]{macfarlane}
{Macfarlane} C.,  et~al., 2021, \mn@doi [MNRAS] {10.1093/mnras/stab1998}, \href {https://ui.adsabs.harvard.edu/abs/2021MNRAS.506.5888M} {506, 5888}

\bibitem[\protect\citeauthoryear{{Mahony} et~al.,}{{Mahony} et~al.}{2016}]{mahony}
{Mahony} E.~K.,  et~al., 2016, \mn@doi [\mnras] {10.1093/mnras/stw2225}, \href {https://ui.adsabs.harvard.edu/abs/2016MNRAS.463.2997M} {463, 2997}

\bibitem[\protect\citeauthoryear{{McConnell} et~al.,}{{McConnell} et~al.}{2020}]{racs}
{McConnell} D.,  et~al., 2020, \mn@doi [\pasa] {10.1017/pasa.2020.41}, \href {https://ui.adsabs.harvard.edu/abs/2020PASA...37...48M} {37, e048}

\bibitem[\protect\citeauthoryear{{Meenakshi}, {Mukherjee}, {Bodo}, {Rossi}  \& {Harrison}}{{Meenakshi} et~al.}{2024}]{meenakshi}
{Meenakshi} M.,  {Mukherjee} D.,  {Bodo} G.,  {Rossi} P.,   {Harrison} C.~M.,  2024, arXiv e-prints, \href {https://ui.adsabs.harvard.edu/abs/2024arXiv240800099M} {p. arXiv:2408.00099}

\bibitem[\protect\citeauthoryear{{Mehdipour} \& {Costantini}}{{Mehdipour} \& {Costantini}}{2019}]{med}
{Mehdipour} M.,  {Costantini} E.,  2019, \mn@doi [\aap] {10.1051/0004-6361/201935205}, \href {https://ui.adsabs.harvard.edu/abs/2019A&A...625A..25M} {625, A25}

\bibitem[\protect\citeauthoryear{{Morabito} et~al.,}{{Morabito} et~al.}{2019}]{morabito}
{Morabito} L.~K.,  et~al., 2019, \mn@doi [\aap] {10.1051/0004-6361/201833821}, \href {https://ui.adsabs.harvard.edu/abs/2019A&A...622A..15M} {622, A15}

\bibitem[\protect\citeauthoryear{Mukherjee, Bicknell, Wagner, Sutherland  \& Silk}{Mukherjee et~al.}{2018}]{mukherjee}
Mukherjee D.,  Bicknell G.~V.,  Wagner A.~Y.,  Sutherland R.~S.,   Silk J.,  2018, \mn@doi [\mnras] {10.1093/mnras/sty1776}, 479, 5544

\bibitem[\protect\citeauthoryear{{Nair} et~al.,}{{Nair} et~al.}{2024}]{dhanya}
{Nair} D.~G.,  et~al., 2024, \mn@doi [arXiv e-prints] {10.48550/arXiv.2409.15587}, \href {https://ui.adsabs.harvard.edu/abs/2024arXiv240915587N} {p. arXiv:2409.15587}

\bibitem[\protect\citeauthoryear{Nims, Quataert  \& Faucher-Giguère}{Nims et~al.}{2015}]{nims}
Nims J.,  Quataert E.,   Faucher-Giguère C.-A.,  2015, \mn@doi [\mnras] {10.1093/mnras/stu2648}, 447, 3612

\bibitem[\protect\citeauthoryear{{Njeri}, {Deane}, {Radcliffe}, {Beswick}, {Thomson}, {Muxlow}, {Garrett}  \& {Harrison}}{{Njeri} et~al.}{2024}]{njeri}
{Njeri} A.,  {Deane} R.~P.,  {Radcliffe} J.~F.,  {Beswick} R.~J.,  {Thomson} A.~P.,  {Muxlow} T.~W.~B.,  {Garrett} M.~A.,   {Harrison} C.~M.,  2024, \mn@doi [\mnras] {10.1093/mnras/stae381}, \href {https://ui.adsabs.harvard.edu/abs/2024MNRAS.528.6141N} {528, 6141}

\bibitem[\protect\citeauthoryear{{Nyland} et~al.,}{{Nyland} et~al.}{2020}]{Nyland2020}
{Nyland} K.,  et~al., 2020, \mn@doi [\apj] {10.3847/1538-4357/abc341}, \href {https://ui.adsabs.harvard.edu/abs/2020ApJ...905...74N} {905, 74}

\bibitem[\protect\citeauthoryear{{O'Dea} \& {Baum}}{{O'Dea} \& {Baum}}{1997}]{odea_1997}
{O'Dea} C.~P.,  {Baum} S.~A.,  1997, \mn@doi [\aj] {10.1086/118241}, \href {https://ui.adsabs.harvard.edu/abs/1997AJ....113..148O} {113, 148}

\bibitem[\protect\citeauthoryear{{O'Dea} \& {Saikia}}{{O'Dea} \& {Saikia}}{2021}]{o_dea_21}
{O'Dea} C.~P.,  {Saikia} D.~J.,  2021, \mn@doi [\aapr] {10.1007/s00159-021-00131-w}, \href {https://ui.adsabs.harvard.edu/abs/2021A&ARv..29....3O} {29, 3}

\bibitem[\protect\citeauthoryear{{O'Dea}, {Baum}  \& {Stanghellini}}{{O'Dea} et~al.}{1991}]{odea_1991}
{O'Dea} C.~P.,  {Baum} S.~A.,   {Stanghellini} C.,  1991, \mn@doi [\apj] {10.1086/170562}, \href {https://ui.adsabs.harvard.edu/abs/1991ApJ...380...66O} {380, 66}

\bibitem[\protect\citeauthoryear{{Orienti}}{{Orienti}}{2016}]{orienti_16}
{Orienti} M.,  2016, \mn@doi [Astronomische Nachrichten] {10.1002/asna.201512257}, \href {https://ui.adsabs.harvard.edu/abs/2016AN....337....9O} {337, 9}

\bibitem[\protect\citeauthoryear{Orienti \& Dallacasa}{Orienti \& Dallacasa}{2013}]{gps}
Orienti M.,  Dallacasa D.,  2013, \mn@doi [MNRAS] {10.1093/mnras/stt2217}, 438, 463

\bibitem[\protect\citeauthoryear{{Panessa}, {Baldi}, {Laor}, {Padovani}, {Behar}  \& {McHardy}}{{Panessa} et~al.}{2019}]{pan}
{Panessa} F.,  {Baldi} R.~D.,  {Laor} A.,  {Padovani} P.,  {Behar} E.,   {McHardy} I.,  2019, \mn@doi [Nature] {10.1038/s41550-019-0765-4}, \href {https://ui.adsabs.harvard.edu/abs/2019NatAs...3..387P} {3, 387}

\bibitem[\protect\citeauthoryear{{Panessa} et~al.,}{{Panessa} et~al.}{2022}]{panessa_22}
{Panessa} F.,  et~al., 2022, \mn@doi [\mnras] {10.1093/mnras/stac1745}, \href {https://ui.adsabs.harvard.edu/abs/2022MNRAS.515..473P} {515, 473}

\bibitem[\protect\citeauthoryear{Patil et~al.,}{Patil et~al.}{2020}]{Patil_2020}
Patil P.,  et~al., 2020, \mn@doi [\apj] {10.3847/1538-4357/ab9011}, 896, 18

\bibitem[\protect\citeauthoryear{{Patil} et~al.,}{{Patil} et~al.}{2022}]{Patil_2022}
{Patil} P.,  et~al., 2022, \mn@doi [\apj] {10.3847/1538-4357/ac71b0}, \href {https://ui.adsabs.harvard.edu/abs/2022ApJ...934...26P} {934, 26}

\bibitem[\protect\citeauthoryear{Perley \& Butler}{Perley \& Butler}{2017}]{Perley_2017}
Perley R.~A.,  Butler B.~J.,  2017, \mn@doi [\apjs] {10.3847/1538-4365/aa6df9}, 230, 7

\bibitem[\protect\citeauthoryear{{Petley} et~al.,}{{Petley} et~al.}{2022}]{petley}
{Petley} J.~W.,  et~al., 2022, \mn@doi [\mnras] {10.1093/mnras/stac2067}, \href {https://ui.adsabs.harvard.edu/abs/2022MNRAS.515.5159P} {515, 5159}

\bibitem[\protect\citeauthoryear{{Petley} et~al.,}{{Petley} et~al.}{2024}]{petley_2024}
{Petley} J.~W.,  et~al., 2024, \mn@doi [\mnras] {10.1093/mnras/stae626}, \href {https://ui.adsabs.harvard.edu/abs/2024MNRAS.529.1995P} {529, 1995}

\bibitem[\protect\citeauthoryear{{Phillips} \& {Mutel}}{{Phillips} \& {Mutel}}{1982}]{phillips}
{Phillips} R.~B.,  {Mutel} R.~L.,  1982, \aap, \href {https://ui.adsabs.harvard.edu/abs/1982A&A...106...21P} {106, 21}

\bibitem[\protect\citeauthoryear{{Planck Collaboration} et~al.,}{{Planck Collaboration} et~al.}{2020}]{planck_2020}
{Planck Collaboration} et~al., 2020, \mn@doi [\aap] {10.1051/0004-6361/201833910}, \href {https://ui.adsabs.harvard.edu/abs/2020A&A...641A...6P} {641, A6}

\bibitem[\protect\citeauthoryear{{Prevot}, {Lequeux}, {Maurice}, {Prevot}  \& {Rocca-Volmerange}}{{Prevot} et~al.}{1984}]{SMC}
{Prevot} M.~L.,  {Lequeux} J.,  {Maurice} E.,  {Prevot} L.,   {Rocca-Volmerange} B.,  1984, \aap, \href {https://ui.adsabs.harvard.edu/abs/1984A&A...132..389P} {132, 389}

\bibitem[\protect\citeauthoryear{{Rakshit}, {Stalin}  \& {Kotilainen}}{{Rakshit} et~al.}{2020}]{rakshit}
{Rakshit} S.,  {Stalin} C.~S.,   {Kotilainen} J.,  2020, \mn@doi [\apjs] {10.3847/1538-4365/ab99c5}, \href {https://ui.adsabs.harvard.edu/abs/2020ApJS..249...17R} {249, 17}

\bibitem[\protect\citeauthoryear{{Rankine}, {Hewett}, {Banerji}  \& {Richards}}{{Rankine} et~al.}{2020}]{rankine}
{Rankine} A.~L.,  {Hewett} P.~C.,  {Banerji} M.,   {Richards} G.~T.,  2020, \mn@doi [\mnras] {10.1093/mnras/staa130}, \href {https://ui.adsabs.harvard.edu/abs/2020MNRAS.492.4553R} {492, 4553}

\bibitem[\protect\citeauthoryear{{Rengelink}, {Tang}, {de Bruyn}, {Miley}, {Bremer}, {Roettgering}  \& {Bremer}}{{Rengelink} et~al.}{1997}]{wenss}
{Rengelink} R.~B.,  {Tang} Y.,  {de Bruyn} A.~G.,  {Miley} G.~K.,  {Bremer} M.~N.,  {Roettgering} H.~J.~A.,   {Bremer} M.~A.~R.,  1997, \mn@doi [\aaps] {10.1051/aas:1997358}, \href {https://ui.adsabs.harvard.edu/abs/1997A&AS..124..259R} {124, 259}

\bibitem[\protect\citeauthoryear{{Richards} et~al.,}{{Richards} et~al.}{2003}]{richards}
{Richards} G.~T.,  et~al., 2003, \mn@doi [\aj] {10.1086/377014}, \href {https://ui.adsabs.harvard.edu/abs/2003AJ....126.1131R} {126, 1131}

\bibitem[\protect\citeauthoryear{{Richards} et~al.,}{{Richards} et~al.}{2006}]{richards_2006}
{Richards} G.~T.,  et~al., 2006, \mn@doi [\aj] {10.1086/503559}, \href {https://ui.adsabs.harvard.edu/abs/2006AJ....131.2766R} {131, 2766}

\bibitem[\protect\citeauthoryear{{Richards} et~al.,}{{Richards} et~al.}{2011}]{richards_2011}
{Richards} G.~T.,  et~al., 2011, \mn@doi [\aj] {10.1088/0004-6256/141/5/167}, \href {https://ui.adsabs.harvard.edu/abs/2011AJ....141..167R} {141, 167}

\bibitem[\protect\citeauthoryear{{Rosario}, {Fawcett}, {Klindt}, {Alexander}, {Morabito}, {Fotopoulou}, {Lusso}  \& {Calistro Rivera}}{{Rosario} et~al.}{2020}]{rosario}
{Rosario} D.~J.,  {Fawcett} V.~A.,  {Klindt} L.,  {Alexander} D.~M.,  {Morabito} L.~K.,  {Fotopoulou} S.,  {Lusso} E.,   {Calistro Rivera} G.,  2020, \mn@doi [\mnras] {10.1093/mnras/staa866}, \href {https://ui.adsabs.harvard.edu/abs/2020MNRAS.494.3061R} {494, 3061}

\bibitem[\protect\citeauthoryear{{Rosario}, {Alexander}, {Moldon}, {Klindt}, {Thomson}, {Morabito}, {Fawcett}  \& {Harrison}}{{Rosario} et~al.}{2021}]{rosario_21}
{Rosario} D.~J.,  {Alexander} D.~M.,  {Moldon} J.,  {Klindt} L.,  {Thomson} A.~P.,  {Morabito} L.,  {Fawcett} V.~A.,   {Harrison} C.~M.,  2021, \mn@doi [MNRAS] {10.1093/mnras/stab1653}, \href {https://ui.adsabs.harvard.edu/abs/2021MNRAS.505.5283R} {505, 5283}

\bibitem[\protect\citeauthoryear{{Rose}}{{Rose}}{2014}]{rose_2014}
{Rose} M.,  2014, in American Astronomical Society Meeting Abstracts \#223. p. 321.03

\bibitem[\protect\citeauthoryear{{Rose}, {Tadhunter}, {Holt}  \& {Rodr{\'\i}guez Zaur{\'\i}n}}{{Rose} et~al.}{2013}]{rose}
{Rose} M.,  {Tadhunter} C.~N.,  {Holt} J.,   {Rodr{\'\i}guez Zaur{\'\i}n} J.,  2013, \mn@doi [\mnras] {10.1093/mnras/stt564}, \href {https://ui.adsabs.harvard.edu/abs/2013MNRAS.432.2150R} {432, 2150}

\bibitem[\protect\citeauthoryear{{Rybicki} \& {Lightman}}{{Rybicki} \& {Lightman}}{1979}]{rybicki}
{Rybicki} G.~B.,  {Lightman} A.~P.,  1979, {Radiative processes in astrophysics}.
{Wiley-VCH}

\bibitem[\protect\citeauthoryear{{Schneider} et~al.,}{{Schneider} et~al.}{2010}]{dr7}
{Schneider} D.~P.,  et~al., 2010, \mn@doi [\aj] {10.1088/0004-6256/139/6/2360}, \href {https://ui.adsabs.harvard.edu/abs/2010AJ....139.2360S} {139, 2360}

\bibitem[\protect\citeauthoryear{{Schwarz}}{{Schwarz}}{1978}]{bic}
{Schwarz} G.,  1978, Annals of Statistics, \href {https://ui.adsabs.harvard.edu/abs/1978AnSta...6..461S} {6, 461}

\bibitem[\protect\citeauthoryear{{Shao}, {Wagg}, {Wang}, {Carilli}, {Riechers}, {Intema}, {Weiss}  \& {Menten}}{{Shao} et~al.}{2020}]{shao_20}
{Shao} Y.,  {Wagg} J.,  {Wang} R.,  {Carilli} C.~L.,  {Riechers} D.~A.,  {Intema} H.~T.,  {Weiss} A.,   {Menten} K.~M.,  2020, \mn@doi [\aap] {10.1051/0004-6361/202038469}, \href {https://ui.adsabs.harvard.edu/abs/2020A&A...641A..85S} {641, A85}

\bibitem[\protect\citeauthoryear{{Shao} et~al.,}{{Shao} et~al.}{2022}]{shao}
{Shao} Y.,  et~al., 2022, \mn@doi [\aap] {10.1051/0004-6361/202142489}, \href {https://ui.adsabs.harvard.edu/abs/2022A&A...659A.159S} {659, A159}

\bibitem[\protect\citeauthoryear{Shen \& Liu}{Shen \& Liu}{2012}]{Shen_2012}
Shen Y.,  Liu X.,  2012, \mn@doi [ApJ] {10.1088/0004-637x/753/2/125}, 753, 125

\bibitem[\protect\citeauthoryear{{Shimwell} et~al.,}{{Shimwell} et~al.}{2017}]{lotss}
{Shimwell} T.~W.,  et~al., 2017, \mn@doi [A\&A] {10.1051/0004-6361/201629313}, 598, A104

\bibitem[\protect\citeauthoryear{{Shimwell} et~al.,}{{Shimwell} et~al.}{2022}]{lotssdr2}
{Shimwell} T.~W.,  et~al., 2022, \mn@doi [\aap] {10.1051/0004-6361/202142484}, \href {https://ui.adsabs.harvard.edu/abs/2022A&A...659A...1S} {659, A1}

\bibitem[\protect\citeauthoryear{{Silpa}, {Kharb}, {Ho}, {Ishwara-Chandra}, {Jarvis}  \& {Harrison}}{{Silpa} et~al.}{2020}]{Silpa2020}
{Silpa} S.,  {Kharb} P.,  {Ho} L.~C.,  {Ishwara-Chandra} C.~H.,  {Jarvis} M.~E.,   {Harrison} C.,  2020, \mn@doi [\mnras] {10.1093/mnras/staa2970}, \href {https://ui.adsabs.harvard.edu/abs/2020MNRAS.499.5826S} {499, 5826}

\bibitem[\protect\citeauthoryear{{Sinha}, {Mangla}  \& {Datta}}{{Sinha} et~al.}{2023}]{sinha}
{Sinha} A.,  {Mangla} S.,   {Datta} A.,  2023, \mn@doi [JA\&A] {10.1007/s12036-023-09978-0}, \href {https://ui.adsabs.harvard.edu/abs/2023JApA...44...88S} {44, 88}

\bibitem[\protect\citeauthoryear{{Snellen}, {Schilizzi}, {de Bruyn}, {Miley}, {Rengelink}, {Roettgering}  \& {Bremer}}{{Snellen} et~al.}{1998}]{snellen_98}
{Snellen} I.~A.~G.,  {Schilizzi} R.~T.,  {de Bruyn} A.~G.,  {Miley} G.~K.,  {Rengelink} R.~B.,  {Roettgering} H.~J.,   {Bremer} M.~N.,  1998, \mn@doi [\aaps] {10.1051/aas:1998281}, \href {https://ui.adsabs.harvard.edu/abs/1998A&AS..131..435S} {131, 435}

\bibitem[\protect\citeauthoryear{Snellen, Schilizzi, Miley, de Bruyn, Bremer  \& Röttgering}{Snellen et~al.}{2000}]{snellen}
Snellen I. A.~G.,  Schilizzi R.~T.,  Miley G.~K.,  de Bruyn A.~G.,  Bremer M.~N.,   Röttgering H. J.~A.,  2000, \mn@doi [\mnras] {10.1111/j.1365-8711.2000.03935.x}, 319, 445

\bibitem[\protect\citeauthoryear{{Stawarz}, {Ostorero}, {Begelman}, {Moderski}, {Kataoka}  \& {Wagner}}{{Stawarz} et~al.}{2008}]{stawarz}
{Stawarz} {\L}.,  {Ostorero} L.,  {Begelman} M.~C.,  {Moderski} R.,  {Kataoka} J.,   {Wagner} S.,  2008, \mn@doi [\apj] {10.1086/587781}, \href {https://ui.adsabs.harvard.edu/abs/2008ApJ...680..911S} {680, 911}

\bibitem[\protect\citeauthoryear{{Stepney}, {Banerji}, {Tang}, {Hewett}, {Temple}, {Wethers}, {Puglisi}  \& {Molyneux}}{{Stepney} et~al.}{2024}]{stepney}
{Stepney} M.,  {Banerji} M.,  {Tang} S.,  {Hewett} P.~C.,  {Temple} M.~J.,  {Wethers} C.~F.,  {Puglisi} A.,   {Molyneux} S.~J.,  2024, \mn@doi [\mnras] {10.1093/mnras/stae1970}, \href {https://ui.adsabs.harvard.edu/abs/2024MNRAS.tmp.1919S} {}

\bibitem[\protect\citeauthoryear{{Sun}, {Greene}  \& {Zakamska}}{{Sun} et~al.}{2017}]{sun}
{Sun} A.-L.,  {Greene} J.~E.,   {Zakamska} N.~L.,  2017, \mn@doi [\apj] {10.3847/1538-4357/835/2/222}, \href {https://ui.adsabs.harvard.edu/abs/2017ApJ...835..222S} {835, 222}

\bibitem[\protect\citeauthoryear{{Swarup}, {Ananthakrishnan}, {Kapahi}, {Rao}, {Subrahmanya}  \& {Kulkarni}}{{Swarup} et~al.}{1991}]{gmrt2}
{Swarup} G.,  {Ananthakrishnan} S.,  {Kapahi} V.~K.,  {Rao} A.~P.,  {Subrahmanya} C.~R.,   {Kulkarni} V.~K.,  1991, Current Science, \href {https://ui.adsabs.harvard.edu/abs/1991CSci...60...95S} {60, 95}

\bibitem[\protect\citeauthoryear{{Temple} et~al.,}{{Temple} et~al.}{2023}]{temple_23}
{Temple} M.~J.,  et~al., 2023, arXiv e-prints, \href {https://ui.adsabs.harvard.edu/abs/2023arXiv230102675T} {p. arXiv:2301.02675}

\bibitem[\protect\citeauthoryear{Urrutia, {Lacy}, {Spoon}, {Glikman}, {Petric}  \& {Schulz}}{Urrutia et~al.}{2012}]{urrutia12}
Urrutia T.,  {Lacy} M.,  {Spoon} H.,  {Glikman} E.,  {Petric} A.,   {Schulz} B.,  2012, \mn@doi [\apj] {10.1088/0004-637X/757/2/125}, \href {https://ui.adsabs.harvard.edu/abs/2012ApJ...757..125U} {757, 125}

\bibitem[\protect\citeauthoryear{{Urry} \& {Padovani}}{{Urry} \& {Padovani}}{1995}]{urry}
{Urry} C.~M.,  {Padovani} P.,  1995, \mn@doi [\pasp] {10.1086/133630}, \href {https://ui.adsabs.harvard.edu/abs/1995PASP..107..803U} {107, 803}

\bibitem[\protect\citeauthoryear{{Venanzi}, {H{\"o}nig}  \& {Williamson}}{{Venanzi} et~al.}{2020}]{marta_20}
{Venanzi} M.,  {H{\"o}nig} S.,   {Williamson} D.,  2020, \mn@doi [\apj] {10.3847/1538-4357/aba89f}, \href {https://ui.adsabs.harvard.edu/abs/2020ApJ...900..174V} {900, 174}

\bibitem[\protect\citeauthoryear{{Venturi} et~al.,}{{Venturi} et~al.}{2021}]{venturi}
{Venturi} G.,  et~al., 2021, \mn@doi [A\&A] {10.1051/0004-6361/202039869}, 648, A17

\bibitem[\protect\citeauthoryear{{Wagner} \& {Witzel}}{{Wagner} \& {Witzel}}{1995}]{wagner2}
{Wagner} S.~J.,  {Witzel} A.,  1995, \mn@doi [\araa] {10.1146/annurev.aa.33.090195.001115}, \href {https://ui.adsabs.harvard.edu/abs/1995ARA&A..33..163W} {33, 163}

\bibitem[\protect\citeauthoryear{Webster, Francis, Petersont, Drinkwater  \& Masci}{Webster et~al.}{1995}]{Webster1995}
Webster R.~L.,  Francis P.~J.,  Petersont B.~A.,  Drinkwater M.~J.,   Masci F.~J.,  1995, \mn@doi [Nature] {10.1038/375469a0}, 375, 469

\bibitem[\protect\citeauthoryear{{Weymann}, {Morris}, {Foltz}  \& {Hewett}}{{Weymann} et~al.}{1991}]{weymann}
{Weymann} R.~J.,  {Morris} S.~L.,  {Foltz} C.~B.,   {Hewett} P.~C.,  1991, \mn@doi [\apj] {10.1086/170020}, \href {https://ui.adsabs.harvard.edu/abs/1991ApJ...373...23W} {373, 23}

\bibitem[\protect\citeauthoryear{{Whiting}, {Webster}  \& {Francis}}{{Whiting} et~al.}{2001}]{whiting}
{Whiting} M.~T.,  {Webster} R.~L.,   {Francis} P.~J.,  2001, \mn@doi [\mnras] {10.1046/j.1365-8711.2001.04287.x}, \href {https://ui.adsabs.harvard.edu/abs/2001MNRAS.323..718W} {323, 718}

\bibitem[\protect\citeauthoryear{{Wilkes}, {Schmidt}, {Cutri}, {Ghosh}, {Hines}, {Nelson}  \& {Smith}}{{Wilkes} et~al.}{2002}]{wilkes}
{Wilkes} B.~J.,  {Schmidt} G.~D.,  {Cutri} R.~M.,  {Ghosh} H.,  {Hines} D.~C.,  {Nelson} B.,   {Smith} P.~S.,  2002, \mn@doi [\apjl] {10.1086/338908}, \href {https://ui.adsabs.harvard.edu/abs/2002ApJ...564L..65W} {564, L65}

\bibitem[\protect\citeauthoryear{{Wo{\l}owska} et~al.,}{{Wo{\l}owska} et~al.}{2021}]{wolowska}
{Wo{\l}owska} A.,  et~al., 2021, \mn@doi [\apj] {10.3847/1538-4357/abe62d}, \href {https://ui.adsabs.harvard.edu/abs/2021ApJ...914...22W} {914, 22}

\bibitem[\protect\citeauthoryear{{Yamada}, {Sakai}, {Inoue}  \& {Michiyama}}{{Yamada} et~al.}{2024}]{yamada}
{Yamada} T.,  {Sakai} N.,  {Inoue} Y.,   {Michiyama} T.,  2024, \mn@doi [\apj] {10.3847/1538-4357/ad3a63}, \href {https://ui.adsabs.harvard.edu/abs/2024ApJ...968..116Y} {968, 116}

\bibitem[\protect\citeauthoryear{{Young}, {Elvis}  \& {Risaliti}}{{Young} et~al.}{2008}]{young}
{Young} M.,  {Elvis} M.,   {Risaliti} G.,  2008, \mn@doi [\apj] {10.1086/592083}, \href {https://ui.adsabs.harvard.edu/abs/2008ApJ...688..128Y} {688, 128}

\bibitem[\protect\citeauthoryear{{Young}, {Turner}, {Shabala}, {Stewart}  \& {Yates-Jones}}{{Young} et~al.}{2024}]{young_24}
{Young} S.~A.,  {Turner} R.~J.,  {Shabala} S.~S.,  {Stewart} G. S.~C.,   {Yates-Jones} P.~M.,  2024, \mn@doi [arXiv e-prints] {10.48550/arXiv.2412.14433}, \href {https://ui.adsabs.harvard.edu/abs/2024arXiv241214433Y} {p. arXiv:2412.14433}

\bibitem[\protect\citeauthoryear{{Yue} et~al.,}{{Yue} et~al.}{2024}]{bohan}
{Yue} B.~H.,  et~al., 2024, \mn@doi [\mnras] {10.1093/mnras/stae725}, \href {https://ui.adsabs.harvard.edu/abs/2024MNRAS.529.3939Y} {529, 3939}

\bibitem[\protect\citeauthoryear{{Zaja{\v{c}}ek} et~al.,}{{Zaja{\v{c}}ek} et~al.}{2019}]{radio_slope}
{Zaja{\v{c}}ek} M.,  et~al., 2019, \mn@doi [\aap] {10.1051/0004-6361/201833388}, \href {https://ui.adsabs.harvard.edu/abs/2019A&A...630A..83Z} {630, A83}

\bibitem[\protect\citeauthoryear{{de Gasperin} et~al.,}{{de Gasperin} et~al.}{2023}]{lolss}
{de Gasperin} F.,  et~al., 2023, \mn@doi [\aap] {10.1051/0004-6361/202245389}, \href {https://ui.adsabs.harvard.edu/abs/2023A&A...673A.165D} {673, A165}

\bibitem[\protect\citeauthoryear{{van Breugel}, {Miley}  \& {Heckman}}{{van Breugel} et~al.}{1984}]{van_bregel}
{van Breugel} W.,  {Miley} G.,   {Heckman} T.,  1984, \mn@doi [\aj] {10.1086/113480}, \href {https://ui.adsabs.harvard.edu/abs/1984AJ.....89....5V} {89, 5}

\bibitem[\protect\citeauthoryear{{van Haarlem} et~al.,}{{van Haarlem} et~al.}{2013}]{lofar}
{van Haarlem} M.~P.,  et~al., 2013, \mn@doi [\aap] {10.1051/0004-6361/201220873}, \href {https://ui.adsabs.harvard.edu/abs/2013A&A...556A...2V} {556, A2}

\makeatother
\end{thebibliography}



\bsp	
\label{lastpage}
\end{document}